\documentclass[11pt]{article}

\usepackage{draft}

\title{Scattering Amplitudes For All Masses and Spins}

\author{Nima Arkani-Hamed$^1$,}

\author{Tzu-Chen Huang$^{2}$,}
\author{Yu-tin Huang$^{3,4}$}

\affiliation{$^1$ School of Natural Sciences, Institute for Advanced Study, Princeton, NJ 08540, USA}

\affiliation{$^2$ Walter Burke Institute for Theoretical Physics
, California Institute of Technology, Pasadena, CA 91125, USA
}
\affiliation{$^3$Department of Physics and Astronomy, National Taiwan University, Taipei 10617, Taiwan}
\affiliation{$^4$Physics Division, National Center for Theoretical Sciences, National Tsing-Hua University,
No.101, Section 2, Kuang-Fu Road, Hsinchu, Taiwan} 

\abstract{We introduce a formalism for describing four-dimensional scattering
amplitudes for particles of any mass and spin. This naturally extends
the familiar spinor-helicity formalism for massless particles to one
where these variables carry an extra $SU(2)$ little group index for
massive particles, with the amplitudes for spin $S$ particles
transforming as symmetric rank $2S$ tensors. We systematically
characterise all possible three particle amplitudes compatible with
Poincare symmetry. Unitarity, in the form of consistent factorization,
imposes algebraic conditions that can be used to construct all
possible four-particle tree amplitudes. This also gives us a
convenient basis in which to expand
all possible four-particle amplitudes in terms of what can be called
``spinning polynomials". Many general results of quantum field
theory follow the analysis of four-particle scattering,  ranging from
the set of all possible consistent theories for massless particles, to
spin-statistics, and the Weinberg-Witten theorem. We
also find a transparent understanding for why massive particles of sufficiently
high spin cannot be ``elementary". The Higgs and Super-Higgs
mechanisms are naturally discovered as an infrared unification of many
disparate helicity amplitudes into a smaller
number of massive amplitudes, with a simple understanding for why this can't be extended to Higgsing for
gravitons. We illustrate a number of applications of the formalism at
one-loop, giving few-line computations of the electron $(g-2)$ as well as the
beta function and rational terms in QCD. ``Off-shell" observables like
correlation functions and form-factors can be thought of as scattering
amplitudes with external ``probe" particles of general mass and spin,
so all these objects---amplitudes, form factors and correlators, can be
studied from a common on-shell perspective.}

\begin{document}
\begin{flushright}
\vspace{10pt} \hfill{NCTS-TH/1714} \vspace{20mm}
\end{flushright}
\maketitle

\section{Scattering Amplitudes in the Real World}

Recent years have seen an explosion of progress in our understanding of scattering amplitudes in gauge theories and gravity. Infinite classes of amplitudes,  whose computation would have seemed unthinkable even ten years ago, can now be  derived with pen and paper on the back of an envelope using a set of ideas broadly referred to as ``on-shell methods"~\cite{TreeRec, BDK}. This has enabled the determination of scattering amplitudes of direct interest to collider physics experiments, while at the same time opening up novel directions of theoretical research into the foundations of quantum field theory,  amongst other things revealing surprising and deep connections of this basic physics with areas of mathematics ranging from algebraic geometry to combinatorics to number theory.

Almost all of the major progress in this field has been in understanding scattering amplitudes for massless particles. There are seemingly good reasons for this, both technically and conceptually. Technically, almost all treatments of the subject, especially in four dimensions, involve the introduction of special variables (such as spinor-helicity, twistor or momentum-twistor variables) to trivialise the kinematical on-shell constraints for massless particles (see~\cite{Reviews} for a comprehensive review). And conceptually, while it is clear that the conventional field-theoretic description of {\it massless} particles with spin, which involves the introduction of huge gauge redundancy,  leaves ample room for improvement---provided by on-shell methods that directly describe particles, eliminating any reference to quantum fields and their attendant redundancies---the advantage of ``on-shell physics" seems to disappear for the case of massive particles where no gauge redundancies are needed.  

As we will see, the technical issue about massless kinematics is just that---the transition to describing massive particles is a triviality---while the conceptual issue is not an obstacle but rather an invitation to understand the both the physics of ``infrared deformation" of massless theories (by the Higgs mechanism and confinement), as well that of UV completion (such as with perturbative string theory), from a new on-shell perspective (see sec.\ref{Higgs}).

But before getting too far ahead of ourselves it suffices to remember that the only exactly massless particles we know of in the real world are photons and gravitons; even the spectacular success of on-shell methods applied to collider physics are for high energy gluon collisions, which are ultimately confined into massive hadrons at long distances. Even if we consider  the weakly coupled scattering amplitudes for Standard Model particles above the QCD scale,  almost all the particles are massive. If the amazing structures unearthed in the study of gauge and gravity scattering amplitudes are indeed an indication of a radical new way of thinking about quantum particle interactions in space-time, they must naturally extend beyond photons, gravitons and gluons to electrons, $W,Z$ particles and top quarks as well.

Keeping this central motivation in mind, in this paper we initiate a systematic exploration of the physics of scattering amplitudes in four dimensions, for particles of general masses and spins. We proceed in sec.\ref{LittleGroup} with an on-shell formalism where the amplitude is manifestly covariant under the massive SU(2) little group. This approach allows us to cleanly categorize all distinct three-couplings for a given set of helicities or masses and spins. When constructing four-point amplitudes, this formalism sharply pinpoints the tension between locality and consistent factorization, which, in turn provides a portal into the difficulty of having higher-spin massive particles that is fundamental. As we will see, everything that is typically taught in an introductory courses on QFT and the Standard Model---including classic computations of the electron $(g-2)$ and the QCD $\beta$ function (sec.\ref{LoopApply})---can be transparently reproduced from an on-shell perspective directly following from the physics of Poincare invariance, locality and unitarity, without ever encountering quantum fields, Lagrangians, gauge and diff invariance,  or Feynman rules.

There are a number of other motivations for developing this formalism. For instance, much of the remarkable progress in our understanding of the dynamics of supersymmetric gauge theories came from exploring their moduli spaces of vacua~\cite{SimplestQFT}. From this point of view the study of massless scattering amplitudes has been stuck on a desert island at the origin of moduli space; we should now be able to study how the $S$-matrix varies on moduli space in general supersymmetric theories, especially beginning with the Coulomb branch of ${\cal N}=4$ SYM in the planar limit (see~\cite{Craig:2011ws} for early surveys).

Another motivation, alluded to above, is the physics of UV completion for gravity scattering amplitudes.  It is easy to show on general grounds that any weakly coupled UV completion for gravity amplitudes must involve an infinite tower of particles with infinitely increasing spins (as of course seen in string theory)~\cite{stringpaper}. This raises the possibility that string theory might be derivable from the bottom-up, as the unique weakly-coupled UV completion of gravity. But it has become clear that consistency conditions for massless graviton scattering alone are not enough to uniquely fix amplitudes---deformations of the graviton scattering amplitudes compatible with all the standard rules have been identified ( eq.(12.6) in~\cite{stringpaper}). This is not surprising, since the most extreme tension in this physics is the coexistence of gravitons with massive higher-spin particles. Indeed (as we will review in \ref{interactionswithother} from an on-shell perspective) the presence of gravity makes the existence of {\it massless} higher-spin particles impossible. We should therefore expect the strongest consistency conditions on perturbative UV completion to involve the scattering of massless gravitons and massive higher-spin particles, the study of which calls for a good general formalism for treating amplitudes for general mass and spin.

Finally, an understanding of amplitudes for general mass and spin removes the distinction between ``on-shell" observables like scattering amplitudes and ``off-shell" observables like correlation functions~\cite{Engelund:2012re}. After all, loosely speaking the way experimentalists actually measure correlation functions of some system is to weakly couple the system to massive detectors, and effectively measure the scattering amplitudes for the detectors thought of as massive particles with general mass and spin! More precisely, as we demonstrate in sex.\ref{Form}, to compute the correlation functions for (say) the stress tensor (in momentum-space), we need only imagine weakly coupling a continuum of massive spin 2 particle to the system with a universal (and arbitrarily weak) coupling; the leading scattering amplitudes for these massive particles is then literally the correlation function for the stress tensor in momentum space. This should allow us to explore both on- and off-shell physics in a uniform ``on-shell" way.

\section{The Little Group}\label{LittleGroup}
Much of the non-trivial physics of scattering amplitudes traces back to the simple question---``what is a particle?"---and the attendant concept of Wigner's ``little group"  governing the kinematics of particle scattering. Let us review this standard story. Following Wigner (and Weinberg's exposition and notation)~\cite{Wigner, Weinberg:1995mt}, we think of ``particles" as irreducible unitary representations of the Poincare group. We diagonalize the translation operator by labelling particles with their momentum $p^\mu$; any other labels a particle state can carry are labelled by $\sigma$. In order to systematically label all one-particle states, we start with some reference momentum $k_\mu$ and the states $|k, \sigma \rangle$. Now, we can write any momentum $p$ as a specified Lorentz-transformation $L(p;k)$ acting on $k$, i.e.
$p_\mu = L_\mu^\nu(p;k) k_\nu$. Note that $L(p;k)$ is not unique since there are clearly Lorentz transformations that leave $p$ invariant---these ``little group" transformations will figure prominently in what follows, for now we simply emphasize that we pick some specific $L(p;k)$ for which $p=L(p;k) k$. We also assume that we have a unitary representation of the Lorentz group, i.e. for every Lorentz transformation $\Lambda$ there is an associated unitary operator $U(\Lambda)$ acting on the Hilbert space, such that $U(\Lambda_1 \Lambda_2) = U(\Lambda_1) U(\Lambda_2)$.  Then we simply {\it define} one-particle states $|p,\sigma \rangle$ as
\begin{equation}
|p,\sigma \rangle \equiv U(L(p;k)) |k,\sigma \rangle\,.
\end{equation}
Note that the $\sigma$ index is the same on the left and the right, this is the sense in which we are {\it defining} $|p,\sigma\rangle$. Having made this definition, we can ask how $|p,\sigma\rangle$ transforms under a general Lorentz transformation
\begin{equation}
U(\Lambda)|p,\sigma\rangle = U(\Lambda) U(L(p;k))|k,\sigma \rangle = U(L(\Lambda p;k)) U(L^{-1}(\Lambda p;k) \Lambda L(p;k)) |k,\sigma \rangle\,.
\end{equation}
Now, $W(\Lambda,p,k) = L^{-1}(\Lambda p;k) \Lambda L(p;k)$ is not in general a trivial Lorentz transformation, it is only a transformation that leave $k$ invariant since clearly $(W k) = k$. This subgroup of the Lorentz group is the ``little group". Thus, we must have that
\begin{equation}
U(W(\Lambda,p;k)) |k,\sigma \rangle = D_{\sigma \sigma^\prime}(W(\Lambda,p;k)) |k,\sigma^\prime \rangle\,,
\end{equation}
where $D_{\sigma \sigma^\prime}(W)$ is a representation of the little group. We have therefore found the desired transformation property
\begin{equation}
U(\Lambda)|p,\sigma \rangle = D_{\sigma \sigma^\prime}(W(\Lambda,p;k))|\Lambda p,\sigma^\prime\rangle\,.
\end{equation}
We conclude that a particle is labeled by its momentum and transforms under some representation of the little group. 

Scattering amplitudes for $n$ particles are thus labeled by $(p_a,\sigma_a)$ for $a=1,\cdots,n$. The Poincare invariance of the S-matrix ---translation and Lorentz invariance---then tells us that 
\begin{eqnarray}
{\cal M}(p_a,\sigma_a) &=& \delta^{D}(p_{a_1}^\mu + \cdots p_{a_n}^\mu) M(p_a,\sigma_a) \nonumber \\
M^{\Lambda}(p_a,\sigma_a) &=&\prod_a \left(D_{\sigma_a \sigma_a^\prime}(W) \right)  M((\Lambda p)_a,\sigma_a^\prime)\,.
\end{eqnarray}
In $D$ spacetime dimensions, the little group for massive particles is $SO(D{-}1)$. For massless particles the little group is the the group of Euclidean symmetries in $(D{-}2)$ dimensions, which is $SO(D{-}2)$ augmented by $(D{-}2)$ translations. Finite-dimensional representations require choosing all states to have vanishing eigenvalues under these translations, and hence the little group is just $SO(D{-}2)$.

So much for the basic kinematics of particle scattering amplitudes. It is when we come to dynamics, and in particular to the crucial question of guaranteeing that the physics of particle interactions is compatible with the most minimal notion of locality encoded in the principle of cluster decomposition, that a fateful decision is made to choose a particular description of particle scattering, introducing the idea of quantum fields. Beyond particles of spin zero (and their associated scalar fields), there is a basic kinematical awkwardness associated with introducing fields: fields are manifestly ``off-shell", and transform as Lorentz tensors (or spinors), while particle states transform instead under the little group. The objects we compute directly with Feynman diagrams in quantum field theory, which are Lorentz tensors, have the wrong transformation properties to be called ``amplitudes". This is why we introduce the idea of ``polarisation vectors", that are meant to transform as bi-fundamentals under the Lorentz and little group, to convert ``Feynman amplitudes" to the actual ``scattering amplitudes". For instance in the case of spin 1 particles, we introduce $\epsilon_\sigma^\mu(p)$, with the property that $\epsilon_\sigma^\mu(\Lambda p) =
\Lambda^\mu_\nu \epsilon^\nu_{\sigma^\prime}(p) D_{\sigma \sigma^\prime}(W)$, so that $\epsilon^\mu_\sigma(p) M_\mu(p, \cdots)$ transforms properly. For massive particles, such polarization vectors certainly exist, though they have to satisfy constraints. For instance we must have $p_\mu \epsilon^\mu_\sigma=0$ for massive spin 1, or for massive spin 1/2, we use a Dirac spinor $\Psi_\sigma^A$ with $(\Gamma^\mu p_\mu - m)_B^A \Psi^B = 0$. These constraints are an artifact of using {\it fields} as auxiliary objects to describe the interactions of the more fundamental {\it particles}. For massless particles with spin $\geq 1$ the  situation is worse, since ``polarisation vectors" transforming as bi-fundamentals under the Lorentz and little groups don't exist. Say for massless particles in four dimensions, if we make some choice for the $\epsilon^\mu_{\pm}$ for photons of helicity $\pm 1$, we find that for Lorentz transformations $(\Lambda p) = p$, $(\Lambda \epsilon_\pm)^\mu = e^{\pm i \theta} \epsilon_\pm^\mu + \alpha(\Lambda,p) p^\mu$. So polarisation vectors don't genuinely transform as vectors under Lorentz transformations, only the ``gauge equivalence class" $\{\epsilon_\pm^\mu | \epsilon_\pm^\mu + \alpha p^\mu\}$ is invariant under Lorentz transformations. This infinite redundancy is hard-wired into the usual field-theoretic description of scattering amplitudes for gauge bosons and gravitons, and is largely responsible for the apparent enormous complexity of amplitudes in these theories, obscuring the remarkable simplicity and hidden infinite-dimensional symmetries actually found in the physics.

The modern on-shell approach to scattering amplitudes departs from the conventional approach to field theory already at this early kinematical stage, by directly working with objects that transform properly under the little group (and so at least kinematically deserve to be called ``scattering amplitudes") from the get-go. Auxiliary objects such as ``quantum fields" are never introduced and no polarization vectors are needed. It is maximally easy to do this in the $D=4$ spacetime dimensions of our world, where the kinematics is as simple as possible. Here the little groups are $SO(2) = U(1)$ for massless particles, and $SO(3) = SU(2)$ for massive particles, which are the simplest and most familiar Lie groups.

In four dimensions, we label massless particles by their helicity $h$. Massive particles transform as some spin $S$ representation of $SU(2)$. The conventional way of labelling spin states familiar from introductory quantum mechanics is by picking a spin axis $\hat{z}$. and giving the eigenvalue of $J_z$ in that direction. This is inconvenient for our purposes, since the introduction of the reference direction $\hat{z}$ breaks manifest rotational (not to speak of Lorentz) invariance. We will find it more convenient instead to label states of spin $S$ as a symmetric tensor of $SU(2)$ with rank $2S$; this entirely elementary group theory is reviewed in appendix \ref{SU(2)App}. Let's illustrate the labelling of states by considering a four-particle amplitudes where particles $1,2$ are massive with spin $1/2$ and $2$, and particles $3,4$ are massless with helicities $+3/2$ and $-1$.
This would be represented as an object
\begin{equation}
M^{\{I_1\},\{J_1,J_2,J_3,J_4\},\{+\frac{3}{2}\},\{-1\}}(p_1,p_2,p_3,p_4)
\end{equation}
where $\{I_1\},\{J_i\}$ are the little group indices of particle $1$ and $2$ respectively, and the amplitude transforms as
\begin{equation}
M^{\{I_1\},\{J_1,J_2,J_3,J_4\},\{+\frac{3}{2}\},\{-1\}} \to (W_{1 K_1}^{I_1}) (W_{2 L_1}^{J_1} \cdots W_{2 L_4}^{J_4}) (w_3)^3 (w_4)^{-2}  M^{\{K_1\},\{L_1,L_2,L_3,L_4\},\{+\frac{3}{2}\},\{-1\}}
\end{equation}
where the $W$ matrices are $SU(2)$ transformation in the spin $1/2$ representation and $w=e^{i \theta}$ is the massless little group phase factor for helicity $+1/2$.

\subsection{Massless and Massive Spinor-Helicity Variables}

Our next item of business is to find variables for the kinematics that hardwire these little group transformation laws, this will be simultaneously associated with convenient representations of the on-shell momenta. As usual we will use the $\sigma^\mu_{\alpha \dot{\alpha}}$ matrices to convert between four-momenta $p^\mu$ and the $2 \times 2$ matrix $p_{\alpha \dot{\alpha}} = p_\mu \sigma^{\mu}_{\alpha \dot{\alpha}}$\footnote{For our conventions of signature and spinor indices, see appendix \ref{Conventions}.}. Note that det$p_{\alpha \dot{\alpha}} = m^2$, so that there is an obvious difference between massless and massive particles.

For massless particles, we have det$p_{\alpha \dot{\alpha}} = 0$ and thus the matrix $p_{\alpha \dot{\alpha}}$ has rank 1. Thus we can write it as the direct product of two, 2-vectors $\lambda, \tilde{\lambda}$ as~\cite{SpinorHelicity}
\begin{equation}
p_{\alpha \dot{\alpha}} = \lambda_\alpha \tilde{\lambda}_{\dot{\alpha}}
\end{equation}
For general complex momenta the $\lambda_\alpha, \tilde{\lambda}_{\dot{\alpha}}$ are independent two-dimensional complex vectors. For real momenta in Minkowski space $p_{\alpha \dot{\alpha}}$ is Hermitian and so we have $\tilde{\lambda}_{\dot{\alpha}} = \pm (\lambda_{\alpha})^*$, (with the sign determined by whether the energy is taken to be positive or negative).

Often the introduction of these ``spinor-helicity" variables is motivated by the desire to explicitly represent the (on-shell constrained) four-momentum $p_{\alpha \dot{\alpha}}$ by the unconstrained $\lambda_\alpha, \tilde{\lambda}_{\dot{\alpha}}$. But the spinor-helicity variables also have another conceptually important role to play: they are the objects that transform nicely under both the Lorentz and Little groups. Thus while amplitudes for massless particles are {\it not} functions of momenta and polarization vectors (or better yet, are only redundantly represented in this way), they {\it are} directly functions of spinor-helicity variables.

The relation to the little group is clearly suggested by the fact that it is impossible to uniquely associate a pair $\lambda_{\alpha}, \tilde{\lambda}_{\dot{\alpha}}$ with some $p_{\alpha \dot{\alpha}}$, since we can always rescale $\lambda_\alpha \to w^{-1} \lambda_\alpha, \tilde{\lambda}_{\dot{\alpha}} \to w \tilde{\lambda}_{\dot{\alpha}}$ keeping $p_{\alpha \dot{\alpha}}$ invariant. The connection can be made completely explicit by attempting to give some specific prescription for picking $\lambda_\alpha^{(p)}, \tilde{\lambda}^{(p)}_{\dot{\alpha}}$, which leads us through an exercise completely parallel to our discussion of the little group. We first choose some reference massless momentum $k_{\alpha \dot{\alpha}}$ and also choose some fixed $\lambda_\alpha^{(k)}, \tilde{\lambda}_{\dot{\alpha}}^{(k)}$ so that $k_{\alpha \dot{\alpha}} = \lambda^{(k)}_\alpha \tilde{\lambda}^{(k)}_{\dot{\alpha}}$. For every other null momentum, we choose a Lorentz transformation ${\cal L}(p;k)_{\alpha}^{\beta}, \tilde{{\cal L}}(p;k)_{\dot{\alpha}}^{\dot{\beta}}$ such that
$p_{\alpha \dot{\alpha}} = {\cal L}(p;k)_{\alpha}^{\beta} \tilde{{\cal L}}(p;k)_{\dot{\alpha}}^{\dot{\beta}} k_{\beta \dot{\beta}}$, and we then {\it define} $\lambda_{\alpha}^{(p)} \equiv {\cal L}(p;k)_{\alpha}^{\beta} \lambda^{(k)}_\beta, \tilde{\lambda}^{(p)}_{\dot{\alpha}} \equiv \tilde{{\cal L}}(p;k)_{\dot{\alpha}}^{\dot{\beta}} \tilde{\lambda}^{(k)}_{\dot{\beta}}$.
Having now picked a way of associating some $\lambda_\alpha^{(p)}, \tilde{\lambda}^{(p)}_{\dot{\alpha}}$ with $p_{\alpha \dot{\alpha}}$, we can ask for the relationship between e.g. $\lambda_\alpha^{(\Lambda p)}$ and $\lambda^{(p)}_\alpha$ for some Lorentz transformation $\Lambda$; what we find is
\begin{equation}
\lambda_\alpha^{(\Lambda p)} = w^{-1}(\Lambda,p,k) \, \Lambda_\alpha^\beta \lambda_\beta^{(p)}
\end{equation}
For general complex momenta $w$ is simply a complex number and we have the action of $GL(1)$, for real Lorentzian momenta we must have $w^{-1} = \pm (w)^*$ so $w = e^{i \theta}$ is a phase representing the $U(1)$ little group. Most obviously we can perform a Lorentz transformation $W$ for which $W k = k$, we simply find $\lambda \to w^{-1} \lambda$. To be explicit, let
\begin{equation}
k_{\alpha \dot{\alpha}} = \left(\begin{array}{cc} 2 E & 0 \\ 0 & 0 \end{array} \right), \, \, \lambda_{\alpha} = \sqrt{2 E} \left(\begin{array}{c} 1 \\ 0 \end{array}\right), \tilde{\lambda}_{\dot{\alpha}} = \sqrt{2E} \left(\begin{array}{c} 1 \\ 0 \end{array} \right)
\end{equation}
represent a massless momentum in the $z$ direction. Then a rotation around the $z$ axis (which leaves $k$ invariant) is
\begin{equation}
\Lambda_{\alpha}^{\beta} = \left(\begin{array}{cc} e^{i \phi/2} & 0 \\ 0 & e^{-i \phi/2} \end{array} \right),
\tilde{\Lambda}_{\dot{\alpha}}^{\dot{\beta}} = \left(\begin{array}{cc} e^{-i \phi/2} & 0 \\ 0 & e^{i \phi/2} \end{array} \right)
\end{equation}
under which obviously $\lambda_{\alpha} \to e^{i \phi/2} \lambda_{\alpha}, \tilde{\lambda}_{\dot{\alpha}} \to e^{-i \phi/2} \tilde{\lambda}_{\dot{\alpha}}$.

To summarize, amplitudes for massless particles are Lorentz-invariant functions of $\lambda_\alpha,\tilde{\lambda}_{\dot{\alpha}}$ with the correct little-group helicity weights, 
\begin{equation}
M(w^{-1} \lambda, w \tilde{\lambda}) = w^{2 h} M(\lambda,\tilde{\lambda})
\end{equation}

We now turn to the case of massive particles. There is no essential difference with the massless case; we simply have the $p_{\alpha \dot{\alpha}}$ has rank two instead of rank one, and so can be written as the sum of two rank one matrices as
\begin{equation}
p_{\alpha \dot{\alpha}} = \lambda_\alpha^I \tilde{\lambda}_{\dot{\alpha} I}
\end{equation}
where $I=1,2$. Note that
\begin{equation}
p^2 = m^2 \rightarrow {\rm det} \lambda \times {\rm det} \tilde{\lambda} = m^2
\end{equation}
We can use this to set ${\rm det} \lambda = M, {\rm det} \tilde{\lambda} = \tilde{M}$ with $M \tilde{M} = m^2$. It is sometimes useful to keep the distinction between $M, \tilde{M}$, but for our purposes in this paper we will simply take $M = \tilde{M} = m$.
Of course $\lambda^{I}, \tilde{\lambda}_I$ can't uniquely be associated with a given $p$, we can perform an $SL(2)$ transformation $\lambda^I \to W^I_J \lambda^J, \tilde{\lambda}_I \to (W^{-1})_I^J \tilde{\lambda}_J$.
Note that we could extend this $SL(2)$ to a $GL(2)$ if we also allowed (opposite) rephrasings of the mass parameters $M, \tilde{M}$, but by making the choice $M=\tilde{M} = m$ does not allow this. This is not a disadvantage for our purposes, since the object $M/\tilde{M}$ transforms only under the $GL(1)$ part of the $GL(2)$ and can be used to uplift any $SL(2)$ invariant into a $GL(2)$ invariant if desired.

For real Lorentzian momenta we have $W$ should be in the $SU(2)$ subgroup of $SL(2)$ and gives us the action of the little group. We can make the connection explicit just as we did for the massless case, by defining $\lambda_\alpha^I, \tilde{\lambda}_{\dot{\alpha}I}$ for a reference momentum $k_{\alpha \dot{\alpha}}$ and boosting to define them for all momenta. A summary of this elementary kinematics is given in appendix B.

We conclude that that amplitudes for massive particles are Lorentz-invariant functions for $\lambda^I, \tilde{\lambda}_I$ which are symmetric rank $2S$ tensors $\{I_1, \cdots, I_{2S}\}$ for spin $S$ particles. Note that we can obviously use $\epsilon^{IJ},\epsilon_{IJ}$ to raise and lower indices so that we can e.g. write $p_{\alpha \dot{\alpha}} = \lambda_\alpha^I \tilde{\lambda}_{\dot{\alpha}}^J \epsilon_{IJ}$. Also note that clearly
\begin{equation}\label{Convert}
p_{\alpha \dot{\alpha}} \tilde{\lambda}^{\dot{\alpha} I} = m \, \lambda^I_\alpha\, , \,  p_{\alpha \dot{\alpha}} \lambda^{\alpha I} = -m \, \tilde{\lambda}_{\dot{\alpha}}^I
\end{equation}
If we combine $(\lambda^I_{\alpha}, \tilde{\lambda}^{\dot{\alpha} I})$ into a Dirac spinor $\Psi^I_A$, this is of course the Dirac equation $(\Gamma^\mu p_\mu - m)_A^B \Psi_B^I = 0$. But there is no particular reason for doing this in our formalism: even the usual (good) reason for introducing Dirac spinors---making parity manifest in theories which have a parity symmetry---can be more easily accomplished without using Dirac spinors in our approach. We will thus not encounter any $\Gamma$ matrices in our discussion. Note also that using $(p_{\alpha \dot{\alpha}}/m)$ allows to freely convert between $\lambda_{\alpha}^I$ and $\tilde{\lambda}_{\dot{\alpha}}^I$ variables. We will sometimes find it useful, especially in the context of the systematic classification of amplitude structures, to use this freedom in order to use e.g. only $\lambda_\alpha^I$ to describe a given massive particle. Then we can write the symmetric tensor as
\begin{equation}
M^{\{I_1 \cdots I_{2S}\}} = \lambda_{\alpha_1}^{I_1} \cdots \lambda_{\alpha_{2S}}^{I_{2S}} M^{\{\alpha_1 \cdots \alpha_{2S}\}}
\end{equation}
where $M^{\{\alpha_1 \cdots \alpha_{2S}\}}$ is totally symmetric in the $\alpha$ indices.\footnote{The amplitude as a function of massive spinors can be viewed as a natural consequence of choosing the space-cone gauge Feynman rules~\cite{Chalmers:2001cy}.}

Let us illustrate our notation for writing amplitudes by returning to the example of a four-particle amplitude with $(1,2)$ being massive with spin $(1/2,2)$, and $(3,4)$ massless with helicity $(+3/2)$ and $(-1)$. Let's give examples of ``legal" expressions for these amplitudes, that is objects with the correct little group transformation properties. Two possible terms are
\begin{equation}
[2^{J_1} 3] [2^{J_2} 3] [2^{J_3} 3] \left( \kappa \langle 1^{I_1} 2^{J_4} \rangle \langle 4| (p_1 p_2) |4 \rangle + \kappa^\prime \langle 4 1^{I_1} \rangle \langle 2^{J_4} 4 \rangle \right) + {\rm symmetrize \, in} \,  \{J_{1,2,3,4}\}
\end{equation}

It would clearly be notationally cumbersome to have our formulas littered with explicit $SU(2)$ little group indices, fortunately it is also entirely un-necessary to do so. We will simply denote the massive spinor helicity variables in {\bf BOLD}, and suppress the $SU(2)$ little group indices. Since these indices are completely symmetrized, putting them back in is completely trivial and unambiguous. In this way, we re-write the above expressions as
\begin{equation}
[{\bf 2} 3]^3 \left(\kappa \langle {\bf 1} {\bf 2} \rangle \langle 4|p_1 p_2 |4 \rangle + \kappa^\prime \langle 4 {\bf 1} \rangle \langle 4 {\bf 2} \rangle \right)
\end{equation}

We stress again that there is no notion of the usual ``helicity weight" little group for the massive particles; we can freely have expressions (as in the above) that from the viewpoint of massless amplitudes look like they are ``illegally" combining terms with different helicity weight. As we will later see this reflects a beautiful feature of this formalism, making it trivial to see how massive amplitudes decompose into the massless helicity amplitudes at very high energies.

We pause to note the relation between our discussion here and a route  to massive spinor-helicity variables taken by a number of other authors~\cite{MassiveSH}. This approach begins by noting that we can always represent $p_{\alpha \dot{\alpha}} = \lambda_\alpha \tilde{\lambda}_{\dot{\alpha}} {-} (m^2/\langle \lambda \eta \rangle[\tilde{\lambda} \tilde{\eta}]) \eta_\alpha \tilde{\eta}_{\dot{\alpha}}$, for some reference spinors $\eta,\tilde{\eta}$.\footnote{The formalism here obviously have some parallels with the 6D spinor-helicity formalism~\cite{6D, 6D1}, but here the little group is a single SU(2) instead of SU(2)$\times$SU(2) as in six-dimensions, and thus there are no ``unnecessary" symmetries. } The states are then labelled by giving the spin in the direction picked out by the lightlike directon $\eta \tilde{\eta}$. Of course this corresponds to a particular choice for our $(\lambda^I_{\alpha}, \tilde{\lambda}^I_{\dot{\alpha}})$, but making this choice at the very outset obscures the Lorentz and little group transformation properties of the amplitude. Practically speaking, given some formula written in terms of the $\lambda, \tilde{\lambda},\eta, \tilde{\eta}$, this makes it difficult to ascertain whether or not it is kinematically a legal expression for an amplitude, and thus the program of systematically classifying and constructing on-shell amplitudes is difficult to pursue in this formalism.

Let us further illustrate our notation by presenting some classic scattering amplitudes in these variables. We will simply state the results here and derive them from first-principles later in the paper; here we are only illustrating the notation and its utility for understanding the physics. Consider for instance the result for tree-level Compton scattering  $(1 2^- 3^+ 4)$ where particles $2,3$ are photons of helicity $(-,+)$ while $1,4$ are charged massive particles of spin $0,1/2,1$. The amplitudes are given by
\begin{equation}\label{ComptonGen}
M(1 2^- 3^+ 4) = \frac{g^2}{(s - m^2)(u- m^2)} \times \left\{\begin{array}{c} \langle 2| (p_1 - p_4)|3]^2 \, \, \quad \quad \quad\quad \quad\quad \quad\quad \quad\quad [{\rm spin}\, 0] \\ \langle 2| (p_1 - p_4)|3] \left(\langle \mathbf{1} 2 \rangle [\mathbf{4} 3] + \langle \mathbf{4} 2 \rangle [\mathbf{1} 3]\right) \, \,  \quad\, [{\rm spin} \, \frac{1}{2}] \\ \left(\langle \mathbf{1} 2 \rangle [\mathbf{4} 3] + \langle \mathbf{4} 2 \rangle [\mathbf{1} 3]\right)^2 \, \, \quad \quad\quad \quad\quad \quad\quad [{\rm spin} \, 1] \end{array} \right\}
\end{equation}
Note the absence of $\gamma$ matrices for the spin $1/2$ case---the common complaint amongst students first doing these computations---``why are we dragging around four-component objects when the electron has only two spin degrees of freedom?"---is entirely absent here. Similarly for the spin 1 case there are no polarization vectors. Indeed these expressions are the most compact representation for these amplitudes possible, directly in terms
of the physical degrees of freedom of the actual particles, with no reference to fields as auxiliary objects.

\subsection{The high-energy limit} 
It is very easy to relate the massive and massless spinor-helicity variables, and especially to take the high-energy limit of scattering amplitudes and see how massive amplitudes for particles with spin decompose into the different helicity components. 
To do so, we note that it is convenient to expand $\lambda_\alpha^I$ in a basis of two-dimensional vectors $\zeta^{\pm I}$ in the little-group space. In other words, we can expand
\begin{eqnarray}\label{HELimitDef}
\lambda_\alpha^I &=& \lambda_\alpha \zeta^{- I} + \eta_\alpha \zeta^{+ I} \nonumber \\ 
\tilde \lambda_{\dot{\alpha}}^I &=& \tilde \lambda_{\dot{\alpha}} \zeta^{+ I} + \tilde{\eta}_{\dot{\alpha}} \zeta^{- I}
\end{eqnarray}
where 
\begin{equation}
\epsilon_{IJ} \zeta^{+ I} \zeta^{- J} = 1, \langle \lambda \eta \rangle = m, [\tilde \lambda \tilde \eta] = m
\end{equation}
Note, as explicitly given in the kinematics Appendix C, in a given frame we naturally have $\zeta^{\pm I}$ as the eigenstates of spin 1/2 in the direction of the spatial momentum $\vec{p}$, and we can identify $\lambda_\alpha = \sqrt{E+p} \zeta^+_\alpha, \eta_\alpha = \sqrt{E - p} \zeta^-_\alpha$ and similarly $\tilde \lambda_{\dot{\alpha}} = \sqrt{E+p} \tilde{\zeta}^-_{\dot{\alpha}}, \tilde{\eta}_{\dot{\alpha}} = \sqrt{E-p}\tilde{\zeta}^+_{\dot{\alpha}}$. Clearly, in the high energy limit $\sqrt{E +
p} \to \sqrt{2E}$ while $\sqrt{E - p} \to m/\sqrt{2 E}$, so that both
$\eta, \tilde{\eta}$ are proportional to $m$ and vanish relative to
$\lambda, \tilde \lambda$. Said in a more Lorentz-invariant way, to take the high-energy limit we take
\begin{equation}
\eta_\alpha = m \hat{\eta}_\alpha,\; \tilde{\eta}_{\dot{\alpha}} = m \hat{\tilde{\eta}}_{\dot{\alpha}};\quad {\rm with}\, \langle \lambda \hat{\eta} \rangle = [\tilde \lambda \tilde{\hat{\eta}}] = 1
\end{equation}
with all dimensionless ratios of the form 
\begin{equation} 
\frac{m}{\langle \lambda_a \lambda_b \rangle}, \;\frac{m}{[\tilde \lambda_a \tilde \lambda_b ]} \to 0
\end{equation}
Note that any scattering amplitude naturally decomposes into different spins states in the spatial direction of motion, via 
\begin{equation} 
M^{I_1 \cdots I_{2S}} = \sum_h \left((\zeta^{+})^{S{+}h} (\zeta^{-})^{S{-}h}\right)^{I_1 \cdots I_{2S}} M_h(\lambda, \tilde \lambda; \eta, \tilde{\eta})
\end{equation}
where trivially
\begin{equation}
M_h(w^{-1} \lambda, w \tilde{\lambda}; w \eta, w^{-1} \tilde{\eta}) = w^{2 h } M_h(\lambda, \tilde \lambda; \eta, \tilde \eta)
\end{equation}
Thus, the different helicity components in the high-energy limit are just given by
\begin{equation}
{\rm Helicity} \, h \, {\rm component} = {\rm Lim}_{m \to 0} M_h(\lambda, \tilde \lambda; \eta = m \hat{\eta}, \tilde \eta= m \hat{\tilde \eta})
\end{equation}

As a simple exercise for taking the high-energy limit, let's consider the coupling of a massive vector to two massless scalars. This amplitude is simply: 
\eq
\frac{\langle\mathbf{3} 1\rangle\langle \mathbf{3} 2\rangle}{\langle 21\rangle}\,.
\eqe
Let us consider the high-energy limit of this amplitude. Substituting eq.(\ref{HELimitDef}), the $(-,0,+)$ component of the vector are separately given as 
\eqa
-:\quad \frac{\langle\mathbf{3} 1\rangle\langle \mathbf{3} 2\rangle}{\langle 21\rangle}\;&\underrightarrow{\rm \quad H. E. \quad}& \; \frac{\langle 3 1\rangle\langle 3 2\rangle}{\langle21\rangle} \nonumber\\
0:\quad \frac{\langle\mathbf{3} 1\rangle\langle \mathbf{3} 2\rangle}{\langle 21\rangle}\;&\underrightarrow{\rm \quad H. E. \quad}&\frac{(\langle\eta_3 1\rangle\langle 3 2\rangle+\langle\eta_3 2\rangle\langle 3 1\rangle)}{2\langle 21\rangle}\nonumber\\
+:\quad \frac{\langle\mathbf{3} 1\rangle\langle \mathbf{3} 2\rangle}{\langle 21\rangle}\;&\underrightarrow{\rm \quad H. E. \quad}&\frac{\langle\eta_3 1\rangle\langle \eta_{3} 2\rangle}{\langle 21\rangle}=\frac{[3|p_2| 1\rangle[3|p_1| 2\rangle}{m^2\langle 21\rangle}=\frac{[32][31]}{[21]}\;
\eqae
We see that only the plus and minus helicity amplitude survives, and as $\eta_3$ scales as $m$, the longitudinal mode is sub-leading in $m$.\footnote{These results can also be obtained by converting  the conventional polarization vector representation of the three particle amplitude to the massive spinor helicity basis. First, being a Lorentz vector and a symmetric tensor in SU(2), the on-shell form of the polarization vector is fixed to (see also~\cite{Guevara:2018wpp})
\eq\label{MassivePol}
\epsilon_{\alpha\dot{\alpha}}=\frac{\lambda_\alpha^{\{I_1}\tilde{\lambda}^{I_2\}}_{\dot{\alpha}}}{m}\,.
\eqe
Contracting with the momenta then converts the polarization vector to pure chiral indices, $\epsilon_{\alpha\beta}=\epsilon_{\alpha\dot{\alpha}}\frac{p^{\dot{\alpha}}\,_\beta}{m}$. Taking the high energy limit, one straight forwardly obtains the  three helicity sectors:
\eq
\epsilon_{\alpha\beta}^{-}=\frac{\la_\alpha\la_\beta}{m},\;\epsilon_{\alpha\beta}^{0}=\frac{\lambda_\alpha\eta_\beta+\eta_\alpha\lambda_\beta}{2m},\;\epsilon_{\alpha\beta}^{+}=\frac{\eta_\alpha\eta_\beta}{m}\,,
\eqe
in the chiral representation.}

Especially in the context of the rather degenerate kinematics of three particle amplitudes, simply setting the $\eta, \tilde{\eta} \to 0$ can give rise to $0/0$ ambiguities, and this proper definition of the high-energy limit we have specified should be used. But for more generic situations, and for any expressions that is manifestly smooth as $m \to 0$, we can simply set $\eta, \tilde \eta \to 0$ to take the high-energy limit.  There is an especially easy way of doing this with the ``{\bf BOLD}" notation we have introduced above, that shortcuts the need for any explicit expansion in terms of $\zeta^{\pm I}$ as we have indicated above. We simply unbold the characters!\footnote{This is analogous to the replacement of $k\rightarrow k^{\flat}$ in the massive spinor helicity formalism of~\cite{Kosower:2004yz}. } Let us illustrate how this works for the case of Compton scattering of a charged spin one particle in eq.(\ref{ComptonGen}), and see how the massive amplitude decomposes into its helicity constituents. Expanding out the square of the numerators we find
\begin{equation}
\begin{array}{ccccc}
\langle 1 2 \rangle^2 [3 4]^2 &+& \langle 4 2 \rangle^2 [3 1]^2 &+& 2 \langle 1 2 \rangle [3 4] \langle 4 2 \rangle [3 1] \\
(1,4) \, {\rm have \, hel.} (-1,+1) & & (1,4) \, {\rm have \, hel.} (+1,-1) & & (1,4) \, {\rm have \, hel.} (0,0) \end{array}
\end{equation}
Note that as helicity amplitudes ``adding" the components in this way would be illegal, but this is exactly how we can pick out the different pieces of the massive amplitude that unifies the different helicity amplitudes together into a single object, in the high-energy limit! Note also that quite nicely the $(0,0)$ helicity components reproduce the HE limit of the scalar Compton amplitude, reflecting the fact that the longitudinal component of the charged massive spin 1 particle is just a charged scalar at high energies.

\section{Massless Three- and Four-Particle Amplitudes}\label{Massless34}
Having dispensed with kinematics, we now move on to determining
dynamics. We will follow a familiar strategy, starting by determining
the structure of all possible three-particle amplitudes:

\begin{equation}
\includegraphics[scale=0.5]{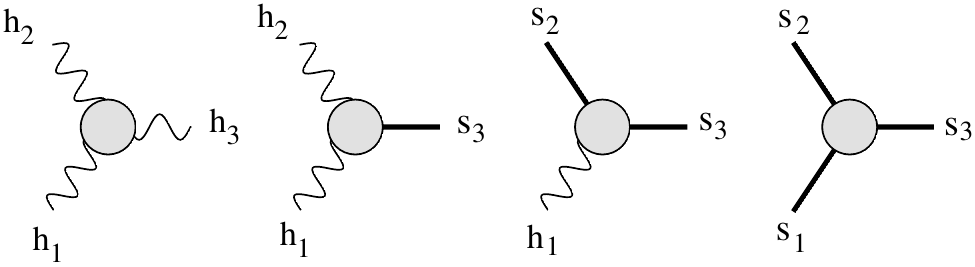}
\end{equation}
When many species $N_{s,m}$ of particle of the identical mass and
spin/helicity, we will label them with an index $``a"$.We will always think of these as real particles, and assume that the
``free propagation" does not change the $a$ index, i.e. that free
propagation has an $SO(N_{s,m})$ symmetry. This choice is hardwiring
the most basic physics of unitarity. Note that it is trivial to have
(non-unitary) Lagrangian theories that violate this rule, for instance
we can have grassmann scalar fields $\psi_a$ with free action $J^{ab}
\partial_\mu \psi_a \partial^\mu \psi_b$ with antisymmetric $J_{ab}$.
Here the free propagation is proportional to $J^{-1}_{ab}$ which
vanishes for $a=b$, and the free theory has an $Sp(N)$ rather than
$SO(N)$ symmetry.

Moving beyond three particles, the central constraint on higher-point
tree amplitudes is unitarity, in the form of consistent factorization.
For massless or massive internal particles goes on shell, spin $s$
goes on-shell, we must have

\begin{equation}
M \to \frac{M_{L}^{a \, h} M_{R}^{a \, -h}}{P^2} \, [{\rm massless}], \, M
\to \frac{M_{L}^{a \, \{I_1 \cdots I_{2s}\}} M^{a}_{R \, \{I_1
\cdots I_{2s}\}}}{P^2 -M^2} \, [{\rm massive}]\,.
\end{equation}
We will impose this consistency condition at 4 points, which must
factorize onto a product of three-particle amplitudes.

As is by now
well-known, these conditions are incredibly restrictive for massless
particles. The kinematics of three-particle momentum conservation forces either $\lambda_1, \lambda_2, \lambda_3$ to be all proportional, or $\tilde{\lambda}_1, \tilde{\lambda}_2, \tilde{\lambda}_3$ to all be proportional. Thus the three-particle amplitudes must either be of the form $[12]^a [23]^b [31]^c$ or $\langle 1 2 \rangle^a \langle 2 3 \rangle^b \langle 3 1 \rangle^c$ in these two cases respectively, and the powers are fixed by the helicities of the three particles. The amplitudes are given by
\begin{equation}\label{Massless3}
M^{h_1 h_2 h_3} = \begin{array}{c} \tilde{g} [12]^{h_1 + h_2 - h_3} [23]^{h_2 + h_3 - h_1} [31]^{h_3 + h_1 - h_2} \, {\rm when} \, h_1 + h_2 + h_3 > 0 \\ g \langle 1 2 \rangle^{h_3 - h_1 - h_2} \langle 2 3 \rangle^{h_1 - h_2 - h_3} \langle 3 1 \rangle^{h_2 - h_3 - h_1} \, {\rm when} \, h_1 + h_2 + h_3 < 0 \end{array}\,.
\end{equation}
Note that only by symmetries we could use either of the two expression regardless of the sign of $h_1 + h_2 + h_3$, but we also demand that the amplitudes have a smooth limit in Minkowski signature where the brackets also go to zero. We see that, up to the overall couplings $g, \tilde{g}$, the three-particle amplitudes are entirely fixed by Poincare symmetry. 

We now move on to determining four-particle amplitudes from consistent factorization. The obvious strategy for doing this is to simply compute the residue in e.g. the $s$-channel by gluing together the three particle amplitudes on the two sides of the channel; then multiply this residue by $1/s$. Adding over the channels should then give us an object that factors correctly in all the channels. This trivially works for $\phi^3$ theory where the coupling is simply a constant $g$, and the residue in each channel is simply $g^2$. Then an object with the correct poles in all channels is $g^2(1/s + 1/t + 1/u)$. Of course in addition to this we may have contact terms with no poles at all, whose form is not fixed by the three-particle amplitudes. But we will only be concerning ourselves with the parts of the four particle amplitudes that are forced to exist by consistent factorization given the three-particle amplitudes.

Let's repeat this exercise for the slightly more interesting case of Yukawa theory, where the three-particle amplitude for fermions 1,2 of helicity $-1/2$ to a scalar 3 is simply $y \langle 1 2 \rangle$. Let us compute the $s$-channel

\begin{equation}
\vcenter{\hbox{\includegraphics[scale=0.5]{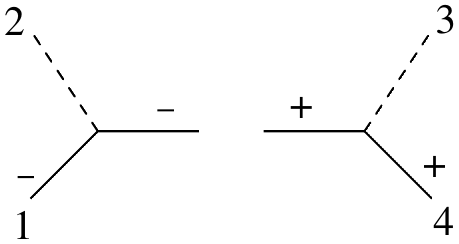}}}, \quad R_s=\langle 1 I \rangle [I 4] = \langle 1|p_I|4]\,,
\end{equation}
where here and in what follows we will suppress the trivial coupling constant dependence.  This can be simplified using
that $p_I = p_{1} + p_{2} = - p_{3} - p_{4}$, to $\langle 1|p_2 |4] = - \langle 1|p_3|4] = \frac{1}{2} \langle 1|(p_2 -
p_3)|4]$. The residue in the $u$ channel is the same swapping $2,3$. So finally the consistently factorizing amplitude is

\begin{equation}
\frac{\langle 1|(p_2 - p_3)|4]}{s} + \frac{\langle 1|(p_3 - p_2)|4]}{u}\,.
\end{equation}
\subsection{Self-interactions}
Let's now try a different example: consider a theory of a single
self-interacting particle of spin ${\bf s}$. The three particle amplitude
for $(1^{-{\bf s}} 2^{- {\bf s}} 3^{+{\bf s}})$ is $\frac{\langle 1 2 \rangle^{3 {\bf s}}}{\langle 1 3
\rangle^{\bf s} \langle 2 3 \rangle^{\bf s}}$. Note a remarkable feature of this expression, which we did not
encounter in either the $\phi^3$ or Yukawa theory cases: already the 3
particle amplitude appears to have poles! Thus in a sense, these
amplitudes are not as ``local" as we might have expected. Now of
course this peculiarity is un-noticed in the usual Minkowski space,
since the three-particle amplitude vanishes in the Lorentzian limit.
It is not a coincidence that this subtle sort of ``non-locality"
appears for precisely the same theories that, in a conventional
Lagrangian description, must introduce gauge redundancies for
consistency. But returning to our problem of determining four-particle
amplitudes by imposing consistent factorization, this feature
introduces an important obstruction. The strategy of computing the
residue in the $s$-channel, multiplying by $1/s$, then summing over
channels, is no longer guaranteed to work; as we will see because of
the poles in the three-particle amplitudes, the residue in the $s$
channel will itself have poles in the the other channels, making it
non-trivial to be able to find an object that consistently factorizes
in all channels. Indeed, while we can define massless three-particle
amplitudes for any helicities, it will be impossible to find
consistent four-point amplitudes for all but the familiar interacting
theories of massless spin $0,1/2,1,3/2$ and $2$ particles. This
exercise has been carried out in systematically in
\cite{LaurentiuDavid}, here we highlight some aspects of this story
before moving on to carrying out the similar analysis with massive
particles.

Let us return to the theory of self-interacting massless particles of spin $\bf{s}$; we will consider the four-particle amplitude
$(1^{- {\bf s}} 2^{+{\bf s}} 3^{-{\bf s}} 4^{+{\bf s}})$. The residue in the $s$-channel, reached when $[12] \to 0$ and $\langle 3 4 \rangle \to 0$, is
\begin{equation}
\vcenter{\hbox{\includegraphics[scale=0.5]{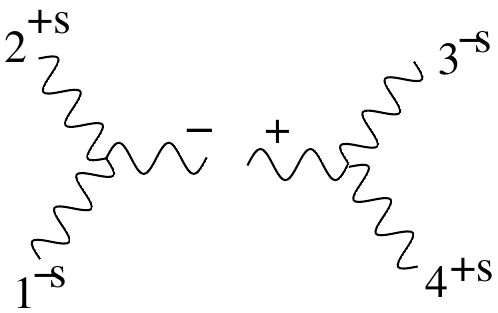}}}, \quad R_s = \left(\frac{\langle  I 1 \rangle^3}{\langle 1 2 \rangle \langle  2 I \rangle}\right)^{\bf s} \left(\frac{[ 4 I]^3}{[I3][34]}\right)^{\bf s} = \left(\frac{\langle 1 3 \rangle^2 [2 4]^2}{t}\right)^{\bf s}
\end{equation}
which, again using that e.g. $\langle 1 I \rangle [I 4] = \langle 1 2 \rangle [2 4] = - \langle 1 3 \rangle [3 4]$, can be simplified to $(\frac{\langle 1 3 \rangle^2 [2 4]^2}{t})^{\bf s}$. We can similarly compute the $t,u$ channel residues, and we find
\begin{equation}
R_s = \left(\frac{\langle 1 3 \rangle^2 [2 4]^2}{t}\right)^{\bf s}, R_t = \left(\frac{\langle 1 3 \rangle^2 [2 4]^2}{s} \right)^{\bf s}, R_u = \left(\frac{\langle 1 3 \rangle^2 [2 4]^2}{t}\right)^{\bf s}
\end{equation}
For ${\bf s} \geq 1$, we encountered the challenge alluded to above: the residue in one channel itself has a pole in another channel. Let us start with ${\bf s} = 1$. Given the structure of the residues, any consistent amplitude must have the form
\begin{equation}
\langle 1 3 \rangle^{2} [2 4]^{2} \left(\frac{A}{s t} + \frac{B}{t u} + \frac{C}{u s} \right)
\end{equation}
Note that as $s \to 0$, we have $t = - u$, e.g. the residue in $s$ is $A/t + C/u = (A - C)/t$. In this way, we find that matching the residues in $s,t,u$ demands that $(A-C) = 1, (B-A)=-1,(B-C)=1$, which is impossible since the sum of the three terms would have to vanish. We conclude that it is impossible to a single self-interacting massless spin 1 particle!
But suppose we have many of these particles labelled by the index $a$; thus the self-interaction of $a_1,a_2,a_3$ is further proportional to a coupling constant $f^{a_1 a_2 a_3}$. Note that for ${\bf s}=1$ the three particle amplitude $(1^{-{1}} 2^{- {1}} 3^{+{1}})= \frac{\langle 1 2 \rangle^{3}}{\langle 1 3
\rangle\langle 2 3 \rangle}$ is anti-symmetric in exchanging $1\leftrightarrow 2$, implying $f^{a_1 a_2 a_3}$ taking on the same property. Extending to all helicity configurations one can conclude that $f^{a_1 a_2 a_3}$ must be totally anti-symmetric.  Next consider the four particle amplitude with labels $a_1,a_2,a_3,a_4$, the residues in the $s,t,u$ channels have additional factors of $f^{a_1 a_2 e} f^{e a_3 a_4}$ and similarly in the $t,u$ channels. Now the ansatz for the four-particle amplitude has the form
\begin{equation}
\langle 1 3 \rangle^{2} [2 4]^{2} \left(\frac{A^{a_1 a_2 a_3 a_4}}{s t} + \frac{B^{a_1 a_2 a_3 a_4}}{t u} + \frac{C^{a_1 a_2 a_3 a_4}}{u s} \right)
\end{equation}
and matching the residues in $s,t,u$ tells us that
\begin{eqnarray}
C^{a_1 a_2 a_3 a_4} - A^{a_1 a_2 a_3 a_4} &=& f^{a_1 a_2 e} f^{e a_3 a_4} \nonumber \\
A^{a_1 a_2 a_3 a_4} - B^{a_1 a_2 a_3 a_4} &=& f^{a_2 a_3 e} f^{e a_4 a_1} \nonumber \\
B^{a_1 a_2 a_3 a_4} - C^{a_1 a_2 a_3 a_4} &=& f^{a_1 a_3 e} f^{e a_4 a_2}
\end{eqnarray}
and now, we can solve for $A^{a_1 a_2 a_3 a_4}, B^{a_1 a_2 a_3 a_4}, C^{a_1 a_2 a_3 a_4}$ if and only if the $f^{a_1 a_2 a_3}$ satisfies the Jacobi identity
\begin{equation}
f^{a_1 a_2 e} f^{e a_3 a_4} + f^{a_2 a_3 e} f^{e a_1 a_4} + f^{a_1 a_3 e} f^{e a_4 a_2} = 0
\end{equation}

Let's now move on to a single particle with ${\bf s}=2$. Naively, since the residue in the $s-$channel is proportional to $1/u^2$, we might think that it is impossible for the four-particle amplitude to have crucial properties of having only single poles! However, this $1/u^2$ is the residue just as $s \to 0$, and so it could also be represented as $-\frac{1}{t u}$. Thus there is a unique possibility for the four-particle amplitude for a single massless spin two particle:
\begin{equation}
- \frac{\langle 1 3 \rangle^4 [24]^4}{s t u}
\end{equation}
which evidently has all the correct residues in all three channels! We can further investigate the possibility on several massless spin two particles, with a coupling constant $g^{a_1 a_2 a_3}$; the same analysis as for spin one then gives us quadratic constraints on the $g^{a_1 a_2 a_3}$ that are solved only by $g$'s that, up to change of basis, are only non-vanishing for $a_1 = a_2 = a_3$, i.e. which are mutually non-interacting.

We have thus seen that the only consistently interacting massless spin one particles must have a Yang-Mills structure, and the only consistent massless spin 2 particles does not non-trivially allow more than one such particle, and gives us the standard gravity amplitude. Of course we have done more than simply show the amplitudes are consistent, we have computed them!

For spin ${\bf s} > 2$, the residue in the $s$-channel is at least $1/u^3$, and so there is no way to have a consistent four particle amplitude with only simple poles in $s,t,u$. We thus conclude that there are no consistent theories of self-interacting massless particles of spin higher than two.
\subsection{Interactions with other particles}\label{interactionswithother}
Let's move on to determine what sorts of self-consistent interactions other particles can have with massless spin 1, 2 particles. Let's start with the coupling of a spin ${\bf s}$ particles to spin one particle, for which the three particle amplitude is $\langle 1 2 \rangle^{2 {\bf s} + 1} \langle 2 3 \rangle^{1 - 2 {\bf s}} \langle 1 3 \rangle^{-1}$.
Let us now consider the residues for the $(1^{-s} 2^+ 3^- 4^{+s})$ amplitude; we get residues in the $s$ and $u$ channels from gluing these three-particle amplitudes together. These residues are trivially computed to be
\begin{equation}
R_s = \frac{1}{u} (\langle 1 3 \rangle [2 4])^{2 {\bf s}} [2|(p_1 - p_4)|3 \rangle^{2 - 2 {\bf s}}, \, R_u = \frac{1}{s} (\langle 1 3 \rangle [2 4])^{2 {\bf s}} [2|(p_1 - p_4)|3 \rangle^{2 - 2 {\bf s}}
\end{equation}
We see there is a qualitative difference between ${\bf s} \leq 1$ and ${\bf s} \geq 3/2$. For ${\bf s} = 0,1/2,1$, while the residues in one channel have poles in the other, we can write down a consistently factorizing four-particle amplitude:
\begin{equation}
\frac{(\langle 1 3 \rangle [2 4])^{2 {\bf s}} [2|(p_1 - p_4)|3 \rangle^{2 - 2 {\bf s}}}{s u}
\end{equation}
But for ${\bf s} \geq 3/2$, the residues have (increasing powers of) the spurious pole in $[2|(p_4 - p_1)|3 \rangle$, and so no consistent four particle amplitude is possible. Thus we recover the correct Compton-scattering expressions for particles of spin $0,1/2,1$ scattering off photons, while also seeing that it is impossible to have a consistent theory of massless charged particles with spin $\geq 3/2$.

When there are several species of spin ${\bf s}$ particles $i$ coupling with several spin one particles $a$, we attach an extra coupling $T^a_{ij}$ to the vertex. Consider $(1^-_i 2^+_a 3^-_b 4^+_j)$ scattering; writing the residues $R$ in any channel as $R = (\langle 1 3 \rangle [2 4])^{2 {\bf s}} [2|p_1|3 \rangle^{2 - 2 {\bf s}} \times r$, we have
\begin{equation}
\vcenter{\hbox{\includegraphics[scale=0.5]{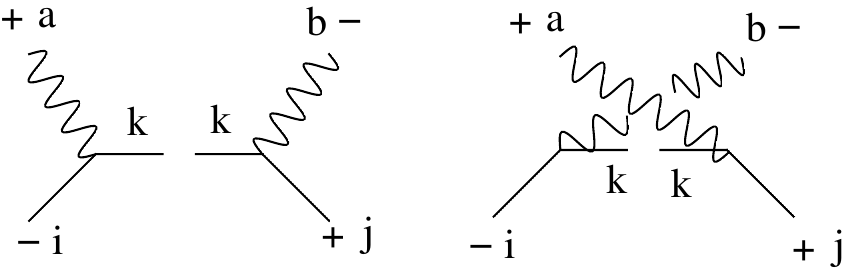}}}, \quad  \, r_s = \frac{1}{u} \, (T^a T^b)_{ij},\quad r_u = \frac{1}{s}\, (T^b T^a)_{ij}\,,
\end{equation}
where $(r_s, r_u)$ satisfies $s=0$ and $u=0$ kinematics respectively. 
Note that if $(T^a T^b)_{ij}=(T^b T^a)_{ij}$, or the commutator $[T^a,T^b]$ vanishes, we can get a consistent amplitude as with our Compton scattering example, with poles only in these $s$ and $u$ channels, but this is not possible if $[T^a,T^b] \neq 0$. This means that the $1/u$ in $r_s$ and the $1/s$ in $r_u$ must secretly be $1/t$ instead, i.e.  must also include a pole in the $t$ channel. Of course fortunately we can have a residue in the $t$ channel, using the cubic self-interaction for gluons.  Quite nicely the same kinematical factor appears in $R_t$, and we find (writing this residue in an $s,u$ symmetric way):
\begin{equation}
\vcenter{\hbox{\includegraphics[scale=0.5]{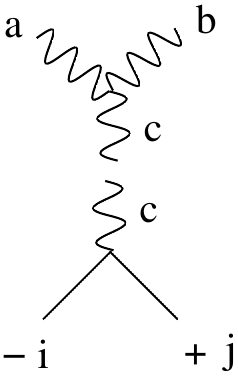}}}, \quad \, r_t = (\frac{1}{s}) \times f^{a b c} T^c_{ij}
\end{equation}
Thus, if we have 
\begin{equation}
 f^{a b c} T^c_{ij}=[T^a,T^b]_{ij} 
\end{equation}
and using the fact that when $t=0$, $s=-u$, we find that the following amplitude indeed consistently factorizes in all channels:
\begin{equation}
(\langle 1 3 \rangle [2 4])^{2 {\bf s}} [2|p_1|3 \rangle^{2 - 2 {\bf s}} \times \left(\frac{(T^a T^b)_{ij}}{t s}{+} \frac{(T^b T^a)_{ij}}{t u} \right)\,.
\end{equation}
This agrees with the result in~\cite{Johansson:2015oia}. Also, clearly once again no consistent amplitudes are possible for spin ${\bf s} \geq 3/2$. Thus we have discovered the familiar structure of Yang-Mills theories for particles of spin $0,1/2,1$.

The same sort of analysis extends to gravity, since the details are virtually identical we will leave them as enjoyable exercises for the reader.  We can consider the coupling of two particles of spin ${\bf s}$ to a graviton, with strength $g$. The residues in the $s,u$ channels are no longer equal, and the only way to make a consistent four particle amplitude is to also have a pole in the $t$ channel, using the graviton self-interaction $\kappa = \frac{1}{M_{Pl}}$. Thus once again the poles for the amplitude is forced to come in the combination $1/stu$. This implies that the coupling constant appearing in the spin-${\bf s}$ exchange channel must be identified with that of the graviton exchange. That is, consistency between the three factorization channel forces the universality of couplings to gravity, $g = \kappa$, with the following form for Compton scattering:
\begin{equation}
\kappa^2\frac{(\langle 1 3 \rangle [2 4])^{2 {\bf s}} [2|(p_1 - p_4)|3 \rangle^{4 - 2 {\bf s}}}{s t u }\,.
\end{equation}
Now we see that for ${\bf s} \geq 2$ one again develops spurious pole, and one reaches the conclusion that for spin greater than 2, the particle cannot consistently couple to gravity. In other words, even if higher spin particles are non self-interacting and free, the moment one turns on gravity it ceases to be consistent in flat space.  Thus we find that the only possible consistent theories that can couple to gravity can only have spins $(0,1/2,1,3/2)$.\footnote{As we remarked in our discussion above on self-interacting spin 2, via a basis change it is always possible to say that the spin 2 particles are effectively in different universes with no mutual interactions; in each one of these decoupled sectors the gravitons can be coupled to their own spectrum of particles with spin $(0,1/2,1,3/2)$.}

We can also discover the need for supersymmetry when massless particles of spin $3/2$ are present. Consider for simplicity the case with a single spin $3/2$ particle $\psi$.  Now let's imagine we also have a massless scalar $\phi$. Both of these particles have a universal coupling to gravity, so there is inevitably an amplitude for $\psi_1 \psi_1 \phi_2 \phi_2$ scattering mediated by gravity.  We can again compute the residue in the $s$-channel, and find that it has a pole in the $t$ channel. But since there is no $(\psi,\phi,$ graviton) coupling (amplitudes must be grassmann even), we can't have any $t$-channel poles, and so this theory is inconsistent. The only way to have a consistent amplitude is if we {\it also} introduce a massless fermion $\chi$, now we can have a $(\psi,\phi, \chi)$ interaction with the same gravitational strength $1/M_{Pl}$, which provides the needed pole in the $t$-channel. The full amplitude is then given as:
\eq
(1, 2 ,3^{-\frac{3}{2}}, 4^{+\frac{3}{2}})=\kappa^2\frac{\langle 3|(p_1{-}p_2)|4]^3}{st}\,.
\eqe
Thus we see that we must have a bose-fermi degenerate spectrum, with the couplings of the ``gravitino" $\psi$ to particles and their superpartners of universal gravitational strength.

We have given a lightning tour of some of the arguments leading to the determination of all consistent theories of massless particles via the ``four-particle scattering" test. It is remarkable to see the architecture of fundamental physics emerge from these concrete algebraic consistency conditions in such a simple way. A more complete and systematic treatment can be found in \cite{LaurentiuDavid}.

Before moving on to considering massive amplitudes, let us briefly comment the (in)consistency of theories with three-particle amplitudes for helicities satisfying $h_1 + h_2 + h_3 = 0$.  Apart from the case of all scalars $h_1 = h_2 = h_3 = 0$, we have ``phase" singularities in the couplings, for instance we have a coupling of the form $\langle 1 3 \rangle/\langle 1 2 \rangle$ or $[12]/[13]$ for a spin zero particle 1 to particles 2, 3 of helicity $\pm 1/2$. This peculiar interaction is unfamiliar, and does not arise from Lagrangian couplings. But, as expected, it is also impossible to find a correctly factorizing four-particle amplitudes with these couplings \cite{LaurentiuDavid}, so consistency forces the couplings to vanish.

\section{General Three Particle Amplitudes}
In this section we will categorize the most general three-point amplitude with arbitrary masses. As discussed in section~\ref{LittleGroup}, the amplitude will be labeled by the spin-$S$ representation of the SU(2) little group for massive legs and helicities for the massless legs. For amplitudes involving massive legs, it will be convenient to expand in terms of $\lambda^I_\alpha$, since any dependence on $\tilde{\lambda}^I_{\dot{\alpha}}$ can be converted using eq.(\ref{Convert}). For example for a general one massive two massless amplitude, with leg $3$ being a massive spin-$S$ state, we have:
\begin{equation}
M_3^{\{I_1 \cdots I_{2S}\},h_1,h_2} = \lambda_{3, \alpha_1}^{I_1} \cdots \lambda_{3, \alpha_{2S}}^{I_{2S}} M_3^{\{\alpha_1 \cdots \alpha_{2S}\},h_1,h_2}\,,
\end{equation}
where $(h_1,h_2)$ are the helicity. We will be interested in the most general form of the stripped $M_3^{\{\alpha_1 \cdots \alpha_{2S}\},h_2,h_3}$, which is now a tensor in the $SL(2,C)$ Lorentz indices. The problem thus reduces to finding two linear independent 2-component spinors that span this space, which we will denote as $(v_\alpha, u_\alpha)$. The convenient choice of $(v_\alpha, u_\alpha)$ will depend on the number of massive legs in a given set up and we will analyze each case separately. We note that a similar classification of three-point interactions using a different basis can be found in~\cite{Conde:2016vxs, Conde:2016izb}.
\subsection{Two-massless one-massive}
Let's first begin with the two massless and one massive interaction:
$$\vcenter{\hbox{\includegraphics[scale=0.5]{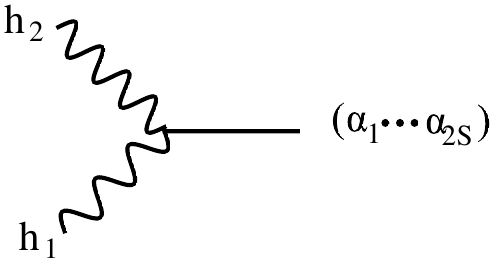}}} \quad M^{h_1h_2}\,_{\{\alpha_1\alpha_2\cdots\alpha_{2S}\}}$$
Since both legs $1,2$ are massless, their spinors can serve as a natural basis:
\eq
(v_\alpha, u_\alpha)=(\lambda_{1\alpha},\lambda_{2\alpha})
\eqe
The helicity weight $(h_1, h_2)$ then completely fixes the degree-$2S$ polynomial in $\lambda_1,\lambda_2$ up to an overall coupling constant:
\eq\label{1M2ML}
M^{h_1h_2}\,_{\{\alpha_1\alpha_2\cdots\alpha_{2S}\}}=\frac{g}{m^{2S+h_1+h_2-1}}\left(\lambda^{S+h_2-h_1}_{1}\lambda^{S+h_1-h_2}_{2}\right)_{\{\alpha_1\alpha_2\cdots\alpha_{2S}\}}[12]^{S+h_1+h_2}\,,
\eqe
where with appropriate factors of $m$ such that it has the correct mass-dimension. Note that we can trade $[12]$ for $\langle12\rangle$ using $[12]=\frac{m^2}{\langle 21\rangle}$. When the massive leg is a fermion, i.e. $S\in \frac{1}{2}\mathbb{Z}$, we must then require precisely one of the massless legs to be a fermion as well.

The fact that the structure of this three-point amplitude is unique implies no go theorems for certain interactions. For example, for identical helicities the factor $[12]^{S+2h_1}$ will attain an extra factor of $(-1)^{1+2h_1}$ under $1,2$ exchange for odd spins. This will result in the wrong spin-statistics, thus a particle of odd spin $S$ cannot decay to identical particles with the same helicity. Now suppose the particles have opposite helicity, namely $h_1=-h_2=h$. If we take into account that the exponents of $\lambda_1$ and $\lambda_2$ must both be positive, we conclude that the amplitude vanishes if $|h|>S/2$. For massive spin one states, this is Yang's theorem---that a massive spin one particle cannot decay to a pair of photons. We also learn that a massive spin three particle cannot decay to a pair of gravitons. Note that we have invoked spin-statistics without giving its on-shell origin. As we will see in the coming subsection \ref{Secxfactor}, when considering the three-point amplitude of identical massive spin-$S$ states to gravity, spin-statistics is immediately forced upon us.

\subsection{One-massless two-massive}
For two massive legs, the three-point amplitude is now labeled by $(h, S_1, S_2)$ 
\eq
\vcenter{\hbox{\includegraphics[scale=0.5]{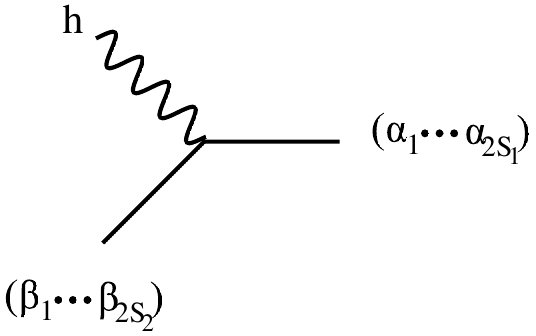}}} \quad M^{h}\,_{\{\alpha_1\alpha_2\cdots\alpha_{2S_1}\},\,\{ \beta_1\beta_2\cdots\beta_{2S_2}\}}
\eqe
The analysis depends on whether or not the masses are identical. For equal mass, the kinematics becomes degenerate and one expects some form of superficial non-locality. The reason is that the equal mass kinematics occurs precisely for minimal coupling, where its massless limit contain inverse power of spinor brackets as discussed in the previous section. As we will see, for this case we need to introduce a new variable $x$ that encodes this non-locality.

\subsubsection{Unequal mass}
For unequal mass, one of the basis spinor can be $\lambda$ of the massless leg, while the remaining can be chosen to be $\tla$ contracted with one of the massive momentum. For example one can choose: 
\eq\label{uv}
\left(v_\alpha, u_\alpha\right)=\left(\la_{\alpha}, \; \frac{p_{1\alpha\dot{\beta}}}{m_1}\tla^{\dot\beta}\right)
\eqe
Unlike the one massive case, here the amplitude is not unique. The helicity constraint only fixes the polynomial degree in $u$ and $v$ to differ by $2h$.  For $S_1\neq S_2$ there are then a total of $C=S_1{+}S_2{-}|S_1{-}S_2|{+}1$ different tensor structures, and the general three-point amplitude is given by:
\eq
M^{h}\,_{\{\alpha_1\alpha_2\cdots\alpha_{2S_1}\},\,\{ \beta_1\beta_2\cdots\beta_{2S_2}\}}
=\sum_{i=1}^{C}g_i(u^{S_1+S_2+h}v^{S_1+S_2-h})^{(i)}_{\{\alpha_1\alpha_2\cdots\alpha_{2S_1}\},\{\beta_1\beta_2\cdots\beta_{2S_2}\}}
\eqe
where $i$ labels the different structure and $g_i$ is the coupling constant for the different tensor structures. Note that the number of possible tensor structures is determined by the lowest spin. For example for one $S_1=1$ $S_2=2$, we have three tensor structures. For a minus helicity photon these are given by:
\eq\label{possibleT}
(vvvv)(uu),\;(vvvu)(vu),\;(vvuu)(vv)\,.
\eqe
where the parenthesis indicates the grouping of the symmetrized SU(2) little group index. One can also compare this with a Feynman diagram vertex $F_{3,\mu\nu}\epsilon_2^{\nu\rho}\partial_\rho\epsilon_1^\mu$, where $\epsilon_1, \epsilon_2$ are the polarization vectors for the massive particles. Again,  substituting the on-shell form of the massless polarization  vectors $\epsilon^-_i=\frac{|i\rangle[\tilde{\mu}|}{[i\tilde{\mu}]}$, $\epsilon^+_i=\frac{|\mu\rangle[i|}{\langle\mu i\rangle}$, where $|\tilde{\mu}], |\mu\rangle$ are reference spinors, and massive ones in eq.(\ref{MassivePol}), one finds:
\eqa\label{FFFF}
{M_3}_{\{\alpha_1\alpha_2\}\{\beta_1\beta_2\beta_3\beta_4\}}&=&\frac{m_1^2}{m_2^4}\frac{1}{m^2_1{-}m^2_2}\left[m_1(uu)_{\{\alpha_1\alpha_2\}} (uuvv)_{\{\beta_1\beta_2\beta_3\beta_4\}}\right.\nonumber\\
&-&\left.m_2(uv)_{\{\alpha_1\alpha_2\}} (uuuv)_{\{\beta_1\beta_2\beta_3\beta_4\}}\right].
\eqae
Indeed the three-point amplitude for the vertex can be expanded on the basis in eq.(\ref{possibleT}), as it should.

\subsubsection{Equal mass: the $x$-factor}\label{Xsection}
If the masses are identical, then $u$ and $v$ are no longer independent, since:
\eq
v^{\alpha}u_{\alpha}=\frac{\langle3|p_1|3]}{m}=0\,.
\eqe
Thus $(u^\alpha, v^\alpha)$ are parallel to each other and pick out just one direction in the SL(2,C) space. There is however a crucial piece of additional data in the constant of proportionality between $u$ and $v$, which we will call ``$x$":
\eq\label{xDef}
x\lambda_{3\alpha}=\frac{p_{1\alpha\dot{\alpha}}}{m}\tla_3^{\dot{\alpha}}\,,\quad \frac{\tla_3^{\dot{\alpha}}}{x}=\frac{p_1^{\dot{\alpha}\alpha}\lambda_{3\alpha}}{m}\,.
\eqe
Note that $x$ carries $+1$ little group weight of the massless leg. Furthermore, $x$ cannot be expressed in a manifestly local way. Indeed contracting both sides of the above equation with a reference spinor $\zeta$ yields:
\eq
x=\frac{\langle \zeta |p_{1}|3]}{m\langle \zeta 3\rangle},  
\eqe
so while $x$ is independent of $\zeta$, any concrete expression for it has an apparent, spurious pole in $\zeta$. In the next section, as we glue the three-point amplitudes to get the four-point, it will be convenient to choose $\zeta$ to be the spinor of the external legs on the other side. The denominator then yields a pole in other channels! This yields non-trivial constraint for the four-point amplitude to have consistent factorisation in all channels.

Now the only objects we have carrying SL(2,C) indices are $\lambda_3$, as well as the  the anti-symmetric tensor $\varepsilon_{\alpha\beta}$.\footnote{Note in the unequal mass case, since $u,v$ provided a basis, we didn't need to separately introduce $\varepsilon_{\alpha\beta}$ since $(u_\alpha v_\beta-u_\beta v_\alpha)=\langle uv\rangle\varepsilon_{\alpha\beta}$. However as $m_1\rightarrow m_2$ these invariants vanish. This also shows the absence of a singularity in eq.(\ref{FFFF}) as $m_1\rightarrow m_2$.} We can then express the three-point amplitude as:
\eqa\label{Massive3}
M^{h}\,_{\{\alpha_1\alpha_2\cdots\alpha_{2S_1}\},\,\{ \beta_1\beta_2\cdots\beta_{2S_2}\}}
&=&\sum_{i=|S_1-S_2|}^{(S_1+S_2)}g_ix^{h+i}(\lambda^{2i}_3\varepsilon^{S_1+S_2-i})_{\{\alpha_1\alpha_2\cdots\alpha_{2S_1}\},\{\beta_1\beta_2\cdots\beta_{2S_2}\}}\nonumber\\
&=&\sum_{i=|S_1-S_2|}^{(S_1+S_2)}g_i\,x^{h}\left[\lambda^{i}_3\left(\frac{p_1\tla_3}{m}\right)^i\varepsilon^{S_1+S_2-i}\right]_{\{\alpha_1\alpha_2\cdots\alpha_{2S_1}\},\{\beta_1\beta_2\cdots\beta_{2S_2}\}}\,,\nonumber\\
\eqae
where the superscript on $\lambda, \varepsilon, p\tla/m$ indicates its power. For later purpose we present it in two equivalent representations.

\subsection{Minimal Coupling for Photons, Gluons, Gravitons}\label{Secxfactor}
We have seen that while there is a unique structure for massless three-particle amplitudes once the helicities are specified, for couplings of e.g. two equal mass particles of spin $S$ to a massless particle there are $(2S{+}1)$ independent structures, each term with $n$ factors of $\varepsilon$ with $n=0,\cdots,2S$. Let us take the massless particle to be a graviton. Note that $\varepsilon$ is antisymmetric with respect to the exchange  $1\leftrightarrow 2$. Furthermore while the definition of $x$ in eq.(\ref{xDef}) implies that it picks up a minus sign under the $1\leftrightarrow 2$, this is irrelevant for gravitational couplings which are proportional to $x^2$. Thus we see that one gravitation two identical spin $S$ amplitude will have a factor of $(-)^{2S{+}1}$ under the exchange of the spin-$S$ states. This is nothing but the spin-statistic theorem! 

Now one of the $(2S{+}1)$ structures is special, and corresponds to what we usually think of as ``minimal coupling" to photons, gluons and gravitons. The defining characteristic of ``minimal coupling" is physically very clear. For massless particles, the mass dimension of the couplings is given by $1 - |h_1 + h_2 + h_3|$, and so the leading low-energy interactions with photons, gluons and gravitons---those with dimensionless gauge couplings $e,g$ or gravitational coupling $1/M_{Pl}$, involve massless particles of opposite helicity. The definition of ``minimal coupling" for massive particles is then simply the interaction whose leading high-energy limit is dominated by precisely this helicity configuration. As we will see the remaining $(2S + 1) - 1 = 2S$ interactions represent the various multipole-moment couplings (such as the magnetic dipole moment in the coupling to photons.)

In our undotted SL(2,C) basis, the amplitude with a positive helicity state can be viewed as an expansion in $\lambda$. The leading piece in this expansion, namely that where the SL(2,C) indices are completely carried by the Levi-Cevita tensors, precisely corresponds to minimal coupling! It is instructive to see why this is the case. Using the simplest example, a photon coupled to two fermions, we find:  
\eq
x m \varepsilon_{\alpha_1\alpha_2}\rightarrow x\langle \mathbf{1}\mathbf{2}\rangle = \langle \mathbf{1}\mathbf{2}\rangle\frac{\langle \zeta |p_{1}|3]}{m\langle \zeta 3\rangle}=\frac{\langle \mathbf{2}\zeta\rangle[3\mathbf{1}]+\langle \mathbf{1}\zeta\rangle[3\mathbf{2}]}{\langle \zeta3\rangle}
\eqe
Taking the high energy limit, we see that the leading term indeed correspond two possible pairs of opposite helicity fermion, 
\eq
\frac{\langle \mathbf{2}\zeta\rangle[3\mathbf{1}]+\langle \mathbf{1}\zeta\rangle[3\mathbf{2}]}{\langle \zeta3\rangle}\underrightarrow{\rm \quad H. E. \quad} \frac{[13]^2}{[12]}+\frac{[23]^2}{[12]}+\mathcal{O}(m)\,.
\eqe
In general the the minimal coupling between photon and two spin-$S$ states is simply:
\eqa\label{GeneralMin}
M^{\rm min, +1}\,_{\{\alpha_1\cdots\alpha_{2S}\}, \{\beta_1\cdots\beta_{2S}\}} = xm \left(\prod_{i=1}^{2S}\varepsilon_{\alpha_i\beta_i}+sym\right),\nonumber\\
M^{\rm min, -1}\,_{\{\dot{\alpha}_1\cdots\dot{\alpha}_{2S}\}, \{\dot\beta_1\cdots\dot\beta_{2S}\}} = \frac{m}{x} \left(\prod_{i=1}^{2S}\varepsilon_{\dot\alpha_i\dot\beta_i}+sym\right)\,,
\eqae
where we've also included the negative helicity photon in its simplest dotted representation. The proper amplitude (with little group indices) is then given as:
\eq
M^{\rm min, +1}=x\frac{\langle \mathbf{1}\mathbf{2}\rangle^{2S}}{m^{2S-1}},\quad M^{\rm min, -1}=\frac{1}{x}\frac{[\mathbf{1}\mathbf{2}]^{2S}}{m^{2S-1}}
\eqe
For gravitons, we simply introduce an extra power of $\frac{m}{M_{pl}}x$. The fact that in this formalism, minimal coupling is as simple as $\lambda\phi^3$ heralds its potential for simplification. It is also instructive to see how such simple representation emerges from the usual vertices in Feynman rules.  Here we present examples for scalar, spinor and vector at three points: 
\eqa
{\rm Scalars}: \vcenter{\hbox{\includegraphics[scale=0.4]{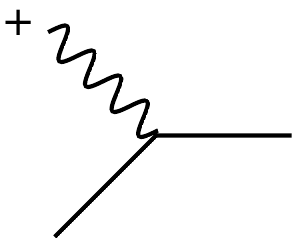}}}&&\epsilon_3\cdot p_1=\frac{\langle \xi|p_1|3]}{\langle 3\xi\rangle}=-mx\,,
\eqae
where we've used the identity $xm\lambda_3=p_1|3]$. Similarly for spin-$\frac{1}{2}$ and $1$, we have:
\eqa
{\rm Fermons}: \vcenter{\hbox{\includegraphics[scale=0.4]{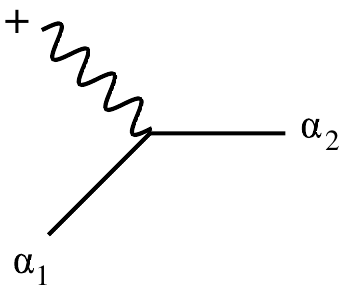}}}&&\bar{u}_1\displaystyle{\,\not}\epsilon_3v_2=\left(\frac{p_2^{\dot{\gamma}\alpha_2}}{m}, \delta^{\alpha_2}_\gamma\right) \left(\begin{array}{cc}  0& \frac{\tla_{3\dot{\gamma}}\xi_{\beta}}{\langle 3\xi\rangle}  \\ -\frac{\tla_{3}^{\dot{\beta}}\xi^{\gamma}}{\langle 3\xi\rangle} & 0 \end{array}\right)\left(\begin{array}{c}  \frac{p_{2\dot{\beta}}\,^{\alpha_1}}{m}\\ \varepsilon^{\alpha_1\beta}\end{array}\right)\nonumber\\
&=&x\varepsilon^{\alpha_1\alpha_2}
\eqae
\eqa
{\rm Vectors}: \vcenter{\hbox{\includegraphics[scale=0.4]{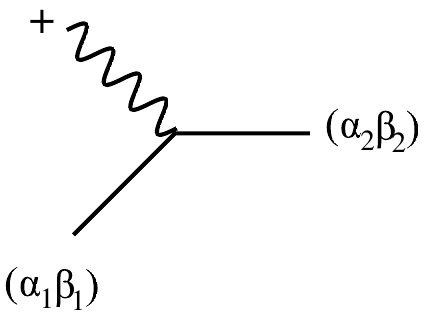}}}&&\frac{p_1^{\beta_1\dot{\alpha}_1}}{m}\left[\epsilon_3\cdot p_1\varepsilon_{\alpha_1\alpha_2}\varepsilon_{\dot\alpha_1\dot\alpha_2}+p_{2\alpha_1\dot{\alpha}_1}\epsilon_{3\alpha_2\dot{\alpha}_2}-p_{1\alpha_2\dot{\alpha}_2}\epsilon_{3\alpha_1\dot{\alpha}_1}\right]\frac{p_2^{\beta_2\dot{\alpha}_2}}{m}\nonumber\\
&=& -mx\left(\varepsilon^{\alpha_1\alpha_2}\varepsilon^{\beta_1\beta_2}+sym(\alpha\leftrightarrow\beta)\,\right)\,.
\eqae

The fact that minimal coupling is literally the ``minimal" interaction in the undotted SL(2,C) representation indicates the $\lambda$ expansion should directly correspond to the presence of couplings through higher-dimensional operators. These precisely are the magnetic and electric moments. Let us begin with the magnetic dipole moment. Since this corresponds to a coupling of the particle with $F^{\mu\nu}$, it can only occur for particles with spin. Thus we can extract the electric dipole moment by separating the minimal coupling into a piece that is universal, and pieces that only exists for spinning particles. 

Recall that the field strength in momentum space becomes $F_{\mu\nu}\rightarrow \lambda_{\alpha}\lambda_{\beta}\varepsilon_{\dot\alpha\dot\beta}+\tilde\lambda_{\dot\alpha}\tilde\lambda_{\dot\beta}\varepsilon_{\alpha\beta}$. This implies that couplings through the field strength will be transparent in the undotted frame for negative helicity photon, and dotted frame for the positive photon. With this in mind we convert the minimal coupling for spin-$\frac{1}{2}$ and negative helicity photon into the dotted frame:
\eq\label{p1p2}
\frac{p_1^{\alpha\dot{\alpha}}}{m}\left(\frac{m}{x}\varepsilon_{\dot{\alpha}\dot{\beta}}\right)\frac{p_2^{\beta\dot{\beta}}}{m}=\frac{m}{x}\left(\varepsilon^{\alpha\beta}{+}x\frac{\lambda_3^\alpha\lambda_3^\beta}{m}\right)\,.
\eqe
Here the piece $\frac{m}{x}\varepsilon^{\alpha\beta}$ is the same as that for scalars, sans the $\varepsilon^{\alpha\beta}$ factor which is necessary to carry the SL(2,C) indices, and thus a universal term.  The extra piece $\lambda_3^\alpha\lambda_3^\beta$ then represents the magnetic moment coupling, with the amplitude given by 
\eq
\frac{\langle\mathbf{1}3\rangle\langle3\mathbf{2}\rangle}{m}.
\eqe
Thus we immediately see that $g=2$ for the magnetic dipole moment.\footnote{As a comparison, for the positive helicity and insisting on the undotted frame, we can make the separation after contracting $\lambda^I$s. More precisely:
$$x\varepsilon_{\alpha\beta}\;\rightarrow \;\; x\frac{\langle\mathbf{1}\mathbf{2} \rangle}{m} = \frac{1}{m}\left(x[\mathbf{1}\mathbf{2}]{+}\frac{[\mathbf{1}3][3\mathbf{2}]}{m}\right)\,.$$  } Thus for minus helicity photon, the general spin-$\frac{1}{2}$ amplitude has the simple expansion:
\eq
M^{\rm -1}\,_{\dot{\alpha}_1\dot{\alpha}_2} = \frac{1}{x}m\varepsilon_{\dot{\alpha}_1\dot{\alpha}_2}{-}\frac{(g{-}2)}{4}\frac{(\tilde\lambda_3\tilde\lambda_3)_{\dot{\alpha}_1\dot{\alpha}_2}}{x^2}\,,
\eqe 
where we've manifestly separated the minimal coupling and the $(g-2)$ part of the magnetic dipole moment. It is straight forward to see that $\frac{(\tilde\lambda_3\tilde\lambda_3)_{\dot{\alpha}_1\dot{\alpha}_2}}{x^2}$ in the undotted frame, is simply $\lambda_3\lambda_3$. For the plus helicity, one has:
\eq\label{G2Def}
M^{\rm +1}\,_{\alpha_1\alpha_2} = mx\varepsilon_{\alpha_1\alpha_2}{+}\frac{(g{-}2)}{4}x^2(\lambda_3\lambda_3)_{\alpha_1\alpha_2}\,.
\eqe

One can trivially extend this to higher spin. For example for spin-1, the minimal coupling now contains both the magnetic dipole moment and electric quadrupole moment. The minimal coupling yields:
\eqa
&&\frac{m}{x}\left(\varepsilon^{\alpha_1\alpha_2}-x\frac{\lambda_3^{\alpha_1}\lambda_3^{\alpha_2}}{m}\right)\left(\varepsilon^{\beta_1\beta_2}-x\frac{\lambda_3^{\beta_1}\lambda_3^{\beta_2}}{m}\right)+(\alpha_1\leftrightarrow \beta_1)\,\nonumber\\
&=&-\frac{m}{x}\varepsilon^{\alpha_1\{\alpha_2}\varepsilon^{\beta_2\}\beta_1}-\varepsilon^{\alpha_1\{\alpha_2}\lambda_3^{\beta_2\}}\lambda_3^{\beta_1}-\varepsilon^{\beta_1\{\beta_2}\lambda_3^{\alpha_2\}}\lambda_3^{\alpha_1}+2x\frac{\lambda_3^{\alpha_1}\lambda_3^{\alpha_2}\lambda_3^{\beta_1}\lambda_3^{\beta_2}}{m}\,.
\eqae
We again see that the first term is the universal piece, the terms quadratic in $\lambda$ is the dipole moment where as the terms quartic in $\lambda$ is the electric quadrupole moment. Thus the general three point amplitude for the charged vector and a photon is:
\eqa
M^{\rm -1}\,_{\{\dot{\alpha}_1\dot{\beta}_1\}\{\dot{\alpha}_2\dot{\beta}_2\}} &=& \frac{1}{x}m\varepsilon_{\{\dot{\alpha}_1\dot{\alpha}_2}\varepsilon_{\dot{\beta}_1\}\dot{\beta}_2}+(g{-}2)\left(\frac{\varepsilon_{\dot{\alpha}_1\{\dot{\alpha}_2}\tilde\lambda_{3\dot{\beta}_2\}}\tilde\lambda_{3\dot{\beta}_1}}{x^2}+\frac{\varepsilon_{\dot{\beta}_1\{\dot{\beta}_2}\tilde\lambda_{3\dot{\alpha}_2\}}\tilde\lambda_{3\dot{\alpha}_1}}{x^2}\right)\,\nonumber\\
&+&2(g'+1)\frac{\tilde\lambda_{3\dot{\alpha}_1}\tilde\lambda_{3\dot{\alpha}_2}\tilde\lambda_{3\dot{\beta}_1}\tilde\lambda_{3\dot{\beta}_1}}{mx^3}
\eqae
where $(g-2)$ and $(g'+1)$ is the anomalous magnetic dipole and electric quadrupole moment respectively.

\subsection{Three massive }
For all massive legs, we no longer have massless spinors to span the $SL(2,C)$ space. This implies that the space has to be spanned by tensors instead. The fundamental building blocks are now 
\eq
\mathcal{O}_{\alpha\beta}=p_{1\{\alpha\dot{\beta}}p_{2\beta\}}\,^{\dot{\beta}},\quad \varepsilon_{\alpha\beta}\,. 
\eqe
Note that since $\mathcal{O}_{\alpha\beta}\mathcal{O}_{\gamma\delta}-\mathcal{O}_{\gamma\beta}\mathcal{O}_{\alpha\delta}\sim\ve_{\alpha\gamma}\ve_{\beta\delta}$, pairs of $\varepsilon_{\alpha\beta}$ can be traded for products of $\mathcal{O}_{\alpha\beta}$. The general form of the three-point amplitude is:
\eqa
\vcenter{\hbox{\includegraphics[scale=0.4]{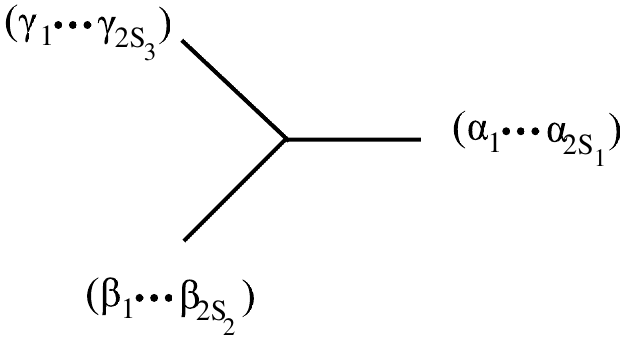}}}&&\nonumber\\
M_{\alpha_1\cdots\alpha_{2S_1}, \beta_1\cdots\beta_{2S_2},\gamma_1\cdots\gamma_{2S_3}}&=&\sum_{i=0}^1 \sum_{\sigma_i}g_{\sigma_i}\left(\mathcal{O}^{S_1+S_2+S_3-i}\ve^i\right)^{{\sigma_i}}_{\{\alpha_1\cdots\alpha_{2S_1}\}, \{\beta_1\cdots\beta_{2S_2}\},\{\gamma_1\cdots\gamma_{2S_3}\}}\nonumber\\
\eqae
where $i=0,1$ represents the number of $\varepsilon$s and  $\sigma_i$ labels all distinct ways the SU(2) indices can be distributed on $\mathcal{O}$s and should be summed over. It will be interesting to see whether the higher spin interactions from string theory, see~\cite{Fotopoulos:2010ay} for recent results, span the space of all interaction allowed. 
\section{Four Particle Amplitudes For Massive Particles} 
Now that we have determined the structure of all possible three-particle interactions, we would like to proceed to investigating the consistency of four-particle amplitudes. Just as we did for all massless particles, we ask: given a spectrum of particles, and a set of three-particle interactions, is it possible to find a four-particle amplitude that consistently factorizes in all possible channels?
We stress that this is a completely sharply defined and straightforward algebraic problem. To be maximally pedantic, suppose we have a set of particles with masses (zero or non-zero) given by $m_i$. Then the most general ansatz for the four-particle amplitude has the form
\begin{equation}
\frac{N}{\prod_i (s-m^2_i)(t - m^2_i)(u - m^2_i)}
\end{equation}
and we simply wish to determine whether there is a consistent numerator $N$ that allows this function to factorize correctly in the $s,t,u$ channels\footnote{Of course the amplitude cannot be uniquely determined in this way, since we can always simply have contact terms that are simply polynomials with no poles at all (corresponding to piece in $N$ that cancels all the poles). To avoid clutter, we will suppress the possible contact terms in what follows.}
\begin{equation}
 M
\to \frac{M^{a}_{L, \, \{I_1 \cdots I_{2s}\}}\varepsilon^{I_1J_1}\cdots\varepsilon^{I_{2s}J_{2s}} M^{a}_{R \, \{J_1
\cdots J_{2s}\}}}{P^2 -M^2} \,.
\end{equation}
As we've shown before, it is convenient to expand the amplitude on the $\lambda^I_\alpha$ basis, in which case the contraction of little group indices now translates to the contraction of undotted SL(2,C) indices:
\eq
\frac{\lambda^I_{\alpha}\lambda^J_{\beta}}{m}\varepsilon_{IJ}=\varepsilon_{\alpha\beta}\,.
\eqe
To make contact with the usual Feynman rules, the numerator of the vector propagator is $G_{\mu\nu}\equiv\eta_{\mu\nu}-\frac{p_{\mu}p_{\nu}}{m^2}$, which in SL(2,C) undotted representation is: 
\eq
 G_{\alpha\dot{\alpha},\beta\dot{\beta}}=2\varepsilon_{\alpha\beta}\varepsilon_{\dot{\alpha}\dot{\beta}}-\frac{p_{\alpha\dot{\alpha}}p_{\beta\dot{\beta}}}{m^2}\;\rightarrow\; \frac{p_{\gamma}^{\dot{\alpha}}p_{\delta}^{\dot{\beta}}}{m^2}G_{\alpha\dot{\alpha},\beta\dot{\beta}}=\varepsilon_{\alpha\{\beta}\varepsilon_{\gamma\delta\}}\,,
\eqe
as expected. This is not surprising, as we've discussed in the introduction, the transverse traceless-ness, which determines the numerator of the propagator, simply translates to symmetrization of the SL(2,C) indices. 

In practice, we don't need to work with this slavishly systematic ansatz for the amplitude with the giant denominator consisting of all possible simple poles. Instead, following the same steps as in the all massless case, given the spectrum and the three-particle amplitudes, we will first simply compute the residues $R(i)_s,R(i)_t,R(i)_u$ in the $s,t,u$ channels from the exchange of the $i$'th particle. If these residues are local, we are trivially done, since the object
\begin{equation}
\sum_i \left(\frac{R_s^{(i)}}{s-m^2_i} + \frac{R_t^{(i)}}{t-m^2_i} + \frac{R_u^{(i)}}{u-m^2_i}\right) 
\end{equation}
manifestly matches the poles in all the channels. This is the case for the massive $g \phi^3$ theory where these residues are all simply $R_s = R_t = R_u = g^2$.
But as we already saw in the massless case, there are more interesting cases where the residues in one channel themselves have poles in another channel.
With massive particles this will occur whenever we have minimal coupling and the $``x"$ factor.
In this case an ansatz separately summing the channels cannot work, and we must use building blocks that have simple poles in more than one channel.
For massless particles, the requirement of four-particle consistency was so strong as to simply make certain theories (of high-enough spin charged or gravitating massless particles) impossible. It also enforced universality of the couplings to gravitons and the usual Yang-Mills structure for coupling to photons and gluons. We will see the analogue of these statements for massive amplitudes. Once again, consistent factorization will demand the standard couplings to photons, gluons and gravitons, will also see that any self-interactions have to be invariant under the (global part) of the gauge symmetry. But with these restriction met, it {\it is} possible to find consistently factorizing four-particle amplitudes for any masses and spins. This is of course expected, since almost all interesting objects in the real world are massive particles of high spin! But of course as we will also see, the impossibility of consistent amplitudes for massless particles of high spin shows up in a singularity of the massive high spin amplitudes in the high-energy (or $m \to 0$) limit, giving a very concrete sense in which particles of high spin cannot be ``elementary".

\subsection{Manifest local gluing}
We first begin with the construction of amplitudes without any $x$-factor non-localities. Let's begin with Yukawa amplitude, i.e. one massless scalar two massive fermion amplitude. The three-point amplitude is simply
\eq
\frac{g}{m_f}\langle\mathbf{1}3\rangle[\mathbf{2}3]+\frac{g'}{m_f}\langle\mathbf{2}3\rangle[\mathbf{1}3]
\eqe
where $m_f$ is the mass of the fermion. The gluing in the $s$- and $u$-channel yields:
\eqa
&&\vcenter{\hbox{\includegraphics[scale=0.6]{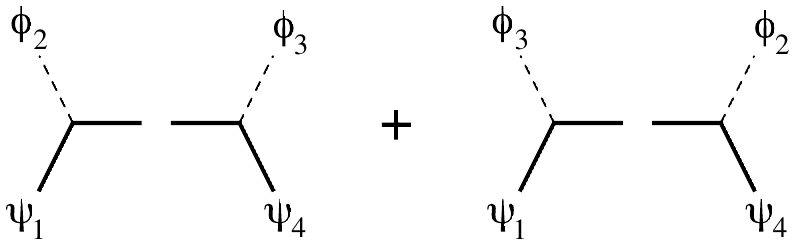}}}\nonumber\\
&=&\frac{g^2\langle\mathbf{1}2\rangle\langle3\mathbf{4}\rangle[32]+g'g[\mathbf{1}2]\langle\mathbf{4}3\rangle\langle2|p_4|3]+c.c.}{m_f(s-m^2_f)}+(2\leftrightarrow 3)\,,
\eqae 
 where by $c.c.$ we are exchanging $\lambda\leftrightarrow \tilde\lambda$ and $g\leftrightarrow g'$. As one can see, since the three-point amplitude was local, the resulting four-point amplitude can be written in a manifest local way with two separate channels. 

A ``slightly" more complicated example would be the process $\gamma^- + t\,\rightarrow gra^++ t$, via a massive spin-$\frac{3}{2}$ exchange:
\eq
\includegraphics[scale=0.5]{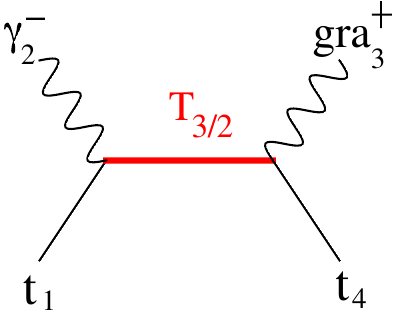}\,.
\eqe
Here, $t_{1,4}$ are the massive top quarks with their mass denoted by $m_t$. The three-point amplitude on both sides are:
\eqa
V_{L}&=&\frac{g}{m^3_t}\langle \mathbf{p}2\rangle^3[\mathbf{1}2]+\frac{g'}{m^3_t}\langle \mathbf{1}2\rangle\langle \mathbf{p}2\rangle^2[\mathbf{p}2],\nonumber\\
V_{R}&=&\frac{g''}{m^3_t}[\mathbf{4}3][\mathbf{p}3]^3\,.
\eqae
There are two tensor structures for $V_L$, reflecting the two distinct way the SL(2,C) indices can distribute. The resulting four-point amplitude is then, 
\eq
\frac{gg''[\mathbf{4}3][2\mathbf{1}]\langle2|p_4|3]^3+g'g''[\mathbf{4}3]\langle\mathbf{1}2\rangle\langle2|p_4|3]^2[32]m_t}{(s-m^2_T)m_t^6}+ (2\leftrightarrow 3)\,,
\eqe 
where $m_T$ is the mass of the spin-$3/2$ particle.

In the above examples, the residues are manifestly local as it is inherited from the three-point amplitude. The only place potential non-locality can occur is when factors of $x$ appear for the three-point amplitude, for example the minimal coupling. Thus in the next section we will focus on minimal coupling for massless spin-1 and 2 particles.

\subsection{Minimal Coupling}
In this subsection we will consider the gluing of minimally coupled higher spin particles. We will first begin with charged particles, which entails the three-point coupling of two massive spin-$S$ state and a positive or negative helicity photon. The three point amplitude is given in eq.(\ref{GeneralMin}), which after dressing with external spinors, the complete amplitude is: 
\eqa
M^{+h}_S=x^h\frac{\langle\mathbf{1}\mathbf{2}\rangle^{2S}}{m^{2S{-}1}},\quad M^{-h}_S=\frac{1}{x^h}\frac{[\mathbf{1}\mathbf{2}]^{2S}}{m^{2S{-}1}}\;.
\eqae 

\subsubsection{Compton Scattering For S $\leq$ 1}
Let us begin with scalar. Here one simply has:
\eq
\vcenter{\hbox{\includegraphics[scale=0.5]{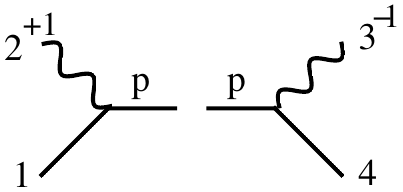}}}\sim m^2\frac{x_{12}}{x_{34}}.
\eqe 
Here the subscripts on $x$ serve to distinguish between different three point vertices. Now since 
\eq
x_{12}\lambda_2=\frac{p_1|2]}{m}\rightarrow x_{12}=\frac{\langle3|p_1|2]}{\langle32\rangle m},\quad \frac{p_4|3\rangle}{m}=\frac{\tla_3}{x_{34}}\rightarrow \frac{1}{x_{34}}=\frac{\langle3|p_4|2]}{[23]m}\,,
\eqe
we see that the residue is given by:
\eq\label{x1x2}
m^2\frac{x_{12}}{x_{34}}=-\frac{\langle3|p_1|2]^2}{t}\,.
\eqe
Again the $s$-channel residue is non-local and must be interpreted as a pole from the other channel! We now have a choice, it can either be interpreted as a massless particle in the $t$-channel, or an $u$-channel massive particle since $-t=u-m^2$ when $s=m^2$. For there to be a $t$-channel massless pole, the vectors must be gluons instead of photons, and we leave this possibility to the later part of this subsection. For the case where one has a $u$-channel massive pole, the amplitude is simply:
\eq
M(\phi_1\gamma_2^+\gamma_3^-\phi_4)=\frac{\langle3|p_1-p_4|2]^2}{4(s-m^2)(u-m^2)}\,,
\eqe
As the amplitude is symmetric under $1\leftrightarrow 4$ exchange, it is guaranteed to be consistent with the $u$-channel factorisation. It is straight forward to see that at H.E. one obtains the usual two adjoint-scalar two gluon, and two charged scalar two photon amplitude.

Let us now consider Compton scattering for general spin. The $s$-channel gluing yields, 
\eq
\vcenter{\hbox{\includegraphics[scale=0.5]{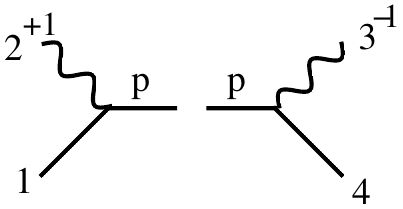}}}\sim\frac{1}{m^{2(S{-}1)}}\frac{x_{12}}{x_{34}}(\langle\mathbf{1}P^I\rangle[P_I\mathbf{4}])^{2S}\,.
\eqe  
Recall that $\frac{x_{12}}{x_{34}}m^2=-\langle3|p_1|2]^2/t$, if we rewrite $t$ as $u-m^2$ and put back the $s$-channel propagator, this has the property that it is symmetric under $1\leftrightarrow4$ (it is the scalar amplitude after all). This means that if $\frac{P^{2S}}{m^{2S}}$ matches to the $u$-channel residue then we are done! Finally using the identity:\footnote{This identity can be derived as follows: $|P^I\rangle[P_I|$ is the internal momentum that satisfies the $s$-channel on-shell constraint,  
\eq
P_{\alpha\dot\alpha}\tilde\lambda^{\dot{\alpha}}_{2}=-mx_{12}\la_{2\alpha},\quad P_{\alpha\dot\alpha}\tilde\lambda^{\dot{\alpha}}_{3}=mx_{34}\la_{3\alpha},\quad P^2=m^2
\eqe 
The solution is given by: 
\eq\label{PIDef}
P_{\alpha\dot\alpha}=\frac{-m^2\lambda_{3\alpha}\tla_{2\dot\alpha}+(p_{1\alpha\dot\beta}\tla_2^{\dot{\beta}})(p_{4\dot\alpha\beta}\la^\beta_3)}{\langle 3|p_1|2]}\,.
\eqe 
Contracting with $\lambda^I_4$ and $\tilde\lambda^I_1$ yields eq.(\ref{FootnoteEq}). }
\eq\label{FootnoteEq}
\frac{\langle \mathbf{1}| P_{I}|\mathbf{4}]}{m}=m\frac{\langle \mathbf{4}3\rangle[\mathbf{1}2]+\langle \mathbf{1}3\rangle[\mathbf{4}2]}{\langle 3|p_1|2]}\,,
\eqe
one derives the following ansatz for the four-point amplitude of minimally coupled general spin-$S$ amplitude,  
\eq\label{GenS}
\frac{\langle3|p_1|2]^{2-2S}}{(s-m^2)(u-m^2)}\left(\langle \mathbf{4}3\rangle[\mathbf{1}2]+\langle \mathbf{1}3\rangle[\mathbf{4}2]\right)^{2S}\,.
\eqe 
Note that the final result has an extra $(-)^{2S}$ sign for spin-$S$ under $1\leftrightarrow4$ exchange. This tells us that charged half integer spins must be fermions, while integer spins are bosons. Thus we've recovered spin-statistics from the principles of Poincare symmetry and unitarity. The result in eq.(\ref{GenS}) contains spurious singularities which cancel for $S\leq1$. This signals that there is something fundamentally different for charged particles of $S\leq1$ and $S>1$. For $S=1/2, 1$ we recover the Compton scattering: 
\eqa\label{GenSCom}
M(\mathbf{1}^{\frac{1}{2}},\gamma_2^{+1},\gamma_3^{-1},\mathbf{4}^{\frac{1}{2}})&=&\frac{\langle3|p_1{-}p_4|2]}{2(s-m^2)(u-m^2)}\left(\langle \mathbf{4}3\rangle[\mathbf{1}2]+\langle \mathbf{1}3\rangle[\mathbf{4}2]\right)\nonumber\\
M(\mathbf{1}^{1},\gamma_2^{+1},\gamma_3^{-1},\mathbf{4}^{1})&=&\frac{(\langle \mathbf{4}3\rangle[\mathbf{1}2]+\langle \mathbf{1}3\rangle[\mathbf{4}2])^2}{(s-m^2)(u-m^2)}\,.
\eqae 
In appendix \ref{FeynDia} we reproduce this result using Feynman diagrams for fermions. 
By studying the H. E. limit, one can easily verify that this is correct. At H.E. for $S=1$ one obtains three terms, two of which are contributions where legs $1$ and $4$ are opposite helicity gluons, and a final one which is when they are both scalars, which are the Goldstone bosons that were eaten in the Higgs mechanism! Note that this is telling us that the Higgs mechanism provides a way to ``unify" the independent massless amplitudes in the IR. We will discuss this phenomenon in more detail in section \ref{Higgs}.

Now in the above discussion the result from the $s$-channel gluing can be matched to the $u$-channel if we have a single species of spin-$S$. If there are multiple species, then similar to the massless discussion in section \ref{Massless34}, we should assign a matrix  $T^{a}_{ij}$ to each vertex, and due to $[T^a,T^b]\neq0$, the matching to the $u$-channel will be off by a piece that is proportional to $f^{abc}T^{c}_{ij}$. This mismatch is a sign that the $t$-channel pole from the $s$-channel factorisation should be assigned into a physical massless pole, i.e. revealing the presence of an non-abelian vector. For this to hold we should show that the $s$-channel residue admits this interpretation. Indeed taking a scalar for example, $\langle3|p_1|2]^2/(s-m^2)t$ can be matched to the $t$-channel residue since
\eq
\vcenter{\hbox{\includegraphics[scale=0.5]{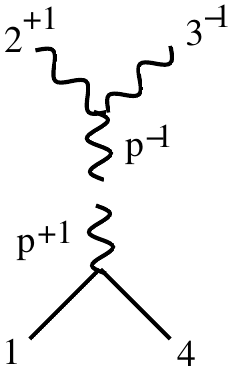}}} x_{14} \frac{\langle 3P\rangle^3}{\langle P2\rangle\langle23\rangle}=-f^{ab}\,_cT^c\frac{\langle3|p_1|3]\langle3|p_1|P]}{x_{14}m\langle P2\rangle}=\langle3|p_1|2]^2\left(\frac{T^aT^b}{s-m^2}{+}\frac{T^bT^a}{u-m^2}\right)\,.
\eqe 
The last equality utilizes the fact that when $t=0$, $s{-}m^2={-}(u{-}m^2)$. Thus the final amplitude is given by:
\eqa
&&\langle3|p_1|2]^{2-2S}\left(\langle \mathbf{4}3\rangle[\mathbf{1}2]{+}\langle \mathbf{1}3\rangle[\mathbf{4}2]\right)^{2S}\frac{1}{t}\left[\frac{(T^aT^b)_{ij}}{s-m^2}{+}\frac{(T^bT^a)_{ij}}{u{-}m^2}\right]\,.
\eqae

\subsubsection{Compton scattering for $S>1$}

The ansatz for general minimal coupling in eq.(\ref{GenS}) appears to contain non-physical poles for $S>1$. Of course this cannot be the final story since there's an abundance of charged higher spin-states in nature, and although we know that they are not fundamental, it has no bearing on the existence of S-matrix for low energy scattering. In deriving eq.(\ref{GenS}), we started from the $s$-channel residue and analytically continued $P_I$ to a form that is manifestly $2\leftrightarrow 3$ and $+\leftrightarrow -$ symmetric, and thus can be directly matched to $u$-channel residues. This is not entirely necessary, since the full amplitude can contain terms that only contain $s$ and not $u$-channel pole. Thus the very fact that eq.(\ref{GenS}) gives us non-physical poles for $S>1$ is precisely telling us that such terms must be present.

To see this subtlety in detail, let's consider minimal coupling for spin-$3/2$, for which the gluing from $s$-channel yields:
\eq
-\frac{\langle3|p_1|2]^2}{t}\left(\frac{\langle \mathbf{4}3\rangle[\mathbf{1}2]+\langle \mathbf{1}3\rangle[\mathbf{4}2]}{\langle 3|p_1|2]}\right)^{3}\,.
\eqe
First note that by using eq.(\ref{Convert}) one can rewrite the internal propagator in a mostly local form:
\eq
\frac{\langle \mathbf{4}3\rangle[\mathbf{1}2]+\langle \mathbf{1}3\rangle[\mathbf{4}2]}{\langle 3|p_1|2]}=\left(\frac{[\mathbf{1}\mathbf{4}]}{m}+\frac{\langle\mathbf{4}2\rangle[2\mathbf{1}]-\langle\mathbf{1}2\rangle[2\mathbf{4}]}{2m^2}\right)+\frac{t[2\mathbf{1}]\langle3\mathbf{4}\rangle}{2m^2\langle 3|p_1|2]}\equiv A+B
\eqe
The first two terms, denoted as $A$, are local at the expense of introducing extra inverse powers of $m$ and are anti-symmetric under $1\leftrightarrow4$, inheriting the symmetry properties of its parent. This guarantees that the local terms can be combined with the pre-factor and reproduce the correct residue on the $u$-channel pole. The last term, denoted as $B$, while being spurious, does not contribute to the $u$-channel residue and thus we are free to rewrite it in a local form using $s$-channel kinematics:
\eq
\left. \frac{t[2\mathbf{1}]\langle3\mathbf{4}\rangle}{2m^2\langle 3|p_1|2]}\right|_{s=m^2}=-\frac{\langle\mathbf{4}3\rangle[32]\langle2\mathbf{1}\rangle}{2m^3}
\eqe
Now expanding $(A+B)^3$, only the $A^3$ term will contribute to both $s$- and $u$- propagators, while terms with $B$ will contribute solely to $s$-channel propagators. Putting everything together, one finds the following local form for the amplitude:
\eqa
M(\mathbf{1}^{\frac{3}{2}},\gamma_2^{+1},\gamma_3^{-1},\mathbf{4}^{\frac{3}{2}})&=&\frac{\langle3|p_1|2]^2}{(u-m^2)(s-m^2)}A^3-\left\{\frac{\langle3|p_1|2][2\mathbf{1}]\langle3\mathbf{4}\rangle}{2m^2(s-m^2)}\times\right.\nonumber\\
&&\left.\left(3A^2-3A\frac{\langle\mathbf{4}3\rangle[32]\langle2\mathbf{1}\rangle }{2m^3}+\frac{\langle\mathbf{4}3\rangle^2[32]^2\langle2\mathbf{1}\rangle^2 }{4m^6}\right)+(1\leftrightarrow 4)\right\}
\eqae 
We now see that in the final local form, all terms contain $1/m$ factors and becomes singular in the H.E. limit. In other words, the obstruction of taking $m\rightarrow0$ reflects the absence of a consistent massless high energy amplitude. For example the leading term in $1/m$ that will contribute to $M(1^{+\frac{3}{2}},\gamma_2^{+1},\gamma_3^{-1},4^{+\frac{3}{2}})$ at high energies is given by:
 \eqa\label{HigherSpinHE}
 \frac{\langle3|p_1|2]^2}{(u-m^2)(s-m^2)}\frac{[\mathbf{1}\mathbf{4}]^3}{m^3}\rightarrow \frac{\langle31\rangle^2[12]^2}{us}\frac{[14]^3}{m^3}.
 \eqae
As we will elaborate below, this is the concrete sense in which charged particles with spin S $\geq 3/2$ cannot be ``elementary", the same conclusion holds for any particles at all of spin S $\geq 5/2$ that can consistently couple to gravity.

\subsubsection{Graviton Compton Scattering}
Let us again begin with scalars, with the massive scalars are on legs $1,4$, a positive and negative helicity graviton on legs $2,3$ respectively. The $s$-channel residue is given as:
\eq\label{x1x2Grav}
\frac{m^4}{M^2_{pl}}\frac{x^2_{12}}{x^2_{34}}=\frac{\langle3|p_1|2]^4}{t^2M^2_{pl}}\,,
\eqe
where $M_{pl}$ is the Plank mass. As with the massless discussion we now have double pole in $t$, which can be identified as the massive pole $1/(u-m^2)$ and a massless $1/t$ pole. Thus the four-point amplitude is simply 
\eq\label{ScaGrav}
-\frac{\langle3|p_1|2]^4}{(s-m^2)(u-m^2)tM^2_{pl}}\,.
\eqe
It is instructive to verify that the massless pole is correct. Let us take the residue at $t=0$, in the kinematics where $\langle i j\rangle=0$. The residue of eq.(\ref{ScaGrav}) is
\eq
-\frac{\langle3|p_1|2]^4}{\langle3 |p_1|3]\langle2 |p_1|2]M^2_{pl}}\,.
\eqe
Since $\langle i j\rangle=0$, the massless three-point amplitude should be  $\overline{\rm MHV}$, and one has
\eq
\vcenter{\hbox{\includegraphics[scale=0.4]{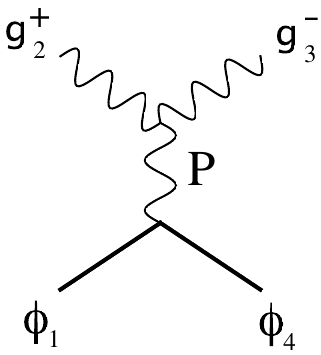}}}\quad\frac{[ 2P]^6}{[ 23]^2[3P]^2}\frac{1}{x^2_{14}M^2_{pl}}=\frac{[ 2P]^4[ 2|p_1|P\rangle^2}{[ 23]^2[3P]^2m^2M^2_{pl}}=\frac{[ 2P]^2[ 2|p_1|3\rangle^2}{[3P]^2M^2_{pl}}
\eqe
where $P$ is the massless internal momenta. Finally using the identity
\eq
\frac{[2P]^2}{[3P]^2}=\frac{[2P][2|p_1|P\rangle}{[3P][3|p_1|P\rangle}=-\frac{[2|p_1|3\rangle^2}{[3|p_1|2\rangle[2|p_1|3\rangle}=-\frac{[2|p_1|3\rangle^2}{\langle3 |p_1|3]\langle2 |p_1|2]}
\eqe
where in the last line, we've applied Schouten on the denominator, keeping in mind that $\langle23\rangle=0$. Thus we see that eq.(\ref{ScaGrav}) yields correct factorization in all channels.

For massive higher-spin particles, we again use the mixed representation. The $s$-channel residue yields:  
\eq
\vcenter{\hbox{\includegraphics[scale=0.4]{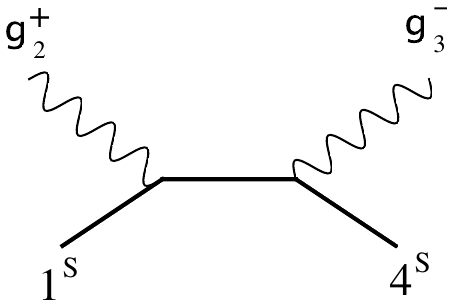}}}\quad\frac{x^2_{12}}{x^2_{34}}\frac{m^4}{M^2_{pl}}\left(\frac{P^{2S}_I}{m^{2S}}\right)_{\alpha_1\cdots\alpha_{2S},\dot{\alpha}_1\cdots\dot{\alpha}_{2S}}\,.
\eqe 
Using the explicit form for $P_I$ in eq.(\ref{PIDef}), we find that the residue, after contracting with the external $(\lambda^I_1,\lambda^I_4)$ is simply  
\eq
\frac{\langle3|p_1|2]^4}{t^2M^2_{pl}}\left(\frac{\langle \mathbf{4}3\rangle[\mathbf{1}2]+\langle \mathbf{1}3\rangle[\mathbf{4}2]}{\langle 3|p_1|2]}\right)^{2S}\,.
\eqe
Thus for $S\leq2$ we find a perfectly local four-point amplitude given by:
\eq
-\frac{\langle3|p_1|2]^4}{(s-m^2)(u-m^2)tM^2_{pl}}\left(\frac{\langle \mathbf{4}3\rangle[\mathbf{1}2]+\langle \mathbf{1}3\rangle[\mathbf{4}2]}{\langle 3|p_1|2]}\right)^{2S}\,.
\eqe
For $S>2$, we see that the formula ceases to be local. Similar to our photon coupling analysis, this indicates that the residue of $s$-channel must be separated into pieces that will combine with other channels and pieces that don't.

\subsection{Massive higher spins cannot be elementary}
We have seen that Compton scattering amplitudes for particles of high
enough spin do not have a healthy high-energy limit, growing as powers
of $(p/m)$. Of course so long as the gauge/gravitational couplings are
small, these amplitudes do not become $O(1)$ till energies
parametrically above the particle mass $m$, so in that sense no
inconsistency is encountered in the effective theory of a single
massive higher spin particle till a cutoff parametrically above its
mass. Nonetheless, the sickness of the $m \to 0$ limit does show that
a single massive higher spin particle cannot be ``elementary", and
that any consistent theory for such particles must also include new
particle states with a mass comparable to $m$. As an example, suppose
we have some strongly-interacting QCD-like gauge theory; can such a
theory have a spectrum consisting of bound states of high spin, with a
parametrically large gap up to higher excited states? Our analysis
suggests that this is impossible. We can imagine weakly gauging a
global symmetry of the theory, or coupling the system to gravity. The
total cross-section for e.g. $\gamma \gamma \to X$ should be bounded
by $\sigma < C \times e^4/s$ for some constant $C$ characterizing the
current four-point amplitude.  But if we have a charged higher spin
particle, just the cross-section for its production would grow as
$e^4/s \times (s/m^2)^n$, and if there is a parametrically large gap
up to other particle states this will exceed the bound when $(s/m^2)^n
> C$. Of course this is a somewhat qualitative argument, but we
believe it captures the essence of why higher-spin massive particles
must be composite.  A sharpening of the argument may be able to give a
more quantitative bound for the scale beneath which new particles must
appear.

We can ask if the presence of new states in the propagator can tame this high-energy behaviour by cancelling the $1/m^6$ singularity in eq.(\ref{HigherSpinHE}). In other words consider the case where one has a new spin $S'$ state with the similar mass as the $S=\frac{3}{2}$, then one can include the contribution: 
\eq
\vcenter{\hbox{\includegraphics[scale=0.5]{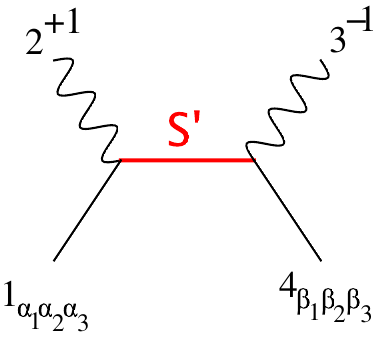}}}\,.
\eqe 
If $S'\neq S$, then in the degenerate mass limit, it is easy to see that the three point amplitude cannot involve the pure $x$ dependent pieces and thus the residue must be local. This then tells us that the contribution of such terms in the high energy limit must take the form $\frac{n_s}{sm^\alpha}+\frac{n_u}{um^\alpha}$ for some $\alpha$, and $n_s, n_u$ is some local function in kinematic invariants. This has a distinct high energy behaviour than eq.(\ref{HigherSpinHE}) which behaves as $1/su$, and thus cannot cancel.\footnote{Strictly speaking, due to our helicity choice eq.(\ref{HigherSpinHE}) really only has an $s$-channel pole at H.E. The bad H.E. behaviour in both channels will be present for other helicity configurations. } For $S'=S$, if the masses are not identical then the residue is again local and we have the same issue. If the masses are the same, then one simply obtains the exact same form as eq.(\ref{HigherSpinHE}) with identical signs, and the H.E. behaviour is again untamed.  

Thus even with finite number of states with comparable mass, the sick H.E. limit still rules out isolated charged higher spin state as a fundamental particle. The above analysis does provide a loop hole: one can have an infinite tower of ever increasing higher spin states. While their presence in the propagator only produces terms with single poles in the H.E. limit, an infinite sum of $n_s/s$ terms can produce poles in $u$ if the degree of polynomial for $n_s$ unbounded. That is, if the exchanged state has unbounded spin. This is precisely what happens for string theories which contain massive higher spin states.

\subsection{All Possible Four Particle Amplitudes}
Having discussed the four-particle amplitudes associated with the most familiar and important three-particle interactions, let us finally turn to computing {\it all} possible four-particle amplitudes. As we have seen when there are no ``$x$" factors involved, we have local residues and the construction of four-particle amplitudes is trivial. We will therefore concentrate on discussing the cases where consistent factorization is non-trivial, which involve having at least one minimal coupling with an $``x"$ factor, but now allowing for the most general set of other couplings. We will see (once again) that consistency demands that the minimal couplings have the standard Yang-Mills/gravitational forms, and that the other interactions have to be (globally) Yang-Mills invariant. But it {\it is} then possible to find consistently factorizing four-point amplitudes for any choice of three-particle interactions satisfying these conditions.

\subsubsection{All Massive amplitude}\label{Scalarxx}
This is the simplest, since we only need to consider the massless exchange. Consider the exchange of a massless-photon, for external scalars we have:
\eq
\vcenter{\hbox{\includegraphics[scale=0.6]{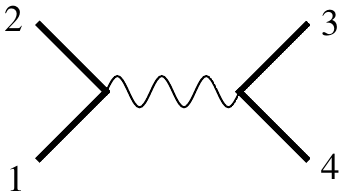}}}:\quad m^2\left(\frac{x_{12}}{x_{34}}+\frac{x_{34}}{x_{12}}\right)\,,
\eqe 
where the two terms correspond to the two different helicities. Using $x_{12}\la_P\equiv \frac{p_1}{m}|P]$, $x_{34}\la_P\equiv \frac{p_3}{m}|P]$ and  $P=p_1+p_2$, we find:
\eq
\frac{1}{s}m^2\left(\frac{x_{12}}{x_{34}}+\frac{x_{34}}{x_{12}}\right)=\frac{1}{s}\left(\frac{\langle \eta|p_1|P]\langle P|p_3|\xi]}{\langle \eta P\rangle[P\xi]}+\frac{\langle \eta|p_3|P]\langle P|p_1|\xi]}{\langle \eta P\rangle[P\xi]}\right)=\frac{2(p_1\cdot p_3)}{s}\,,
\eqe
where one uses the fact that $\langle P|p_i|P]=0$ for any external momenta $p_i$. This is not the complete answer, as one expects $\frac{(p_1\cdot p_3)-(p_2\cdot p_3)}{s}$ from minimal coupling. The difference is $s/s$ and thus have no factorization poles. The correct answer can be inferred from symmetry arguments under $1\leftrightarrow2$ exchange. Thus the correct completion is 
\eq\label{ScalarXC}
\frac{1}{s}m^2\left(\frac{x_{12}}{x_{34}}+\frac{x_{34}}{x_{12}}\right)=\frac{(p_1-p_2)\cdot p_3}{s}\,,
\eqe
For the exchange of a general massless spin $S$ state, we simply get a factor of $((p_1-p_2)\cdot p_3)^{S}$ for the numerator.

Now we let the external particles carry spin. For simplicity we will consider the case where all four particles are of the same spin. Then the residue for the most general coupling is given by: 
\eqa
&&\frac{x_{12}}{m^{2S{-}2}}\left[\langle\mathbf{1}\mathbf{2}\rangle^{2S}+\sum_{i=0}^{2S-1}\left(a_i\langle\mathbf{1}\mathbf{2}\rangle^{i}\left(\frac{\langle\mathbf{1}P\rangle[\mathbf{2}P]}{m}\right)^{2S-i}+\tilde{a}_i\langle\mathbf{1}\mathbf{2}\rangle^{i}\left(\frac{\langle\mathbf{2}P\rangle[\mathbf{1}P]}{m}\right)^{2S-i}\right)\right]\nonumber\\
\times&& \frac{1}{x_{34}}\left[[\mathbf{3}\mathbf{4}]^{2S}+\sum_{i=0}^{2S-1}\left(b_i[\mathbf{3}\mathbf{4}]^{i}\left(\frac{\langle\mathbf{3}P\rangle[\mathbf{4}P]}{m}\right)^{2S-i}+\tilde{b}_i[\mathbf{3}\mathbf{4}]^{i}\left(\frac{\langle\mathbf{4}P\rangle[\mathbf{3}P]}{m}\right)^{2S-i}\right)\right]
\eqae
where $a_i,b_i, \tilde{a}_i,\tilde{b}_i$ parameterize all possible coupling to the photon, and for parity invariant theories we have $a_i=b_i$ and $ \tilde{a}_i=\tilde{b}_i$. Since besides the leading term in the square brackets, each of the terms contains $|P]\langle P|$ which can readily convert $\frac{x_{12}}{x_{34}}$ into local forms, thus we only need to worry about the term
\eq
\frac{1}{m^{2S{-}2}}\left(\frac{x_{12}}{x_{34}}\langle\mathbf{1}\mathbf{2}\rangle^{2S}[\mathbf{3}\mathbf{4}]^{2S}+\frac{x_{34}}{x_{12}}[\mathbf{1}\mathbf{2}]^{2S}\langle\mathbf{3}\mathbf{4}\rangle^{2S}\right)\,,
\eqe 
where we've included the contribution where the photon helicity is flipped. Finally, using the identity:
\eq
[\mathbf{1}\mathbf{2}]=\langle\mathbf{1}\mathbf{2}\rangle+\frac{\langle \mathbf{1}|P|\mathbf{2}]}{m}\,,
\eqe
introduces $|P]\langle P|$ that can again be used to absorb the $x$-factors leaving behind
\eq
\frac{\langle\mathbf{1}\mathbf{2}\rangle^{2S}\langle\mathbf{3}\mathbf{4}\rangle^{2S}}{m^{2S{-}2}}\left(\frac{x_{12}}{x_{34}}+\frac{x_{34}}{x_{12}}\right)=\frac{\langle\mathbf{1}\mathbf{2}\rangle^{2S}\langle\mathbf{3}\mathbf{4}\rangle^{2S}}{m^{2S}}(p_1-p_2)\cdot p_3\,,
\eqe
where we've used eq.(\ref{ScalarXC}). Thus we see that the massless gluing of any three point vertex can be converted into a local form. For more general external spins, the analysis is the same albeit more complicated. 

\subsubsection{Three-massive one-massless}
If we have three-massive legs, the dangerous $x$-factors can occur in two types of diagrams for the $s$-channel residue:
\eq
\vcenter{\hbox{\includegraphics[scale=0.6]{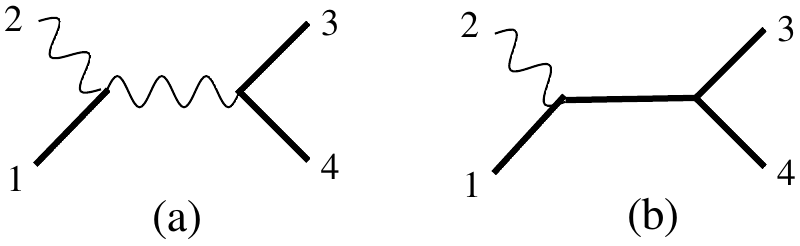}}}\,.
\eqe
Let us first consider the case where the solid lines are massive scalars, and the wavy line is the positive helicity photon. Diagram $(a),(b)$ gives:
\eq
(a)\quad\frac{[2P]^2}{x_{34}}=\frac{[2P][2|p_3|P\rangle}{m}\,,(b)\quad m x_{1P}=\frac{[2|p_1|\xi\rangle}{\langle2\xi\rangle}\,.
\eqe 
The first is manifestly local. For the second, let's consider the all massive vertex being $\phi\phi'\phi''$ vertex, and the photon only couples to $\phi$ and $\phi'$  with coupling $e,e'$. Then gluing leads to:
\eq
e\frac{[2|p_1|\xi\rangle}{\langle2\xi\rangle(s-m^2)}+e'\frac{[2|p_4|\xi\rangle}{\langle2\xi\rangle(u-m^2)}=(e+e')\frac{[2|p_4|\xi\rangle}{\langle2\xi\rangle(u-m^2)}+e\frac{[2|p_4p_1|2]}{(s-m^2)(u-m^2)}\,,
\eqe 
where legs $1,4$ are $\phi,\phi'$ respectively. We see that only when the charge is conserved, i.e. $e+e'=0$ does the $\langle2\xi\rangle$ pole cancels and the amplitude becomes local. If the scalars were all charged with charges $e,e^\prime, e^{\prime \prime}$, the same analysis would tell us that $e + e^\prime + e^{\prime \prime} = 0$. Next suppose the photon was instead a gluon, with the scalars carry indices $i, i^\prime, i^{\prime \prime}$ and the three-point amplitude given by $c_{i i^\prime i^{\prime \prime}}$. We have already seen that consistency demands the couplings to the gluons $T^{a}_{ij}, T^{a}_{i^\prime j^\prime}, T^{a}_ {i^{\prime \prime} j^{\prime \prime}}$ be generators in some representation of the Yang-Mills group. Then we discover that we must have $T^{a}_{i j} c_{j i^\prime i^{\prime \prime}} + T^{a}_{i^\prime j^\prime} c_{i j^\prime i^{\prime \prime}} + T^{a}_{i^{\prime \prime} j^{\prime \prime}} c^{i i^\prime j^{\prime \prime}}=0$, in other words the cubic interaction must be invariant under the (global) Yang-Mills symmetry.  Finally, for graviton, gluing to a $\phi^3$ vertex leads to:
\eq
g_1\frac{[2|p_1|\xi\rangle^2}{M_{pl}\langle2\xi\rangle^2(s-m^2)}+g_3\frac{[2|p_3|\xi\rangle^2}{M_{pl}\langle2\xi\rangle^2(t-m^2)}+g_3\frac{[2|p_4|\xi\rangle^2}{M_{pl}\langle2\xi\rangle^2(u-m^2)}\,,
\eqe
where we've let all three scalars couple to gravity. Again after rearranging the terms, one finds that the auxiliary spinor drops out only if $g_1=g_2=g_3$, and one arrives at:
\eq
\frac{g_1}{M_{pl}}\frac{[2|p_1p_3|2]^2}{(s-m^2)(u-m^2)(t-m^2)}\,.
\eqe
Thus we see that coupling to photons, the consistency of the four-point amplitude requires charge to be conserved, for a gluon it requires the particles to be in the adjoint representation, and finally for a graviton, it leads to the equivalence principle. Note that this discussion does not refer to any gauge redundancy and the independence there of. On the other hand, the astute reader will recognize that the factor $\frac{[2|p_1|\xi\rangle}{\langle2\xi\rangle}$ can be identified with $\epsilon_2\cdot p_1$ from Feynman rules,  where $\lambda_\xi$ is the reference spinor for the polarization vector $\epsilon_2$. Indeed from the photon and graviton soft-theorem~\cite{WeinbergSoft}, it is precisely this factor whose gauge invariance (Ward identity) demands the conservation of charges and equivalence principle. Here, there's no gauge redundancy, the auxiliary spinor $\lambda_\xi$ is simply a projection of eq.(\ref{xDef}), and the independence thereof is the requirement that factorization is consistent to all solutions of $x$ defined through eq.(\ref{xDef}).

Again the same applies if we consider external spinning particles. For example for massive spin-1, diagram $(a)$ yields, 
\eqa
(a)&&\frac{[2P]^3\langle\mathbf{1}2\rangle\langle{1}P\rangle}{m^5}\frac{1}{x_{34}}\left(\langle\mathbf{3}\mathbf{4}\rangle-x_{34}\frac{\langle\mathbf{3}P\rangle\langle\mathbf{4}P\rangle}{m}\right)^2\nonumber\\
&=&[2P]^2\langle\mathbf{1}2\rangle\langle{1}P\rangle\frac{[2|p_3|P\rangle}{m^6}\left(\langle\mathbf{3}\mathbf{4}\rangle-\frac{\langle\mathbf{4}|p_3|P]\langle\mathbf{3}P\rangle}{m^2}\right)^2
\eqae
where again the residue is local. For diagram $(b)$ the only non-locality originates from the minimal coupling piece, and hence one recovers the same condition as before.  
\subsubsection{One-massive three-massless}
So far we have found that all potential non-localities can be converted into local expressions, and hence the residue of one-channel does not encode information with respect to other channels. For three massless particles things are more interesting. The potential $s$-channel factorization diagrams are:
\eq
\vcenter{\hbox{\includegraphics[scale=0.5]{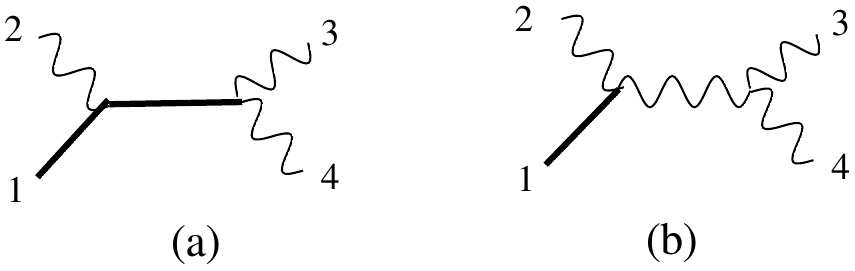}}}\,.
\eqe 
For our purpose, only minimal coupling is relevant for the two massive one massless vertex in (a). We will consider a massive scalar coupled to abelian and non-abelian vectors.  

First for the abelian case we only need to consider diagram $(a)$. Taking all vectors to be plus helicity, one finds the $s$-channel residue given by 
\eq\label{smassive}
(a)\quad x_{12}[34]^2=\frac{\langle3|p_1|2][34]^2}{m\langle23\rangle}=\frac{m[42][34]}{\langle23\rangle}\,.
\eqe
The appearance of $\langle23\rangle$ seems to indicate an illegal massless pole. However since $s=m^2$, this can be identified as a $u$-channel massive pole, $u-m^2=-t$. Thus one can write the amplitude as
\eq
\frac{m[42][34][23]}{(s-m^2)(u-m^2)}
\eqe 
Note however the extra $-$ sign under the $2\leftrightarrow 3$ exchange will lead to the violation of spin-statistics for identical vectors. Thus we see that minimally coupled scalars are incompatible with a di-photon coupling. Indeed from the action point of view, this is simply the statement that the U(1) symmetry of a charged scalar forbids the appearance of $\phi F^2$ coupling. Thus there is no such four particle amplitude for the abelian theory. For the non-abelian case, one must also consider diagram (b), which yields
\eq
(b):\quad\frac{[2P]^2}{m}\frac{[34]^3}{[3P][P4]}=\frac{m^3[34]}{\langle23\rangle\langle24\rangle}\,.
\eqe
We gain find the illegal pole $1/\langle24\rangle$. Since we are considering the non-abelian theory we can consider the colour stripped amplitude and convert the spurious pole into a legal $t$-channel massive pole. This suggest us to write
\eq
M_4(\mathbf{1},2^+,3^+,4^+)=\frac{m^3[24][23][34]}{st}\left(\frac{1}{(t-m^2)}+\frac{1}{(s-m^2)}+\frac{1}{m^2}\right)\,,
\eqe
where we've added the massless $t$-channel image, and the extra $1/m^2$ is to guarantee that both massless channels factorises correctly. One can check the $s$-channel massive residue, which was given in eq.(\ref{smassive}), matches when taken into account that $\langle34\rangle=\frac{m^2}{[43]}$. Note that the amplitude vanishes as $m\rightarrow0$ as it should.

Now let's move on to the case where there are external spins. For example, one can consider a massive spin-1 particles couple to three massless vectors. If the vector is abelian, Yang's theorem tells us that there is no $\includegraphics[scale=0.3]{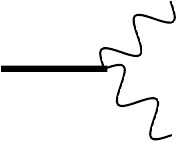}$ vertex to consider, and thus there are no factorizable four-point amplitude to consider. We instead begin with a massive vector and three gluons. We will start with colour stripped all plus-helicity gluons,  whose residue for the massless $s$-channel is given as: 
\eqa
\vcenter{\hbox{\includegraphics[scale=0.5]{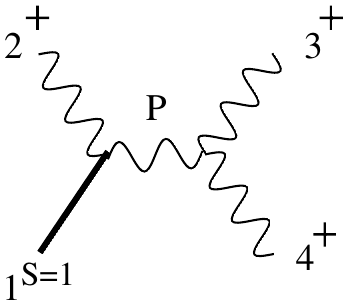}}} &\rightarrow&\langle\mathbf{1}2\rangle\langle\mathbf{1}P\rangle\frac{[P2]^3}{m^4}\times\frac{[34]^3}{[4P][P3]}\,.
\eqae
Now since the vertex on one side contains high power of momenta, there are different ways of rewriting this residue which are equivalent on the kinematics $\langle34\rangle=0$. We will choose the representation where one separates the various pieces that contain information on other channels. More precisely, we have: 
\eqa\label{SX}
\langle\mathbf{1}2\rangle\langle\mathbf{1}P\rangle\frac{[P2]^3}{m^4}\times\frac{[34]^3}{[4P][P3]}=\langle\mathbf{1}2\rangle\langle\mathbf{1}P\rangle[P2][34]\frac{([P3][42]+[P4][23])^2}{m^4[4P][P3]}\nonumber \\
=[34]\left(\frac{2[\mathbf{1}|p_2|\mathbf{1}\rangle[42][23]}{m^3}+\frac{[42]\langle\mathbf{1}|p_4p_2|\mathbf{1}\rangle}{m^2\langle23\rangle}+\frac{[23]\langle\mathbf{1}|p_2p_3|\mathbf{1}\rangle}{m^2\langle24\rangle}\right)
\eqae
where the last equality sign is understood to hold on $\langle34\rangle=0$ kinematics. We see that unavoidably there is an $1/\langle24\rangle$ pole in the $s$-channel massless residue, which is spurious unless it can be interpreted as a t-channel pole $1/(t-m^2)$. Thus the massless residue for the amplitude tells us that there must be a two massive vector, one gluon matrix element that must be present to explain the apparent spurious singularity. The contribution of this matrix element for the $s$-channel is given by:
\eq
\vcenter{\hbox{\includegraphics[scale=0.5]{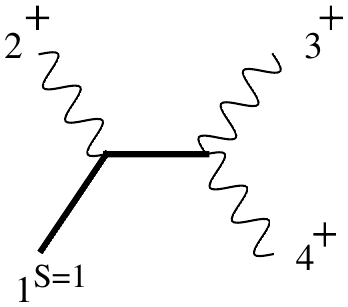}}} \rightarrow \left(x_{12}\frac{\langle\mathbf{1}\mathbf{P}\rangle^2}{m}\times \langle \mathbf{P}3\rangle\langle \mathbf{P}4\rangle\frac{[34]^3}{m^4}\right)=\frac{[42][34]\langle\mathbf{1}|p_3 p_4|\mathbf{1}\rangle}{\langle32\rangle m^2}\,.
\eqe
This suggests that we begin with the following piece which factorises correctly on the $s$ and $t$-channel massive pole:
\eq\label{MassiveForm}
\frac{[42][34]\langle\mathbf{1}|p_3p_4|\mathbf{1}\rangle}{\langle32\rangle (s-m^2)m^2}-\frac{[42][32]\langle \mathbf{1}|p_3p_2|\mathbf{1}\rangle}{\langle34\rangle (t-m^2)m^2}\,.
\eqe 
Note that the above is symmetric in $(2\leftrightarrow4)$ and contains $\langle34\rangle, \langle23\rangle$ poles as well. Taking $\langle34\rangle \rightarrow0$, only the second term in eq.(\ref{MassiveForm}) contributes to its residue:
\eqa
Res\left[-\frac{[42][32]\langle \mathbf{1}|p_3p_2|\mathbf{1}\rangle}{\langle34\rangle (t-m^2)m^2}\right]\bigg|_{\langle34\rangle=0}&=&\frac{[23]\langle \mathbf{1}|p_2p_3|\mathbf{1}\rangle}{\langle24\rangle  m^2}\,.
\eqae
This is nothing but the spurious residue appearing in eq.(\ref{SX})!

Putting the information built from the $s$- and $t$-channel massive, and $s$-channel massless residue together, leads to:
\eqa\label{CurrentPlus}
M(\mathbf{1}^{S=1}2^+3^+4^+)&=&\frac{[42][23][34]}{m^2}\left\{\frac{1}{t}\left(\frac{\langle\mathbf{1}|p_3p_4|\mathbf{1}\rangle}{ (s-m^2)}+\frac{2\langle\mathbf{1}|p_1p_4|\mathbf{1}\rangle}{ m^2}\right)+\frac{\langle\mathbf{1}|p_4p_2|\mathbf{1}\rangle}{st}\right.\nonumber\\
&+&\frac{1}{s}\left(\frac{\langle\mathbf{1}|p_3p_2|\mathbf{1}\rangle}{ (t-m^2)}+\frac{2\langle\mathbf{1}|p_1p_2|\mathbf{1}\rangle}{m^2}\right)\,.
\eqae
The matching to the massless $t$-channel is straight forward given that the above is symmetric in (2$\leftrightarrow$3). Note that unlike the uniqueness of the one massive two massless amplitude, a priori the coupling between the two massive and one massless vector does not have to match that of minimal coupling. It is the consistency between the massless and massive factorisation that fixes this choice. A quick recap: beginning with the massless residue, for which the three-point coupling involving the massive spin-1 is unique, the anti-symmetric property with respect to the massless legs tells us that the massive state must be in adjoint rep of the color group. Then the presence of an $1/\langle24\rangle$ singularity becomes spurious unless it arises from the massive propagators evaluated on degenerate kinematics. Thus the massless residue in one channel encodes the massive residue in the other.  

For the other helicity components, the derivation is simpler as one can construct the full amplitude from the residue of the massive channel, and we simply list the results:
\eqa
\vcenter{\hbox{\includegraphics[scale=0.5]{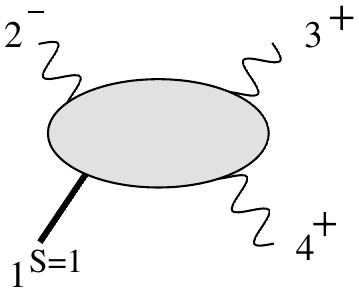}}} &\rightarrow& \frac{[3|p_1|2\rangle\langle23\rangle[34][\mathbf{1}3][\mathbf{1}4]}{m^4t(s-m^2)}\nonumber\\
\vcenter{\hbox{\includegraphics[scale=0.5]{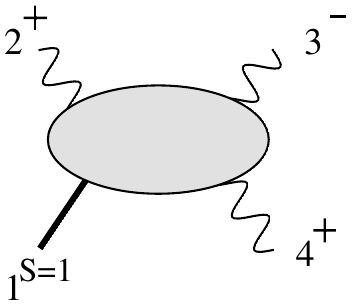}}} &\rightarrow&\frac{[\mathbf{1}2][\mathbf{1}4][24]^2\langle32\rangle\langle43\rangle}{stm^2}\,.
\eqae
In the first line, we've listed the amplitude in the dotted frame for simplicity. One can check that the leading contribution for the H.E. limit of this amplitude yields the amplitude generated by the $tr(F^3)$ extension of Yang-Mills theory. 

As a final example, let's consider a possible singlet massive spin-2 particle that interacts with gluons via a higher-dimensional operator $RF^2$. For the one massive three positive-helicity gluon amplitude, we   expect that the final result is cyclic invariant in $(2,3,4)$. The massless $s$-channel residue can now be written as, 
\eqa
&&\vcenter{\hbox{\includegraphics[scale=0.5]{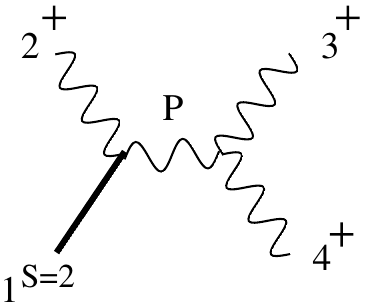}}} \rightarrow\langle\mathbf{1}2\rangle^2\langle\mathbf{1}P\rangle^2\frac{[P2]^4}{m^7}\times\frac{[43]^3}{[4P][P3]}\nonumber\\
\;&=& \frac{[43][\mathbf{1}| p_2|\mathbf{1}\rangle^2}{m^3}\left(\frac{[42]}{\langle23\rangle}+\frac{[32]}{\langle24\rangle}\right)\,,
\eqae
where we've suppressed the symmetrized SL(2,C) indices, keeping in mind they are distributed amongst the $(p_i\cdot p_j)$s. Putting back the massless propagator, this suggest that we start with:
\eq\label{MasslessMatch}
\frac{[\mathbf{1}| p_2|\mathbf{1}\rangle^2}{m^3\langle34\rangle}\left(\frac{[42]}{\langle23\rangle}+\frac{[32]}{\langle24\rangle}\right)\,.
\eqe
The above result contains other additional poles, which under cyclic rotation $(2,3,4)$, will generate terms that will modify the original $\frac{1}{\langle34\rangle}$ residue. Thus before summing over its cyclic image, we should augment eq.(\ref{MasslessMatch}) with terms that kill the extra poles in  $\langle24\rangle$ and $\langle23\rangle$. Putting everything together, we find:
\eqa
M(\mathbf{1}^{S=2}2^+3^+4^+)&=&\frac{[\mathbf{1}| p_2|\mathbf{1}\rangle^2}{m}\frac{[43][32][42]}{stu}-\frac{[2\mathbf{1}]^2}{m^3}\left(\frac{\langle 24\rangle[43][32]}{st}[4\mathbf{1}]^2+\frac{\langle 23\rangle[43][42]}{su}[3\mathbf{1}]^2\right)\nonumber\\
&+&cyclic(2,3,4)\,.
\eqae

We give further examples of massive amplitudes involving one massive higher spin and non-identical spin massless particles in appendix~\ref{FurtherExp}.

\subsection{The spinning polynomial basis}
The fact our on-shell formalism provides a convenient basis to classify distinct three-point couplings lends itself to another important application: construction of a basis polynomial to expand the four-point amplitude. A well known example for such a polynomial is the Gegenbauer polynomial, or its four dimensional representation the Legendre polynomial, as a basis for the four-point scalar amplitude. The Gegenbauer polynomials arises from the exchange of a spin-$S$ particle for a four scalar amplitude. Note that we have one polynomial for a given $S$ because the three-point coupling between two scalars and a spin-$S$ particle is fixed.

As we've seen in the previous discussion, the three-point amplitude for one massive, two massless particles is also unique. This implies that we can similarly construct ``spinning" Gegenbauer polynomials for massless scattering amplitude, where each polynomial correspond to a different spin exchange. To see how this works let's consider the residue for a spin-$S$ exchange in the $s$-channel for $ M(1^{-h}2^{+h}3^{-h}4^{+h})$. We can write down the unique three-point amplitudes on both sides: 
\begin{eqnarray*}
\frac{\lambda_1^{S+2h}\lambda_2^{S-2h}[12]^S}{m^{2S-1}},\quad \frac{\lambda_3^{S+2h}\lambda_4^{S-2h}[34]^S}{m^{2S-1}}
\end{eqnarray*}
Such coupling only exists for $S\geq 2h$. Now when we glue the two tensor structures together the indices on $\lambda_1,\lambda_2$ must be fully contracted with those on $\lambda_3,\lambda_4$. This can be done in many ways, each with its own pre-factor counting the number of equivalent contractions. The gluing procedure is thus a sum over all possible contractions with suitable combinatoric factors:
\begin{equation}
\frac{[12]^S[34]^S}{m^{4S-2}}\sum_a c_{2S,S+2h,a}\langle 13\rangle^a \langle 14\rangle^{S+2h-a} \langle 23\rangle^{S+2h-a} \langle 24\rangle^{a-4h}
\end{equation}
where the summation $a$ ranges from $4h$ to $S+2h$, and 
\begin{equation}
c_{2S,S+2h,a}=\f{(S+2h)!^2(S-2h)!^2}{a!(S+2h-a)!^2(a-4h)!}\,,.
\end{equation}
It would be useful to convert this polynomial into a function of the scattering angle in the center of mass frame for particles 1 and 2. We write $\lambda_1=m^{\frac{1}{2}}\begin{pmatrix}1\\0 \end{pmatrix}, \lambda_2=m^{\frac{1}{2}}\begin{pmatrix}0\\1 \end{pmatrix},\lambda_3=im^{\frac{1}{2}}\begin{pmatrix}\cos\f{\theta}{2}\\ \sin\f{\theta}{2} \end{pmatrix},\lambda_4=im^{\frac{1}{2}}\begin{pmatrix}\sin\f{\theta}{2}\\ -\cos\f{\theta}{2} \end{pmatrix}$. The spinning Gegenbauer polynomial is then given as:
\begin{align}
P_S^h(\cos\theta)&=\f{1}{(S!)^2}\sum_a \f{(S+2h)!^2(S-2h)!^2}{a!(S+2h-a)!^2(a-4h)!}\left(\f{\cos\theta -1}{2}\right)^{a-2h}\left(\f{\cos\theta +1}{2}\right)^{S+2h-a}
\end{align}
As a few example (with $x=\cos\theta$):
\eqa
P_2^1(x)&=&\frac{3}{2}(x-1)^2, \quad P_3^1(x)=\frac{5}{6}(x-1)^2(2+3x)\nonumber\\
P_4^1(x)&=&\frac{5}{8} (x-1)^2 (1 + 7 x (1 + x))\,.
\eqae
The universal prefactor $(x-1)^2$ can be identified with $\langle13\rangle^2[24]^2$ which takes care of the overall helicity weights of this amplitude. Taking $\ell=0$ and we indeed recover the Legendre polynomials $P_S^0(x)=P_S(x)$.

For completely general helicities $h_1,h_2,h_3,h_4$ of external massless particles, we have:
\begin{align*}
P_S^{h_i}(x) &=\f{1}{(S!)^2}\sum_a \f{(S+h_4-h_3)!(S+h_3-h_4)!(S+h_1-h_2)!(S+h_2-h_1)!}{a!(S+h_4-h_3-a)!(S+h_2-h_1-a)!(a+h_1+h_3-h_2-h_4)!}\\
&\times\left(\f{x -1}{2}\right)^{a+\f{h_1+h_3-h_2-h_4}{2}}\left(\f{x +1}{2}\right)^{S-a-\f{h_1+h_3-h_2-h_4}{2}}
\end{align*}
This reduces to equal spin polynomial if we take all $|h_i|$ to be equal.

Three-point couplings with more than one massive leg are no longer unique. This means that for a given spin-exchange, one instead has a symmetric matrix where the rows and the columns label the independent three-point vertices on both sides of the factorization channel. We illustrate this for the two massive spin-$1$  and two massless spin-1 amplitude.

Now the three-point coupling involved in the factorization involves a massive spin-1 spin-$S$ and massless spin-1 amplitude. The number of such coupling is determined by the lowest spin massive particle, which in this case is $1$ and there are three independent coupling. To give an explicit example, consider $S$=2
\begin{center}
\begin{tikzpicture}[scale=0.7]
\draw[thick] (-1, 1.73) node[above] {$(\alpha_1\alpha_2)$}-- (0, 0);
\draw[photon] (-1, -1.73) node[left]{$1$} -- (0, 0);
\draw[thick] (0, 0) -- (2, 0)node[right] {$(\beta_1\beta_2\beta_3\beta_4)$};
\end{tikzpicture}
\end{center}
The building blocks of tensor structures will be $\{\lambda_1, P_2\tilde{\lambda}_1\}=\{v, u\}$. If the massless particle has $-$ helicity, we have three tensor structures listed in eq.(\ref{possibleT}). Now imagine gluing the two three-point amplitude:

\begin{center}
\begin{tikzpicture}[scale=0.7]
\draw[thick] (-1, 1.73) node[above]{$(\alpha_1\alpha_2)$}-- (0, 0);
\draw[photon] (-1, -1.73) node[left]{$1^-$} -- (0, 0);
\draw[thick] (0, 0) -- (2, 0)node[above] {$(\beta_1\beta_2\beta_3\beta_4)$} ;

\draw[thick] (4, 0) node[below] {$(\beta_1\beta_2\beta_3\beta_4)$} -- (6,0);
\draw[thick] (6,0) -- (7, 1.73) node[above]{$(\gamma_1\gamma_2)$};
\draw[photon] (6,0) -- (7, -1.73) node[right]{$4^-$};
\end{tikzpicture}
\end{center}

The residue will be a polynomial of $(u_L, v_L, u_R, v_R)$ with
\begin{align*}
u_L^\alpha=\epsilon^{\alpha\beta}(P_2)_{\beta\dot{\beta}}\tilde{\lambda}_1^{\dot{\beta}},
\quad
&v_L^\alpha=\lambda_1^{\alpha}\\
u_R^\alpha=\epsilon^{\alpha\beta}(P_3)_{\beta\dot{\beta}}\tilde{\lambda}_4^{\dot{\beta}},
\quad
&v_R^\alpha=\lambda_4^{\alpha}.
\end{align*}
By gluing them we contract the internal indices in all possible ways, then sum them up with appropriate combinatoric factors. We can distribute indices carried by exchanged particle into a bunch of $u$'s and $v$'s:
\begin{eqnarray*}
\#(u_L)+\#(v_L)&=2S\\
\#(u_R)+\#(v_R)&=2S
\end{eqnarray*}
where $S$ is the spin of exchanged particle.
For a contraction with $(u_L)^{k_1}$ and $(u_R)^{k_2}$ on exchanged leg, suppose $u_L$ and $u_R$ are contracted together $k_3$ times. Then we have
\begin{equation}
\langle u_Lu_R\rangle^{k_3}\langle u_Lv_R\rangle^{k_1-k_3}\langle v_Lu_R\rangle^{k_2-k_3}\langle v_Lv_R\rangle^{2N-k_1-k_2+k_3}
\end{equation} which means a factor of
\begin{equation}
\binom{k_2}{k_3}\binom{2N-k_2}{k_1-k_3}(k_1)!(2N-k_1)!.
\end{equation}
The first two factors come from choosing which $u_L$s and $u_R$s are to be contracted together. Since we can always redefine coupling constants for interactions, the $k_3$-independent factors shall not concern us here. Summing this factor over $k_3$ one gets $(2N)!$, the total number of permutations on $2N$ indices.

Assigning a coupling constant $g_i$ for each three-point vertex, the residue of the four-point amplitude can then be expanded as $g_iM_{ij}g_j$ where each element in $M_{ij}$ is a polynomial given by the contraction of the corresponding three-point amplitudes. Since we have two external spin-1 particles, $M_{ij}$ is a $3\times3$ symmetric matrix irrespective of the exchanged spin. For the case where one exchanges a spin-$2$, the matrix elements are given by:
\begin{align}
M_{11}&=24\langle v_Lv_R\rangle^4\langle u_L\mathbf{1}\rangle^2\langle u_R\mathbf{4}\rangle^2\nonumber\\
M_{12}&=24\langle v_Lu_R\rangle\langle v_Lv_R\rangle^3\langle u_L\mathbf{1}\rangle^2\langle u_R\mathbf{4}\rangle\langle v_R\mathbf{4}\rangle\nonumber\\
M_{13}&=24\langle v_Lu_R\rangle^2\langle v_Lv_R\rangle^2\langle u_L\mathbf{1}\rangle^2\langle v_R\mathbf{4}\rangle^2\nonumber\\
M_{22}&=\left(18\langle u_Lv_R\rangle\langle v_Lu_R\rangle\langle v_Lv_R\rangle^2+6\langle u_Lu_R\rangle\langle v_Lv_R\rangle^3\right)\langle u_L\mathbf{1}\rangle\langle v_L\mathbf{1}\rangle\langle u_R\mathbf{4}\rangle\langle v_R\mathbf{4}\rangle\nonumber\\
M_{23}&=\left(12\langle u_Lv_R\rangle\langle v_Lu_R\rangle^2\langle v_Lv_R\rangle+12\langle u_Lu_R\rangle\langle \langle v_Lu_R\rangle\langle v_Lv_R\rangle^2\right)\langle u_L\mathbf{1}\rangle\langle v_L\mathbf{1}\rangle\langle v_R\mathbf{4}\rangle^2\nonumber\\
M_{33}&=\left(4\langle u_Lv_R\rangle^2\langle v_Lu_R\rangle^2+16\langle u_Lu_R\rangle\langle u_Lv_R\rangle\langle v_Lu_R\rangle\langle v_Lv_R\rangle+4\langle u_Lu_R\rangle^2\langle v_Lv_R\rangle^2\right)\langle v_L\mathbf{1}\rangle^2\langle v_R\mathbf{4}\rangle^2\,.
\end{align}
where we've contracted each entry with the external $\lambda^I_1, \lambda^I_4$s.

For convenience, we will also give the representation in terms of scattering angle. We can parameterize the kinematics as
\begin{align*}
p_1&=(x,0,0,x)\\
p_2&=(\sqrt{x^2+m_2^2},0,0,-x)\\
p_3&=(-\sqrt{y^2+m_3^2},-y\sin\theta,0,-y\cos\theta)\\
p_4&=(-y,y\sin\theta,0,y\cos\theta)
\end{align*}
where
\begin{equation*}
x=\sqrt{\frac{(m_2^2+t)^2}{-4t}},\quad
y=\sqrt{\frac{(m_3^2+t)^2}{-4t}}.
\end{equation*}
One can explicitly check that $\sum_ip_i=0$, $p_i^2=m_i^2$. In this parametrization, the matrix elements then take the form, where we've stripped the external spinor dependent terms: 
\begin{equation}
\begin{pmatrix}
6+12x+6x^2 & 12(1+x)\sqrt{1-x^2} & 24-24x^2\\
12(1+x)\sqrt{1-x^2} & 12-12x-24x^2 & -48x\sqrt{1-x^2} \\
24-24x^2 & -48x\sqrt{1-x^2} & -32+96x^2 \\
\end{pmatrix}.
\end{equation}

\section{(Super)Higgs Mechanism as IR Unification}\label{Higgs}
Our exploration of consistent four-particle amplitudes has given us an
almost complete understanding of the broad architecture of particle
physics. Theories of massless particles are incredibly constrained,
allowing only helicities (0,1/2,1,3/2,2), and limited to the
(super)gravity coupled to (super)Yang-Mills theories. Massless higher
spins are made impossible by the mere presence of gravity. We have
also seen that the amplitudes for massive particles of sufficiently
high spin have sick high-energy limits---as expected, since there is no
consistent theory of massless high-spin particles they can match to at
high-energies---so such particles cannot be ``elementary".

The final case to consider is then that of massive particles of low
spin $S \leq 2$. Here of course there {\it is} in principle a
consistent high-energy theory to match to, but as we will see in this
section, doing so puts non-trivial restrictions on the particle
content and interactions of the theory. This investigation will lead
to the on-shell discovery of the Higgs and Super-Higgs mechanisms.

Note that we will not simply be rephrasing well-known ``bottom-up"
facts, such as the high-energy growth of scattering amplitudes for
longitudinal components of massive spin one particles, and the
attendant need for the Higgs particle to tame this growth, in an
on-shell language. It is of course perfectly possible to do this, and
the on-shell methods do simplify the explicit computations, but the
advantage is purely technical and does not add anything conceptually
new to this standard textbook discussion.

We will instead take a different, ``top-down" point of view, where as
described above we  insist that  massive amplitudes manifestly match
to consistent massless amplitudes in the high-energy limit. As we will
see this gives us a satisfying understanding of the Higgs mechanism
that is at least psychologically quite opposite to the usual picture
of gauge symmetry ``breaking". Indeed in textbook  language, the gauge
symmetry is ``broken" or ``hidden", and becomes more manifest only at
high energies. By contrast in the on-shell picture, the massive
``Higgsed" amplitudes do not ``break" or ``hide" the (non-existent in
this formalism!) gauge redundancies. Instead, they unify the different
helicity components of massive amplitudes, the Higgs mechanism can be
thought of as an {\it infrared unification} of massless amplitudes,
and this unification is more disguised at high energies!

We will see this beginning already at the level of three-particle
amplitudes. Here, the non-locality associated with the poles in
massless three-particle amplitudes gets IR-deformed to $1/m$ poles.
Such $1/m$ poles non-trivially disappear in the high-energy limit
while the massive amplitudes unify different helicity components
together. Matching the high-energy limit enforces all the usual
consistency conditions associated with the Higgs mechanism. Moving on
to four-particle amplitudes, we will obtain them both by gluing the
three-particle amplitudes as usual, but also in a novel way, starting
with the massless helicity amplitudes, simply adding them so they fit
into massive multiplets, then shifting the poles  and ``{\bf BOLD}"ing
the spinor-helicity variables to make massive amplitudes! This will
highlight the Higgs mechanism as an ``IR unification" in an even more
vivid way.

Rather than present a completely systematic analysis of all possible
``Higgsings", in this section we will content ourselves with
illustrating this physics in three standard examples: the Abelian
Higgs model, the Super-Higgs mechanism in a simple model with ${\cal
N}=1$ SUSY, and the general structure of the non-Abelian Higgs
mechanism for a model with enough scalars so that all the spin one
particles are massive. As alluded to above we will also discuss why gravity
cannot be Higgsed in this way.

\subsection{Abelian Higgs}
Let us start with the simplest example - a theory with a massless photon and a charged scalar; we'll call the scalar's two real degrees of freedom ``$H$" and ``$E$".

The three-particle amplitudes are
\eq
\vcenter{\hbox{\includegraphics[scale=0.45]{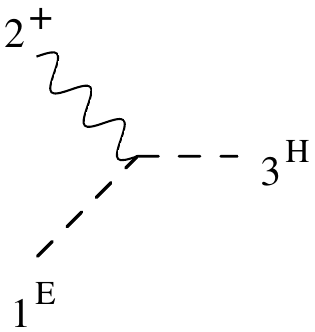}}}\quad g\frac{[12][32]}{[13]},\quad\quad \vcenter{\hbox{
\includegraphics[scale=0.45]{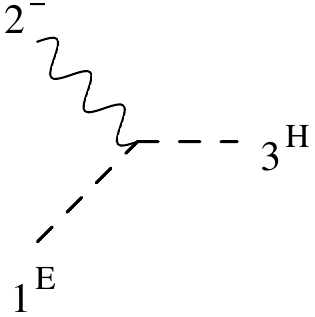}}}\quad g\frac{\langle 12\rangle \langle 32\rangle}{\langle 13\rangle}
\eqe
We now want to see how to introduce masses as an ``infrared deformation". The first step is a trivial kinematical one. We declare that $(+,-,E)$ are to become the $3$ components of a massive spin $1$ particle, leaving $H$ as an additional scalar. Now, the two massive vector (with $m^2_\gamma$) and one massive scalar (with $m^2_H$) amplitude can only be,
\eq\label{Abelian3}
\vcenter{\hbox{\includegraphics[scale=0.45]{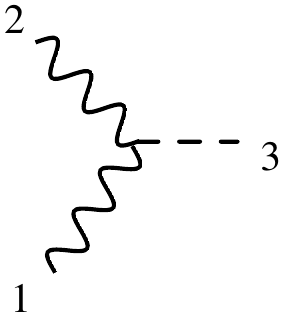}}}\quad \frac{g}{m_\gamma}\langle \bold{1}\bold{2}\rangle [\bold{2}\bold{1}]\,.
\eqe
The coefficient is fixed by the requirement that this $3$ particle amplitude matches the massless amplitude in the high-energy limit. It is illuminating to see how this happens explicitly. Recall that to take the HE limit we put
\begin{align}
\lambda_\alpha^I&=\lambda_\alpha\xi_+^I+\eta_\alpha\xi_-^I\nonumber\\
\tilde{\lambda}_{\dot{\alpha}}^I&=\tilde{\lambda}_{\dot{\alpha}}\xi_-^I+\tilde{\eta}_{\dot{\alpha}}\xi_+^I\,,
\end{align}
where we scale each of $\eta,\tilde{\eta}$ as $\sim m$. We are looking for pieces that survive in the $m_\gamma, m_H\to 0$ limit. The leading piece in the numerator are those with zero  $\eta,\tilde{\eta}$'s which is given as:
\eq
\vcenter{\hbox{\includegraphics[scale=0.45]{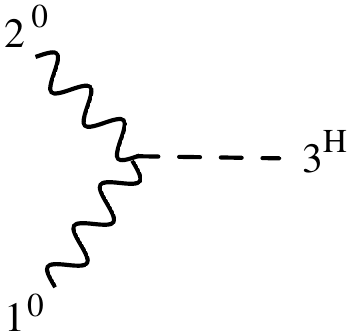}}}
=\quad\frac{g}{m_\gamma}\langle 12\rangle [21]=\frac{g}{m_\gamma}(m_H^2)\,.
\eqe
This indeed vanishes as $m_H \to 0$, as expected since we don't have an $(EEH)$ coupling in the UV. For the order $\tilde{\eta}$ piece, we have
\eq
\vcenter{\hbox{\includegraphics[scale=0.45]{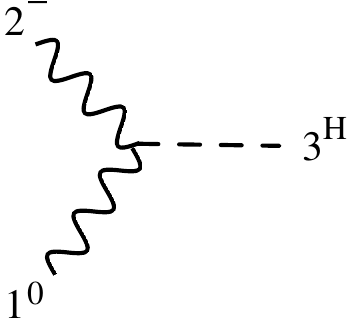}}}\quad=\quad\frac{g}{m_\gamma}\langle 12\rangle [1\tilde{\eta}_2]\,.
\eqe
This term is more interesting. To compute it, note that in the UV we have our usual restrictions on $3$ particle kinematics --- either $\lambda_1\propto\lambda_2\propto \lambda_3$ or $\tilde{\lambda}_1\propto \tilde{\lambda}_2\propto \tilde{\lambda}_3$. This $3$-particle amplitude vanishes in the first case. On the other hand, in the second case, we have by momentum conservation that
\begin{equation}
\tilde{\lambda}_1=\langle 23\rangle\tilde{\xi},\quad \tilde{\lambda}_2=\langle 31\rangle\tilde{\xi},\quad \tilde{\lambda}_3=\langle 12\rangle\tilde{\xi}\,,
\end{equation}
and so
\begin{equation}
[1\tilde{\eta}_2]=[2\tilde{\eta}_2]\frac{\langle 23\rangle}{\langle 31\rangle}=m_\gamma \frac{\langle 23\rangle}{\langle 31\rangle}\,. 
\end{equation}
So this amplitude becomes
\eq
\vcenter{\hbox{\includegraphics[scale=0.45]{figures/Higgs4}}}=\frac{g}{m_\gamma}\langle 12\rangle\times m_\gamma\frac{\langle 23\rangle}{\langle 31\rangle}=g\frac{\langle 12\rangle \langle 23\rangle}{\langle 31\rangle}\quad=\vcenter{\hbox{\includegraphics[scale=0.45]{figures/Higgs1b}}}\eqe
\noindent exactly as desired. Obviously we get the analogous result for $2^+$. Thus quite beautifully the massive three-particle amplitude \begin{minipage}{.1\textwidth}\includegraphics[scale=0.4]{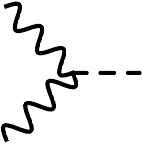}\end{minipage} reproduces the component helicity amplitudes and unifies them into a single object. Note also that despite appearances, there is no singularity as $m_\gamma \to 0$.

Here we see an interesting counterpart to the purely massless $3$pt amplitudes - which are not manifestly local due to the presence of poles. Healthy theories of massless particle (which we should reproduce in the UV) do not have such non-local poles at $4$pts and higher. When we perform this "IR deformation", we have removed the non-local poles but are left with seeming factors of $\frac{1}{m_\gamma}$ in the amplitude. But as we have seen the $3$pt amplitude is --- by design --- chosen to match the correct massless helicity amplitudes and thus be smooth as $m_\gamma \to 0$, and this will be inherited at higher points.

Indeed let us compute the $4$-particle amplitude with all massive spin $1$ particles consistent with factoring into the three-point amplitude in eq.(\ref{Abelian3}). Since we have no ``$x$" factors to worry about, we can proceed in the most naive possible way, simply gluing the $3$-pt amplitudes in the $s,t$ and $u$ channels, and we find:
\eq
\vcenter{\hbox{\includegraphics[scale=0.5]{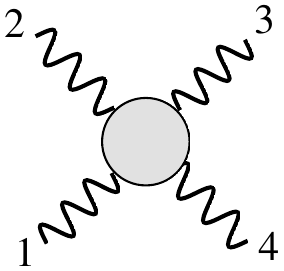}}}\quad \frac{g^2}{m_\gamma^2}\left[\frac{\langle \bold{1}\bold{2}\rangle [\bold{1}\bold{2}]\langle \bold{3}\bold{4}\rangle [\bold{3}\bold{4}]}{s-M_h^2}{+}\frac{\langle \bold{2}\bold{3}\rangle [\bold{2}\bold{3}]\langle \bold{1}\bold{4}\rangle [\bold{1}\bold{4}]}{t-M_h^2}{+}\frac{\langle \bold{1}\bold{3}\rangle [\bold{1}\bold{3}]\langle \bold{2}\bold{4}\rangle [\bold{2}\bold{4}]}{u-M_h^2}\right]\,.
\eqe
Since there are no three-point massive spin-1 amplitude, there is no poles involving $m_\gamma$. Note that all possible contact terms here can be eliminated since they give growing amplitudes for some of the helicity components in the UV, which we are assuming not to have. Now again, despite appearances this amplitude is guaranteed (by construction!) to be smooth in the high-energy (or $m_\gamma, m_H \to 0$) limit. Let us first show this directly for some of the helicity components. For instance, the all-longitudinal amplitude is
\begin{equation}
\frac{g^2}{m_\gamma^2}\left[\frac{(\hat{p}_1\cdot \hat{p}_2)(\hat{p}_3\cdot \hat{p}_4)}{s-M_h^2}+\frac{(\hat{p}_1\cdot \hat{p}_4)(\hat{p}_2\cdot \hat{p}_3)}{t-M_h^2}+\frac{(\hat{p}_1\cdot \hat{p}_3)(\hat{p}_2\cdot \hat{p}_4)}{u-M_h^2}\right]\,.
\end{equation}
Where with $p=\lambda\tilde{\lambda}+\eta\tilde{\eta}$ we define $\hat{p}=\lambda\tilde{\lambda}-\eta\tilde{\eta}$. Just to take a first look at the HE limit, which naively goes as $\frac{g^2}{m\gamma^2}$, we drop the $\eta$'s and find at $O(\frac{1}{m_\gamma^2})$
\begin{equation}
\frac{g^2}{m_\gamma^2}\times \left[s+t+u\right]=0\,,
\end{equation}
and so as expected there is no $(\frac{s,t,u}{m_\gamma^2})$ singularity as $m_\gamma\to 0$. In order to find the leading high-energy limit, let us define $q\equiv\eta\tilde{\eta}$. Note that $p\cdot q=\frac{m_\gamma^2}{2}$ so $q=O(m_\gamma^2)$, and we will work to first order in $q$. Using $2p_1\cdot p_2=s-2m_\gamma^2$, and also $\hat{p}=p-2q$, we find in the HE limit
\begin{equation}
\frac{4(\hat{p}_1\cdot \hat{p}_2)(\hat{p}_3\cdot \hat{p}_4)}{s-M_h^2}=s-4m_\gamma^2+M_h^2-4(q_1\cdot p_2+q_2\cdot p_1+q_3\cdot p_4+q_4\cdot p_3){+}\mathcal{O}(M^4_h,m^4_\gamma)\,.
\end{equation}
So summing over channels gives
\begin{align}
&s+t+u-3\times 4m_\gamma^2+3M_h^2-4(q_1\cdot (p_2+p_3+p_4)+...)\nonumber\\
&=4m_\gamma^2-3\times 4m_\gamma^2+3M_h^2+2\times 4m_\gamma^2\nonumber\\
&=3M_h^2
\end{align}
Hence the all-longitudinal amplitude is fixed to be $\frac{3}{4}g^2\frac{M_h^2}{m_\gamma^2}$. This tells us we must have a quartic coupling in the UV, and by the $U(1)$ invariance it must be $\lambda(E^2+H^2)^2$ with
\begin{equation}
\frac{\lambda}{M_H^2}=\frac{g^2}{m_\gamma^2}\,.
\end{equation}

Let's see how some of the other component amplitudes work. Consider $(1^02^-3^+4^0)$, which should match $(1^E2^-e^+4^E)$ in the high-energy limit. This is
\begin{equation}
\frac{g^2}{m_\gamma^2}\left[\frac{\langle 12\rangle [1\tilde{\eta}_2]\langle 4\eta_3\rangle [43]}{s}+\frac{\langle 2\eta_3\rangle [\tilde{\eta}_2 3]\langle 14\rangle [14]}{t}+\frac{\langle 1\eta_3\rangle [13]\langle 24\rangle [\tilde{\eta}_24]}{u}\right]
\end{equation}
Note that since all that matters is $[2\tilde{\eta}_2]=m_\gamma, \langle 3\eta_3\rangle=m_\gamma$, $\tilde{\eta}_2,\eta_3$ are defined up to shifts such as $\tilde{\eta}_2\to \tilde{\eta}_2+\alpha \tilde{\lambda}_2$. Not surprisingly the above representation is independent of such shifts. The term in the brackets shifts as
\begin{equation}
\alpha\left[\langle 4\eta_3\rangle[43]+\langle 2\eta_3\rangle[23]+\langle 1\eta_3\rangle[13]\right]=\alpha\langle\eta_3|p_3|3]=0\,.
\end{equation}
Hence we are free to choose $\tilde{\eta}_2=\frac{m_\gamma\tilde{\lambda}_3}{[23]},\eta_3=\frac{m_\gamma\lambda_2}{\langle 23\rangle}$, \footnote{Using this representation for $\tilde{\eta}_2$, one can also show that the $\mathcal{O}(m^{-1}_\gamma)$ term in the amplitude vanishes as well, with $\tilde\lambda_2^I\rightarrow \tilde{\eta}_2$, while all other massive spinors are set to their massless limit. } then only the $s+u$ channel terms contribute and we find
\eq
\vcenter{\hbox{\includegraphics[scale=0.4]{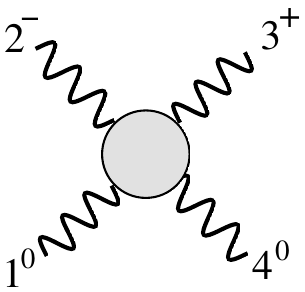}}}{=}g^2\langle 2|p_4|3]^2\left(\frac{1}{st}{+}\frac{1}{tu}\right)={-}g^2\frac{\langle 2|(p_1{-}p_4)|3]^2}{4su}=\vcenter{\hbox{\includegraphics[scale=0.4]{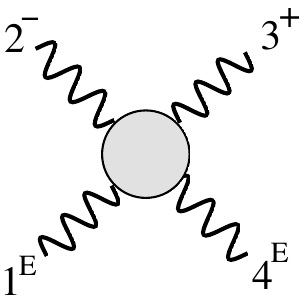}}}
\eqe
Thus we find the correct amplitude for minimally charged scalars in the UV. All other helicity amplitude components vanish as $m_\gamma\to 0$.
We have thus verified that the $4$pt massive amplitudes are an ``infrared deformation" of the massive ones, reproducing and unifying the different helicities in the HE limit.

\subsection{Higgsing as UV Unification $\to$ IR Deformation}

Given that we see the massive amplitudes reproduce the massless ones at high energy, we are motivated to consider directly assembling the high-energy massless amplitudes in a way that one can readily ``IR deform" the amplitude by simply putting in the mass for the propagator and ``\textbf{BOLD}ing" the spinor brackets. We are then guaranteed to have a result that gives the correct high-energy behaviour, and what remains is simply to add in higher order corrections in mass that ensure the massive residue is matched. 

Let's first consider all the different component amplitudes - Compton scattering for $H, E$, and the quartic interaction for $E$. We will first merely group these amplitudes together, ready to be ``\textbf{BOLD}"ed + unified into a massive amplitude. The massive amplitude in the IR will be the four massive vector amplitude, and thus we will need a total of eight spinors to carry the SU(2) Little group indices, these are the objects that will be \textbf{BOLD}ed. Thus the name of the game is to write the massless amplitudes in a form which contains eight spinors, two for each legs, and every thing else can only be expressed as momenta. Note that because of this the $E^4$ quartic must be written in an interesting way. Naively it is just $3\lambda$, but to put it in a form where by \textbf{BOLD}ing we can recognize it as a component of massive spin $1$, we have to write it in the following way:
\begin{equation}
3\lambda=\lambda\frac{s^3{+}t^3{+}u^3}{stu}=\lambda\frac{\left(\langle12\rangle[21][3|p_4|3\rangle[4|p_3|4\rangle{+}\langle34\rangle[43][1|p_2|1\rangle[2|p_1|2\rangle\right)+\{u\}{+}\{t\}}{2stu}\,,
\end{equation}
where $\lambda=\frac{g^2 M^2_h}{m^2_\gamma}$ and $\{t\}$ $\{u\}$ represents its $t,u$ image. This is the only way to represent the ``constant" without introducing double poles. Similarly for the two photons two $E$ amplitudes we write
\begin{equation}
-g^2\frac{\langle 2|p_1{-}p_4|3]^2}{4su}=g^2\frac{[14]\langle14\rangle\langle 2|p_1{-}p_4|3]^2}{4stu}.
\end{equation}
Collecting all the component amplitudes together, we are ready to IR deform: declaring the particles have mass $m_\gamma$ by \textbf{BOLD}ing the spinors, and deforming $s\to s-M_h^2$ etc., giving an IR deformed object:
\begin{equation}
\frac{[\mathbf{1}\mathbf{2}]\langle\mathbf{1}\mathbf{2}\rangle (g^2\langle\mathbf{3}|p_1{-}p_2|\mathbf{4}]^2+g^2\langle\mathbf{4}|p_1{-}p_2|\mathbf{3}]^2+2\lambda[\mathbf{3}|p_4|\mathbf{3}\rangle[\mathbf{4}|p_3|\mathbf{4}\rangle{+}(1,2\leftrightarrow3,4)}{4(s-M_h^2)(t-M_h^2)(u-M_h^2)}+\{u\}{+}\{t\}\,.
\end{equation}
The above result by construction gives the correct answer in the High-energy limit, with mismatch at higher order in $m_\gamma^2, M^2_{h}$. Thus we have the identity
\begin{align}
&\frac{g^2}{m_\gamma^2}\left(\frac{\langle\bold{1}\bold{2}\rangle [\bold{1}\bold{2}]\langle\bold{3}\bold{4}\rangle [\bold{3}\bold{4}]}{s-M_h^2}+\frac{\langle\bold{1}\bold{4}\rangle [\bold{1}\bold{4}]\langle\bold{3}\bold{2}\rangle [\bold{3}\bold{2}]}{t-M_h^2}+\frac{\langle\bold{1}\bold{3}\rangle [\bold{1}\bold{3}]\langle\bold{2}\bold{4}\rangle [\bold{2}\bold{4}]}{u-M_h^2}\right)\nonumber\\
=& \frac{[\mathbf{1}\mathbf{2}]\langle\mathbf{1}\mathbf{2}\rangle (g^2\langle\mathbf{3}|p_1{-}p_2|\mathbf{4}]^2+g^2\langle\mathbf{4}|p_1{-}p_2|\mathbf{3}]^2+2\lambda[\mathbf{3}|p_4|\mathbf{3}\rangle[\mathbf{4}|p_3|\mathbf{4}\rangle{+}(1,2\leftrightarrow3,4)}{4(s-M_h^2)(t-M_h^2)(u-M_h^2)}\nonumber\\
+&\{u\}{+}\{t\}+\mathcal{O}(m_\gamma^2, M^2_{h})\,.
\end{align}
But now in this form, the challenge is to check the factorization channels, which will fix the $\mathcal{O}(m_\gamma^2, M^2_{h})$ terms. For example in the limit where $m_\gamma^2=M^2_{h}\equiv m^2$, the remaining term is simply
\eq
\mathcal{O}(m^2)=\frac{m^2(\langle\mathbf{4}\mathbf{3}\rangle^2[\mathbf{1}\mathbf{2}]^2+\langle\mathbf{1}\mathbf{2}\rangle^2[\mathbf{3}\mathbf{4}]^2-\langle\mathbf{4}\mathbf{3}\rangle[\mathbf{4}\mathbf{3}]\langle\mathbf{1}\mathbf{2}\rangle[\mathbf{1}\mathbf{2}])+\{u\}{+}\{t\}}{(s-m^2)(t-m^2)(u-m^2)}\,.
\eqe
We have thus seen the Higgs mechanism very explicitly as an IR deformation.
Note that while it is pleasing to see everything work explicitly, the correct HE limit was guaranteed once we ensured the $3$ particle amplitudes reproduced and unified the helicity amplitudes in the high-energy limit. Again: all the non-trivial physics was in the ``unified packaging" of all the massless helicity amplitudes into the massive amplitudes - everything was guaranteed to work after that point.

We could also consider 
\begin{minipage}{.3\textwidth}
\includegraphics[scale=0.3]{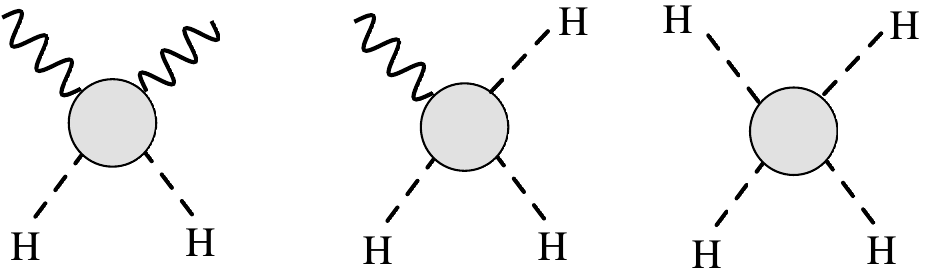}
\end{minipage}
and derive the rest of the physics. For example from the fact that we know there is a coupling $\lambda(E^2+H^2)^2$ in the UV, tells us that we have an $(EEHH)=\lambda$ component that needs to be unified into
$$\includegraphics[scale=0.5]{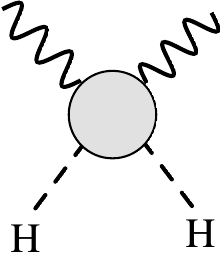}\,$$ 
Naively, one would combine this with $(\gamma\gamma HH)$, however, the bolded version of this amplitude:
\eq
\vcenter{\hbox{\includegraphics[scale=0.5]{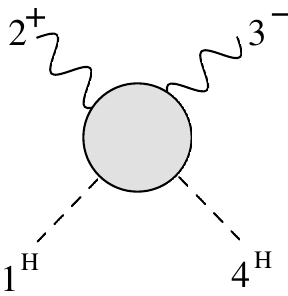}}}=\frac{\langle3|p_1{-}p_4|2]^2}{4st}\;\rightarrow\;\frac{\langle\mathbf{3}|p_1{-}p_4|\mathbf{2}]^2}{4(s-m^2_\gamma)(t-m^2_\gamma)}\,,
\eqe 
will not contain such a high-energy scalar contact piece. This suggests that we should directly IR deform it:
\eq
\lambda=\lambda \frac{\langle23\rangle[23]}{t}\;\rightarrow\;\lambda\frac{\langle\mathbf{2}\mathbf{3}\rangle[\mathbf{2}\mathbf{3}]}{(t-m^2_H)}\,.
\eqe
Thus we see that by IR deforming it, we are forced to have a Higgs propagator, whose residue reveals the presence of a Higgs cubic coupling \begin{minipage}{.1\textwidth}
\includegraphics[scale=1]{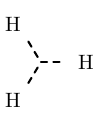}
\end{minipage}.

\subsection{Super-Higgs}
Let us now describe the Super-Higgs mechanism. Again, we will consider the simplest case, and $\mathcal{N}=1$ SUGRA where we have a graviton, gravitino $\psi$ as well as a chiral superfield - a fermion $\chi$ and a scalar $\phi$. First in the massless limit, in addition to the universal couplings to gravity we have
\eqa
\vcenter{\hbox{\includegraphics[scale=0.6]{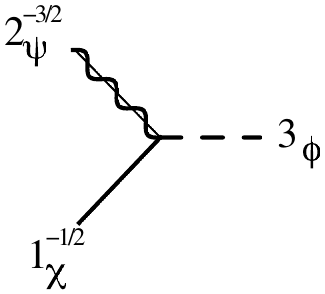}}}&&\quad \frac{1}{M_{pl}}\frac{\langle 12\rangle^2\langle 23\rangle}{\langle 13\rangle}\nonumber\\
\vcenter{\hbox{\includegraphics[scale=0.6]{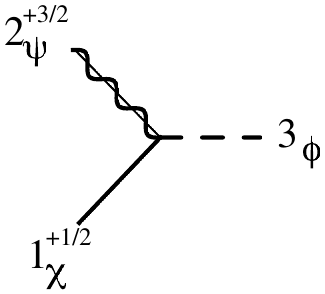}}}&&\quad\frac{1}{M_{pl}}\frac{[12]^2[23]}{[13]}
\eqae

Now, we wish to see whether the $(\psi,\chi)$ amplitudes can be unified into those of a single massive spin $\frac{3}{2}$ multiplet. The logic completely parallels to the Abelian Higgs mechanism we discussed above. Indeed, again we simply have the following massive amplitude for massive spin-$\frac{3}{2}$, spin-$\frac{3}{2}$ and scalar:

$$\vcenter{\hbox{\includegraphics[scale=1]{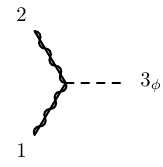}}}\quad\frac{1}{M_{pl}}\frac{1}{m_{3/2}}\langle\mathbf{1}\mathbf{2}\rangle[\mathbf{1}\mathbf{2}]\left([\mathbf{1}\mathbf{2}]+\langle \mathbf{1}\mathbf{2}\rangle\right)$$
The correct HE limit emerges in exactly the same way. For instance the $(1^{-\frac{1}{2}}2^{-\frac{3}{2}}3^0)$
$$
\vcenter{\hbox{\includegraphics[scale=1]{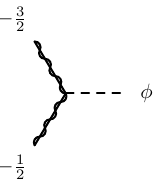}}}\quad \frac{1}{M_{pl}}\frac{1}{m_{3/2}}\langle21\rangle[1\tilde\eta_2]\left([\tilde{\eta}_1\tilde{\eta}_2]+\langle 12\rangle\right)
$$
The first term vanishes as $m_{3/2}\to 0$, while the second term becomes  $\frac{\langle 12\rangle^2[1\tilde\eta_2]}{M_{pl}m_{3/2}}$. Substituting $[1\tilde\eta_2]=m_{3/2}\frac{\langle 23\rangle}{\langle 13\rangle}$ yields the correct massless amplitude in the HE limit. After this point everything is guaranteed to work just as with the Abelian Higgs mechanism, and we omit the details. (We have described spontaneous SUSY breaking with the chiral superfield $X=\phi+\theta\chi+\theta^2F_\phi$ and $W=\mu^2 X$)
\subsection{Non-Abelian Higgs}
Let us now look at the most general case. In the UV we have gluons and scalars in some representation $R$:
\newline
\begin{figure}[h]
\centering
\begin{subfigure}[b]{0.3\textwidth}
\centering
\begin{tikzpicture}[scale=0.5]
\draw[gluon] (-1, 1.73) node[left] {$2_b^-$}-- (0, 0);
\draw[gluon] (-1, -1.73) node[left]{$1_a^-$} -- (0, 0);
\draw[gluon] (0, 0) -- (2, 0)node[right] {$3_c^+$};
\end{tikzpicture}
\caption*{$gf^{abc}\frac{\langle 12\rangle^3}{\langle 13\rangle\langle 23\rangle}$}
\end{subfigure}
\begin{subfigure}[b]{0.3\textwidth}
\centering
\begin{tikzpicture}[scale=0.5]
\draw[gluon] (-1, 1.73) node[left] {$2^a$}-- (0, 0);
\draw[scalardashed] (-1, -1.73) node[left]{$1_I$} -- (0, 0);
\draw[scalardashed] (0, 0) -- (2, 0)node[right] {$3_J$};
\end{tikzpicture}
\caption*{$g(T^a)_{IJ}\frac{\langle 12\rangle \langle 32\rangle}{\langle 13\rangle}$}
\end{subfigure}
\end{figure}

Now, we want to take the $\pm$ component of index $a$, together with some linear combination of the scalars $(u^a_J\phi^J)$, and make the part of a massive vector of mass $m_a$. Here, we are assuming that all the vectors are massive, in particular this means that the number of scalars $N_\phi$ is larger than or equal to the number of massless vectors. Then, what we are doing is considering a big $SO(N_\phi)$ matrix $U_{IJ}$, such that $U_{aJ}\phi_J$  will become the longitudinal component of the massive vector. The remaining scalars are ``Higgses" $U_{iJ}\phi_J$. We can always diagonalise so these have mass $M_i^2$, i.e. $U_{aI}U_{bI}=\delta_{ab}, U_{aI}U_{iI}=0, U_{iI}U_{jI}=\delta_{ij}$. So, we have
\begin{equation}
\begin{gathered}
\begin{tikzpicture}[scale=0.5]
\draw[gluon] (0,0) -- (4,0);
\node at (2, -0.7) {$a$};
\node at (2, 0.7) {};
\end{tikzpicture}
\end{gathered}\quad m_a,\quad
\begin{gathered}
\begin{tikzpicture}[scale=0.5]
\draw[scalar] (0,0) -- (4,0);
\node at (2, -0.7) {$i$};
\node at (2, 0.7) {};
\end{tikzpicture}
\end{gathered}\quad M_i\,.
\end{equation}
The relevant massive amplitudes in question includes 
\begin{equation}\label{massive3}
\begin{gathered}
\begin{tikzpicture}[scale=0.5]
\draw[gluon] (-1, 1.73) node[left] {$b$}-- (0, 0);
\draw[gluon] (-1, -1.73) node[left]{$a$} -- (0, 0);
\draw[gluon] (0, 0) -- (2, 0)node[right] {$c$};
\end{tikzpicture}
\end{gathered}
,\quad
\begin{gathered}
\begin{tikzpicture}[scale=0.5]
\draw[gluon] (-1, 1.73) node[left] {$b$}-- (0, 0);
\draw[gluon] (-1, -1.73) node[left]{$a$} -- (0, 0);
\draw[thick] (0, 0) -- (2, 0)node[right] {$i$};
\end{tikzpicture}
\end{gathered}
,\quad
\begin{gathered}
\begin{tikzpicture}[scale=0.5]
\draw[gluon] (-1, 1.73) node[left] {$b$}-- (0, 0);
\draw[thick] (-1, -1.73) node[left]{$j$} -- (0, 0);
\draw[thick] (0, 0) -- (2, 0)node[right] {$i$};
\end{tikzpicture}
\end{gathered}.
\end{equation}
In particular in the high energy limit we must have, for example:
\begin{align}
\begin{gathered}
\begin{tikzpicture}[scale=0.5]
\draw[gluon] (-1, 1.73) node[left] {$b$}-- (0, 0);
\draw[gluon] (-1, -1.73) node[left]{$a$} -- (0, 0);
\draw[gluon] (0, 0) -- (2, 0)node[right] {$c$};
\end{tikzpicture}
\end{gathered}
&\underrightarrow{\;\quad H.E.\quad \;}\quad
g f^{abc}
\begin{gathered}
\begin{tikzpicture}[scale=0.5]
\draw[gluon] (-1, 1.73) node[left] {$2_b^-$}-- (0, 0);
\draw[gluon] (-1, -1.73) node[left]{$1_a^-$} -- (0, 0);
\draw[gluon] (0, 0) -- (2, 0)node[right] {$3_c^+$};
\end{tikzpicture}
\end{gathered}
+gU_I^aU_J^c(T_{IJ}^b)
\begin{gathered}
\begin{tikzpicture}[scale=0.5]
\draw[gluon] (-1, 1.73) node[left] {$b$}-- (0, 0);
\draw[thick, dashed] (-1, -1.73) node[left]{$J$} -- (0, 0);
\draw[thick, dashed] (0, 0) -- (2, 0)node[right] {$I$};
\end{tikzpicture}
\end{gathered}\\
\begin{gathered}
\begin{tikzpicture}[scale=0.5]
\draw[gluon] (-1, 1.73) node[left] {$b$}-- (0, 0);
\draw[gluon] (-1, -1.73) node[left]{$a$} -- (0, 0);
\draw[thick] (0, 0) -- (2, 0)node[right] {$i$};
\end{tikzpicture}
\end{gathered}
&\underrightarrow{\;\quad H.E.\quad \;}\quad
gU_I^aU_{Ji}(T_{IJ}^b)
\begin{gathered}
\begin{tikzpicture}[scale=0.5]
\draw[gluon] (-1, 1.73) node[left] {$b$}-- (0, 0);
\draw[thick, dashed] (-1, -1.73) node[left]{$I$} -- (0, 0);
\draw[thick, dashed] (0, 0) -- (2, 0)node[right] {$J$};
\end{tikzpicture}
\end{gathered}.
\end{align}

Being able to unify these into massive amplitudes will allow us some interesting interpretations of the $U$ matrix. First, the only possibility for 
the first figure in \eqref{massive3}
is\footnote{This can be verified by noting that $\epsilon_{\alpha\dot\alpha}=\frac{\lambda^{\{I}_\alpha\tilde\lambda^{J\}}_{\dot{\alpha}}}{m}$, and substitute into the usual Feynman rules.} 

\begin{equation}
\begin{gathered}
\begin{tikzpicture}[scale=0.5]
\draw[gluon] (-1, 1.73) node[left] {$2^b$}-- (0, 0);
\draw[gluon] (-1, -1.73) node[left]{$1^a$} -- (0, 0);
\draw[gluon] (0, 0) -- (2, 0)node[right] {$3^c$};
\end{tikzpicture}
\end{gathered}
=
\frac{g f^{abc}}{m_am_bm_c}\left(\langle\bold{1}\bold{2}\rangle [\bold{1}\bold{2}]\langle \bold{3}|p_1{-}p_2|\bold{3}]+\text{cyc.}\right)
\end{equation}

We can again compute the HE limit of the component amplitudes. The details of this limit is given in appendix \ref{HELimit}, and we simply summarise the result:

\begin{align}
\begin{gathered}
\begin{tikzpicture}[scale=0.5]
\draw[gluon] (-1, 1.73) node[left] {$2_b^-$}-- (0, 0);
\draw[gluon] (-1, -1.73) node[left]{$1_a^-$} -- (0, 0);
\draw[gluon] (0, 0) -- (2, 0)node[right] {$3_c^+$};
\end{tikzpicture}
\end{gathered}
&\longrightarrow
g f^{abc}\frac{\abraket{12}^3}{\abraket{23}\abraket{31}}\\
\begin{gathered}
\begin{tikzpicture}[scale=0.5]
\draw[gluon] (-1, 1.73) node[left] {$2_b^-$}-- (0, 0);
\draw[scalardashed] (-1, -1.73) node[left]{$1_a$} -- (0, 0);
\draw[scalardashed] (0, 0) -- (2, 0)node[right] {$3_c$};
\end{tikzpicture}
\end{gathered}
&\longrightarrow
\frac{g f^{abc}}{m_am_c}\frac{\abraket{12}\abraket{23}}{\abraket{31}}\left(m_b^2-m_c^2-m_a^2\right).
\end{align}

From the above we see that in order for the massless amplitudes to be unified into a single massive amplitude, the matrix $U^a_I$ must satisfy
\begin{equation}\label{ColorConstraint}
U^a_IT^b_{IJ}U^c_{J}=f^{abc}\frac{m_b^2-m_a^2-m_c^2}{m_am_c}\,.
\end{equation}
Let's define $\tau^a_I=m_aU^a_I$, then
\begin{equation}
\tau^a_I\tau^b_I=m_a^2\delta^{ab}.
\end{equation}
So, we can re-write the eq.(\ref{ColorConstraint}) as
\begin{equation}\label{ColorConstraint2}
(\tau^a T^b\tau^c)=f^{abd}(\tau^b\tau^d-\tau^a\tau^d-\tau^c\tau^d)
\end{equation}
where we have suppressed the contraction of indices $I,J$. The solution to the constraint for $\tau^a_I$ is simply that
\begin{equation}
\tau^a_I=T^a_{IJ}V_J
\end{equation}
for some constant vector $V_J$ (the ``vev"). Indeed this is precisely what we get in the usual Higgs mechanism. The combination $T^a_{IJ} V_J\phi_I$ is ``eaten", and diagonalising $(M^2)^{ab}=V^TT^aT^bV$.

One can check that after substituting for $\tau$, eq.(\ref{ColorConstraint2}) becomes  
\begin{align}
V^TT^aT^bT^cV&=-V^TT^cT^bT^aV=\frac{1}{2}V^T(T^aT^bT^c-T^cT^bT^a)V
\end{align}
(note we are always writing with real states so $T^a_{IJ}=-T^a_{JI}$). Now, if we assume that the ``coupling tensor" $f^{abc}$ is the structure constant for the Lie group associated with $T^a$, then we can repeatedly use $T^aT^b=f^{abd}T^d+T^bT^a$, and we find,
\begin{align}
T^aT^bT^c&=f^{bcd}T^aT^d+T^aT^cT^b\nonumber\\
&=f^{bcd}T^aT^d+f^{acd}T^dT^b+T^cT^aT^b\nonumber\\
&=f^{bcd}T^aT^d+f^{acd}T^dT^b+f^{abd}T^cT^d+T^cT^bT^a
\end{align}
Using the fact that $V^TT^aT^bV$ is diagonalised, we find: 
\begin{align}
V^T&(T^aT^bT^c-T^cT^bT^a)V\nonumber\\
=&f^{bca}m_a^2+f^{acb}m_b^2+f^{abc}m_c^2\nonumber\\
=&f^{abc}(m_a^2+m_c^2-m_b^2).
\end{align}
Once eq.(\ref{ColorConstraint}) is satisfied, the rest of the story is again the same as our previous examples. Note in particular that we \textit{must} have Higgses! Even if we have $N_{\text{scalar}}=N_{\text{gluon}}$ precisely, the interactions are \textit{not} the correct ones for the full UV theory due to the standard polynomial growth of the longitudinal piece scattering, which is not present for the UV theory. But with the ``uneaten Higgses" included, is simply chosen to match the high energy limit, and we manifestly match to a healthy UV theory.

\subsection{Obstruction for Spin 2} 
We now consider massive spin-$2$ particles, which in the HE limit should yield a graviton, a massless vector and scalar. We would like to see if the massless interactions can be consistently unified into an IR massive amplitude. The three-point massive spin-2 amplitude can be easily written down as:
\eqa
\begin{gathered}
\begin{tikzpicture}[scale=0.5]
\draw[photon] (-1, 1.73) node[left] {$2$}-- (0, 0);
\draw[photon] (-1, -1.73) node[left]{$1$} -- (0, 0);
\draw[photon] (0, 0) -- (2, 0)node[right] {$3$};
\end{tikzpicture}
\end{gathered}
\quad&=&\quad \frac{1}{M_{pl}m^6}\left[\langle\bold{1}\bold{2}\rangle [\bold{1}\bold{2}]\langle \bold{3}|p_1{-}p_2|\bold{3}]+\text{cyc.}\right]^2\,,
\eqae
where $m$ is the mass of the massive graviton. Let us look at the HE limit. We can directly import what was done for non-abelian Higgs, and one finds:
\eq
\frac{1}{M_{pl}m^6}\left[\langle\bold{1}\bold{2}\rangle [\bold{1}\bold{2}]\langle \bold{3}|p_1{-}p_2|\bold{3}]+\text{cyc.}\right]^2\quad\underrightarrow{\;\quad HE\quad \;} \quad\left\{\begin{array}{c}({-}2,{-}2,{+}2):\quad \frac{1}{M_{pl}}\frac{\langle12\rangle^6}{\langle 13\rangle^2\langle 23\rangle^2} 
\\ 
(0,{-}2,0):\quad \frac{3}{M_{pl}}\frac{\langle12\rangle^2\langle 23\rangle^2}{\langle 13\rangle^2}
\end{array}\right.
\eqe
Notice the extra factor of $3$ associated with the minimally coupled scalars. This extra factor is due to the $3$ different combinations $(+,-,-)\times(-,-,+) $, $(-,-,+)\times(+,-,-) $ and $(0,-,0)\times(0,-,0) $. Thus the scalar coupling at high energy is three times what it should be. This is unacceptable since gravitational coupling is universal, and the coupling strength $M_{pl}$ has already been set by the self-interaction. Note that similar difficulties arise for the HE limit that yields the one graviton two minimally coupled vector, where one obtains  $-2\langle12\rangle^4/M_{pl}\langle13\rangle^2$. Again the factor of $2$ is inconsistent with graviton self coupling. Thus we see that there is a fundamental obstruction in organising the massless degrees of freedom into a massive spin-2 particle, in a way such that the massive interactions have HE limit that morphs into a consistent UV theory.    

\section{Loop Amplitudes}\label{LoopApply}
In this section we briefly touch on constructing loop amplitudes by an
on-shell gluing of the tree amplitudes we have found in previous
sections. We will follow the philosophy of ``generalized unitarity"~\cite{BDK, Forde},
where the integrand for loop amplitudes is determined by a knowledge
of its (generalized) cuts, putting internal propagators on-shell. As
is well-known, at one-loop this gives a systematic way of determining
the integrand from gluing together on-shell tree amplitudes.\footnote{There is an obvious  subtlety in this on-shell approach to loop amplitudes, regarding ``wavefunction renormalization". In the unitarity approach where one glues tree amplitude on both sides of the cut, there will be diagrams which correspond to a bubble insertion on the external leg, and hence give rise to an $1/0$ from the on-shell propagator. In the Feynman diagram approach, these are wave function diagrams that are to be amputated, replaced by counter terms. This procedure breaks gauge invariance in the intermediate steps. For massless internal states, these can be side stepped since there will be  UV-IR cancellation for these diagrams. For massive internal particles this is no-longer the case, and we refer the reader to~\cite{Bubbleprob} for unitarity based treatments of this issue. This subtlety will not affect any of the examples we discuss in this section: for (g-2) and rational terms, the 1-loop corrections are leading, while for the beta function the external massive particles are merely probes.} While we
are not adding anything new to this conceptual framework, the
technical advantages offered by our formalism for massive particles
with spin are significant in many cases,  including the dispensation
of complicated gamma matrix algebra, the clear separation of electric
and magnetic moments for charged particles, the extraction of UV
divergent properties without the contamination from IR divergences (by
virtue of using massive external and internal states), and finally
directly obtaining the (internal) mass depending pieces in the small
mass expansion relevant for obtaining rational terms for massless
one-loop amplitudes. In all of these processes, as they do not have tree counterparts, bubbles on external legs do not contribute. It is pleasing to continue seeing directly
the way in which Poincare symmetry and Unitarity fully determines the physics,
not just at tree-level but incorporating the leading quantum loop
corrections as well.
 
\subsection{$g{-}2$ for spin-$\frac{1}{2}$ and $1$}

As seen in previous discussions the simplicity of minimal coupling allows us to straight forwardly separate the magnetic moment pieces. The same simplicity translate to a straightforward computation for the loop level magnetic moment. 

Let's consider the $e^+,e^-\rightarrow \gamma$ at one loop. The diagram we want to build is:
\eq\label{g2Loops}
\vcenter{\hbox{\includegraphics[scale=0.55]{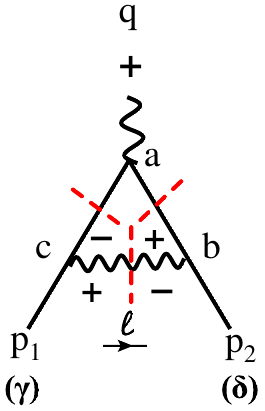}}}\sim e^3m^3x_{a}\ve_{\alpha\beta}\left[\ve^{\beta\gamma}\frac{x_b}{x_{c}}\left(\ve+x_c\frac{\la_{\ell}\la_{\ell}}{m}\right)^{\alpha\delta}+\ve^{\alpha\delta}\frac{x_c}{x_b}\left(\ve-x_b\frac{\la_{\ell}\la_{\ell}}{m}\right)^{\beta\gamma}\right]
\eqe
where we've glued the three-point vertices according to the two possible helicity configurations in the internal photon lines. Notice that here, we are using the three point amplitude in the SL(2,C) undotted basis. This is motivated by eq.(\ref{G2Def}), which yields a clear separation of $(g{-}2)$ factors in this basis. One can also understand this from the fact that anomalous moments should arise only if the particle carries spin. By expanding the integrand in eq.(\ref{g2Loops}), one notices that the $\lambda$ independent terms will be present for charged scalars as well, and thus the piece of the integrand that can contain the magnetic moment is:
\eq
e^2m^2x_{a}(x_b-x_c)\la_{\ell}^\delta\la_{\ell}^\gamma=-mx_{a}q^\delta\,_{\dot{\alpha}}\ell^{\dot{\alpha}\beta}\,.
\eqe
This gives us the following integrand:
\eq
-mx_{a}\int \frac{d^{4}\ell}{(2\pi)^4}\frac{q^\delta\,_{\dot{\alpha}}\ell^{\dot{\alpha}\beta}}{\ell^2((\ell-p_2)^2-m^2)((\ell+p_1)^2-m^2)}=\frac{e^2}{(4\pi)^2}2x_a\frac{q^\delta\,_{\dot{\alpha}}p_1^{\dot{\alpha}\beta}}{m}=\frac{\alpha}{2\pi}x_a^2\la_q^\gamma\la_q^\delta\,.
\eqe
This gives the $(g{-}2)=\frac{\alpha}{2\pi}$ by comparing with eq.(\ref{G2Def}).

Just to give us a little bit more challenge,  let's now consider the $W^+,W^-\rightarrow \gamma$ at one loop involving only photon coupling. The integrand is again built from:
\eqa
\vcenter{\hbox{\includegraphics[scale=0.55]{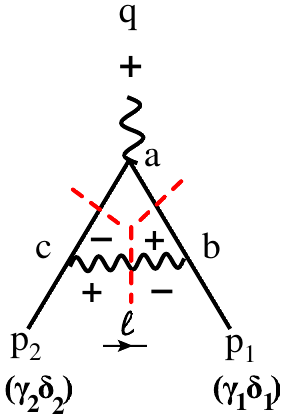}}}\sim &&e^3m^3x_{a}\ve_{\{\alpha_1\alpha_2}\ve_{\beta_1\}\beta_2}\left[\ve^{\beta_1\{\gamma_1}\ve^{\alpha_1\delta_1\}}\frac{x_b}{x_{c}}\left(\ve+x_c\frac{\la_{\ell}\la_{\ell}}{m}\right)^{\alpha_2\{\delta_2}\left(\ve+x_c\frac{\la_{\ell}\la_{\ell}}{m}\right)^{\beta_2\gamma_2\}}\right.\nonumber\\
&&\left.+\ve^{\beta_2\{\gamma_2}\ve^{\alpha_2\delta_2\}}\frac{x_c}{x_b}\left(\ve-x_b\frac{\la_{\ell}\la_{\ell}}{m}\right)^{\alpha_1\{\delta_1}\left(\ve-x_b\frac{\la_{\ell}\la_{\ell}}{m}\right)^{\beta_1\gamma_1\}}\right]\,.
\eqae 
Leaving behind the electric coupling, we now have two structures for the numerator of the integrand:
\eqa
&&e^2x_{a}(x_b-x_c)m^2\left[4\left(\ve^{\delta_1\{\delta_2}\la_{\ell}^{\gamma_1}\la_{\ell}^{\gamma_2\}}+\ve^{\gamma_1\{\delta_2}\la_{\ell}^{\delta_1}\la_{\ell}^{\gamma_2\}}\right)\right]+16e^2x_{a}x_bx_cm\la_\ell^{\delta_1}\la_\ell^{\delta_2}\la_\ell^{\gamma_1}\la_\ell^{\gamma_2}\nonumber\\
&=&\underset{\Large f_1(q)}{\underline{-4e^2x_{a}m\left[\ve^{\delta_1\{\delta_2}q^{\gamma_1}\,_{\dot{\alpha}}\ell^{\dot{\alpha}\gamma_2\}}+\ve^{\gamma_1\{\delta_2}q^{\delta_1}\,_{\dot{\alpha}} \ell^{\dot{\alpha}\gamma_2\}}\right]}}\nonumber \\
&+&\underset{\Large f_2(q)}{\underline{\frac{2e^2x_{a}}{3m} (p_{1\dot{\alpha}}\,^{\{\delta_1} \ell^{\dot{\alpha}\gamma_1\}})(p_{2\dot{\alpha}}\,^{\{\delta_2}\ell^{\dot{\alpha}\gamma_2\}})}}
\eqae 
Here $f_1(q)$ is the same as the electron moment, and leads to:
\eq
F_1(q)= \int \frac{d^{4}\ell}{(2\pi)^4}\frac{f_1(q)}{\ell^2((\ell-p_2)^2-m^2)((\ell+p_1)^2-m^2)}=4\frac{\alpha}{2\pi}x^a\left(\ve^{\delta_1\{\delta_2}\la_q^{\gamma_1}\la_q^{\gamma_2\}}+\ve^{\gamma_1\{\delta_2}\la_q^{\delta_1}\la_q^{\gamma_2\}}\right)\,.
\eqe
For the second tensor structure, one has:
\eq
F_2(q)=\int \frac{d^{4}\ell}{(2\pi)^4}\frac{f_2(q)}{\ell^2((\ell-p_2)^2-m^2)((\ell+p_1)^2-m^2)}=\frac{\alpha}{(4\pi)9m^3}\mathcal{O}^{\{\delta_1\gamma_1\}}_{1,2}\mathcal{O}^{\{\delta_2\gamma_2\}}_{1,2}\,,
\eqe
where we've defined $\mathcal{O}^{\alpha\beta}_{i,j}\equiv p_{i\dot{\alpha}}\,^\alpha p_j^{\dot{\alpha}\beta}$.
\subsection{The beta function}
Let's now turn to the extraction of beta function. For massless amplitudes, these can be obtained by extracting the coefficient for the bubble integrals in the scalar integral basis~\cite{Forde, SimplestQFT}. However, extra care needs to be taken for the subtraction of infrared divergence. Here we will instead consider two massive scalar probes of a photon propagator, and consider the correction to the propagator due to an internal massive scalar, fermion and vector (denoted by $X$):
$$\includegraphics[scale=0.45]{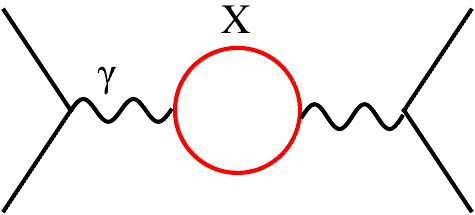}$$ 
The UV divergence of this amplitude contains \textit{the} contribution of a scalar to the beta function, without the IR-contamination. The loop amplitude will be constructed by gluing the 2$\rightarrow$2 amplitude involving the scalar probe particle exchanging a photon with $X$. This will allow us to obtain the beta function for different spins. From the massive vector, we will also be able to extract the contribution for a massless vector by simply subtracting a scalar. Assuming that the mass of $X$ is identical with that of the scalar probe, the relevant tree amplitudes can be easily constructed by generalizing the examples in subsection \ref{Scalarxx}:
\begin{align}
\vcenter{\hbox{\includegraphics[scale=0.45]{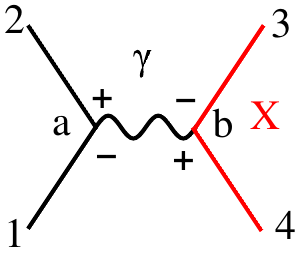}}}:\quad X \in{\rm scalar}\quad &\frac{m^2}{s}\left(\frac{x_a}{x_b}+\frac{x_b}{x_a}\right)=\frac{(p_1-p_2)\cdot p_3}{s}\quad\quad\quad\quad\quad\quad\quad\quad\quad\quad\quad\quad\nonumber \\
\vcenter{\hbox{\includegraphics[scale=0.45]{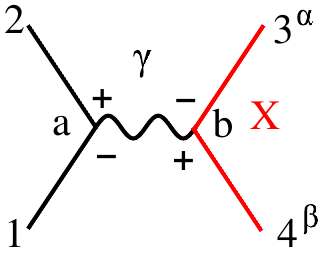}}}:\quad X \in{\rm fermion}\quad  &\frac{m}{s}\left(\frac{x_a}{x_b}[\mathbf{3}\mathbf{4}] {+}\frac{x_b}{x_a}\langle\mathbf{3}\mathbf{4}\rangle\right)\nonumber \\ 
&=\frac{1}{2ms}\left(2(p_1-p_2)\cdot p_3\langle\mathbf{3}\mathbf{4}\rangle{-}\langle\mathbf{3}|p_{1} P-Pp_1|\mathbf{4}\rangle\right)\nonumber\\
\vcenter{\hbox{\includegraphics[scale=0.45]{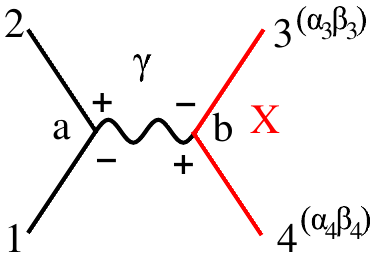}}}:\quad X \in{\rm vector} \quad &\frac{1}{s}\left(\frac{x_a}{x_b}[\mathbf{3}\mathbf{4}]^2 {+}\frac{x_b}{x_a}\langle\mathbf{3}\mathbf{4}\rangle^2\right)\nonumber \\ 
&=\frac{1}{m^2s}\left((p_1-p_2)\cdot p_3\langle\mathbf{3}\mathbf{4}\rangle^2{-}\langle\mathbf{3}\mathbf{4}\rangle\langle\mathbf{3}|p_{1} P-Pp_1|\mathbf{4}\rangle\right.\nonumber\\
&\left.-\frac{\langle\mathbf{3}|p_{1} P-Pp_1|\mathbf{4}\rangle\langle\mathbf{3}|P|\mathbf{4}]}{2m}\right)
\end{align} 
where we've again summed over the two possible photon helicity configuration and $P=p_3+p_4$. The  second equality for each amplitude gives the manifest local form, which can be checked against the H.E. limit where one should find a finite result as $m\rightarrow0$. Note that each term contains a piece which is identical to the scalar contribution. 

We can now glue the tree amplitudes into the one-loop integrand. The beta function can be readily read off by picking out the divergent piece which is proportional to the tree amplitude. For further simplification, we can take the $s\rightarrow0$ limit, and we will be looking for the term that is proportional to $\frac{2(p_1\cdot p_3)}{s}$. Let us use the scalar correction as an example. The one-loop amplitude is now 
\eq
\vcenter{\hbox{\includegraphics[scale=0.45]{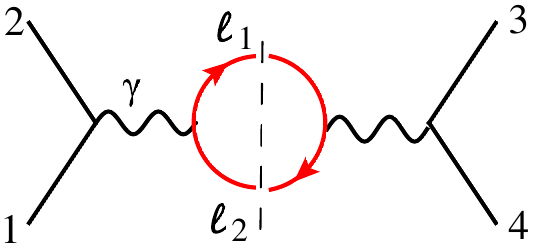}}}= \left.A^{\footnotesize scalar}_4(p_1,\ell_1)A^{\footnotesize scalar}_4(\ell_2,p_3)\right|_{s\rightarrow0}=\frac{4(p_1\cdot \ell_1)(p_3\cdot \ell_2)}{s^2}\,.
\eqe 
The one-loop integrand is then simply:
\eq
\frac{4}{s^2}\int \frac{d^{4-2\epsilon}\ell}{(2\pi)^4}\frac{(p_1\cdot \ell_1)(p_3\cdot \ell_2)}{(\ell^2-m^2)((\ell-P)^2-m^2)}=-\frac{1}{(4\pi)^2\epsilon}\frac{1}{6}\frac{(2p_1\cdot p_3)}{s}+\cdots
\eqe
where $\cdots$ represent terms terms that are purely functions of $s$, or finite. For fermions, there are now two pieces that are relevant, the square of the scalar piece, and the square of the $p_i P$ piece. All other contributions cannot generate the $p_1\cdot p_3$ tensor structure. We find:
\eq
A^{\footnotesize fermion}_4(p_1,\ell_1)A^{\footnotesize fermion}_4(\ell_2,p_3)=\frac{8(p_1\cdot \ell_1)(p_3\cdot \ell_2)}{s^2}-2\frac{(p_1\cdot p_3)}{s}+\cdots\,.
\eqe
The relevant part of the one-loop integrand is then:
\eq
\frac{1}{s}\int \frac{d^{4-2\epsilon}\ell}{(2\pi)^4}\frac{8(p_1\cdot \ell_1)(p_3\cdot \ell_2)/s-2(p_1\cdot p_3)}{(\ell^2-m^2)((\ell-P)^2-m^2)}=-\frac{1}{(4\pi)^2\epsilon}\frac{4}{3}\frac{(2p_1\cdot p_3)}{s}+\cdots\,.
\eqe
Finally, similar analysis for vectors yields:
\eq
A^{\footnotesize vector}_4(p_1,\ell_1)A^{\footnotesize vector}_4(\ell_2,p_3)=\frac{12(p_1\cdot \ell_1)(p_3\cdot \ell_2)}{s^2}+8\frac{(p_1\cdot p_3)}{s}
\eqe
which leads to 
\eq
\frac{1}{s}\int \frac{d^{4-2\epsilon}\ell}{(2\pi)^4}\frac{12(p_1\cdot \ell_1)(p_3\cdot \ell_2)/s+8(p_1\cdot p_3)}{(\ell^2-m^2)((\ell-P)^2-m^2)}=\frac{1}{(4\pi)^2\epsilon}\frac{7}{2}\frac{(2p_1\cdot p_3)}{s}+\cdots\,.
\eqe
Thus we've found that the beta function for a scalar is $\frac{1}{6}$ a Dirac fermion $\frac{4}{3}$ and a massless vector being $-\frac{7}{2}+\frac{1}{6}=-\frac{11}{3}$, where we've subtracted the scalar ``eaten" by the massive vector. 
\subsection{Rational terms}
Another application of massive amplitudes is to derive rational terms for massless amplitudes, that are not constructible via four-dimensional cuts. These terms appear due to the fact that the integrals are regulated and one can encounter $\epsilon/\epsilon\sim\mathcal{O}(1)$ effects. These terms can be obtained by considering the states in the internal loops to be massive~\cite{BadgerRational}, where the mass $m^2$ is identified with the extra $-2\epsilon$ dimension piece of $\ell^2$, denoted as $\mu^2$.\footnote{See~\cite{Fazio:2014xea} for some recent applications.} For QCD, one considers the contribution of a massive adjoint scalar state that is minimally coupled to the external gluons. These ``$\mu$" terms are computed using the tree-level amplitudes in $D$-dimensions~\cite{BernMorgan, 6D} and consider the extra dimension momenta as four-dimensional mass.

Here we will directly use the four-dimensional massive amplitudes to obtain the integral coefficients for $I_4[\mu^{2k}]$, the four-point scalar box integral with $\mu^{2k}$ as its numerator.  For the box-integral coefficient one considers the quadruple cut, where the two solutions for the cut loop momentum are:
\eq\label{CutSol}
\ell_1=\frac{1}{2}\left(c^{\pm}\tilde{\lambda}_1\lambda_4-\frac{m^2}{tc^{\pm}}\lambda_1\tilde\lambda_4\right),\quad c^{\pm}=\frac{\langle 12\rangle}{2\langle 42\rangle}\left(1\pm\sqrt{1+\frac{4m^2u}{st}}\right)\,.
\eqe
The box-coefficient is then obtained by gluing the four tree-amplitudes substituted with the cut loop momenta. 

First consider the four-point all-plus amplitude, where the cut is given by:  
\eqa
\vcenter{\hbox{\includegraphics[scale=0.5]{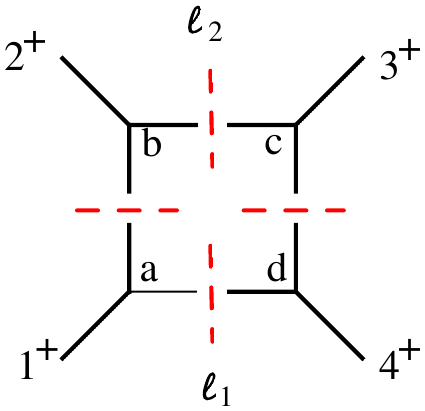}}}\quad \sim m^4 x_ax_bx_cx_d&=&m^4\frac{[41]m}{\langle 4|\ell_1|1]}\frac{\langle 4|\ell_1|1]}{\langle 41\rangle m}\frac{[23]m}{\langle 2|\ell_2|3]}\frac{\langle 2|\ell_2|3]}{\langle 23\rangle m}\nonumber\\
&=&m^4\frac{[41][23]}{\langle 41\rangle\langle 23\rangle}
\eqae
This directly gives the all plus integrand, $\frac{[12][34]}{\langle 12\rangle\langle 34\rangle}I_4[\mu^4]$. For the single minus amplitude, one instead has: 
\eqa
\vcenter{\hbox{\includegraphics[scale=0.5]{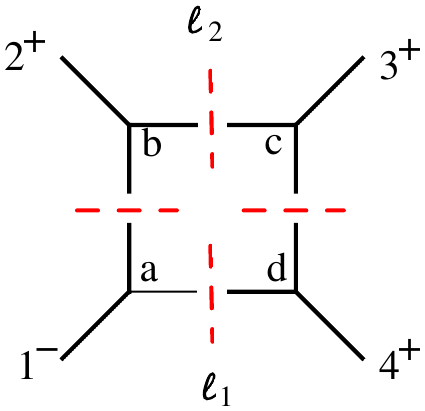}}}\quad \sim m^4 \frac{x_b x_c x_d}{x_a}&=&m^4\frac{\langle 4|\ell_1|1]^2}{\langle 41\rangle[14] m^2}\frac{[23]m}{\langle 2|\ell_2|3]}\frac{\langle 2|\ell_2|3]}{\langle 23\rangle m}\nonumber\\
&=&m^2\frac{\langle 4|\ell_1|1]^2}{t}\frac{[23]}{\langle 23\rangle}\,.
\eqae
Substituting the two solutions for the cut in eq.(\ref{CutSol}) and summing the results, one obtains
\eq
\frac{[23][42]\langle12\rangle}{4\langle23\rangle\langle 42\rangle[12]}\left(\frac{st}{2u}I_4[\mu^2]+I_4[\mu^4]\right)\,.
\eqe
The above rational terms are in agreement  with~\cite{BernMorgan}.
\section{Form Factors and Correlation Functions}\label{Form}
The ability to discuss scattering amplitudes for general mass and spin largely removes the distinction between  amplitudes and ``off-shell" objects such as correlation functions and form-factors. Consider correlation functions for the stress tensor for some theory.  The computations are precisely the same as what we would carry out if we were computing the scattering amplitude for a massive spin two particle, (arbitrarily) weakly coupled to the theory. The scattering amplitude for these massive particles gives us the correlation function in momentum-space, corresponds closely to the experiments that are actually done to measure correlation functions.  Strictly speaking we are coupling a continuum of particles of different masses, and we are getting the correlator in momentum space for the external legs $p_a$ in the timelike Lorentzian region where $p_a^2>0$. But we can then define the correlators for null and spacelike momenta by analytic continuation. At least in perturbation theory---which is what we will largely concern ourselves with here in this subsection---there is no ambiguity for what this means in practice.

It is important to imagine that the massive particle ${\cal O}$ corresponding to the operator is simply an external probe and does not participate in the dynamics. In other words, we should not have any ``internal propagators" associated with cuts that put ${\cal O}$ on-shell. In practice, this means that we should be able to make the coupling of ${\cal O}$ to our system proportional to a parameter $\epsilon$ that we can make as small as we wish. To take an example, consider a 3-point coupling of ${\cal O}$ to a pair of massless particles for the system of interest; making this proportional to $\epsilon$ means that the leading amplitudes will never involve internal ${\cal O}$ particles: 
\begin{equation}
\begin{gathered}
\begin{tikzpicture}[scale=0.5]
\draw[thick, double] (-2, 0) node[left] {$\mathcal{O}$}-- (0, 0);
\draw[photon, red] (0, 0) -- (1, 1.73);
\draw[photon, red] (0, 0) -- (1, -1.73);
\draw[thick, fill=black] (0, 0) circle (.1);
\end{tikzpicture}
\end{gathered}\,: \epsilon,
\qquad
\begin{gathered}
\begin{tikzpicture}[scale=0.5]
\draw[thick, double] (-2, 0) node[left] {$\mathcal{O}$}-- (0, 0);
\draw[photon, red] (1, 0) circle (1);
\draw[thick, double] (2, 0) -- (4, 0) node[right] {$\mathcal{O}$};
\draw[thick, fill=black] (2, 0) circle (.1);
\draw[thick, fill=black] (0, 0) circle (.1);
\end{tikzpicture}
\end{gathered}\,: \epsilon^2.
\end{equation}

In general, the leading amplitude involving $N$ ${\cal O}$'s will be proportional to $\epsilon^N$ and will never involve internal ${\cal O}$ particles.
\subsection{Observables in Gauge Theories and Gravity}
Before moving on to illustrating how this interpretation is useful in concrete calculations, let us pause to interpret some standard and elementary facts about observables in gauge theories and gravity from this on-shell perspective. 

In particular, let us understand the reason for the absence of charged local operators in gauge theory, or any local operators whatsoever in gravity. Consider a charged operator $\Phi$. We know that consistency enforces universal coupling of $\Phi$ to photons/gluons, with strength set by the gauge coupling $g$, and so we {\it can't} arbitrarily weakly couple $\Phi$ to the system. Thus we can't speak of charged local operator. Similarly with gravity, the coupling of any particle to gravity is universal given by $\sqrt{G_N}$, so in the presence of gravity we can't meaningfully talk about any local operators at all. In a conventional Lagrangian description of the physics, this is associated with the impossibility of making local charged operators gauge invariant. Of course we can always fix a gauge and compute correlators for operators in that gauge, but then these are not quite local. If we start with correlators of local operators in the limit as $g^2 \to 0$ or $G_N \to 0$, the weak gauging attaches Wilson lines to the operators in some way. Of course this also has an obvious on-shell meaning, again corresponding closely to physical experiments that measure these Wilson-line dressed correlators. 

Consider again a charged scalar $\Phi$ of charge +1 in an abelian gauge theory, and let's consider the correlator $\langle \Phi^*(x) \Phi(y) \rangle$ first in the limit where we turn off the gauge coupling. We may have $U(1)$ invariant self-interactions for $\Phi$ of the form e.g. $(\Phi^* \Phi)^2$, and we can also turn on the gauge-interactions. But we also couple $\Phi$ to some heavy external probe particles $X^{(q)}$, $Y^{(q{+}1)}$ and $A^{(Q)}, B^{(Q+1)}$ via the couplings $\epsilon X^{(q)} Y^{(q{+}1) *} \Phi$, $\epsilon^\prime A^{({-}Q)} B^{({-}Q{-}1) *} \Phi^*$. Let's now look at the $(X Y^* B^* A)$ scattering amplitude. Since this breaks the global particle number symmetries acting separately on $X,Y,A,B$ as $\epsilon, \epsilon^\prime \to 0$, this amplitude is proportional to the product $\epsilon \epsilon^\prime$; some of the diagrams contributing to the amplitude are shown below:
\begin{equation}
\vcenter{\hbox{\includegraphics[scale=0.5]{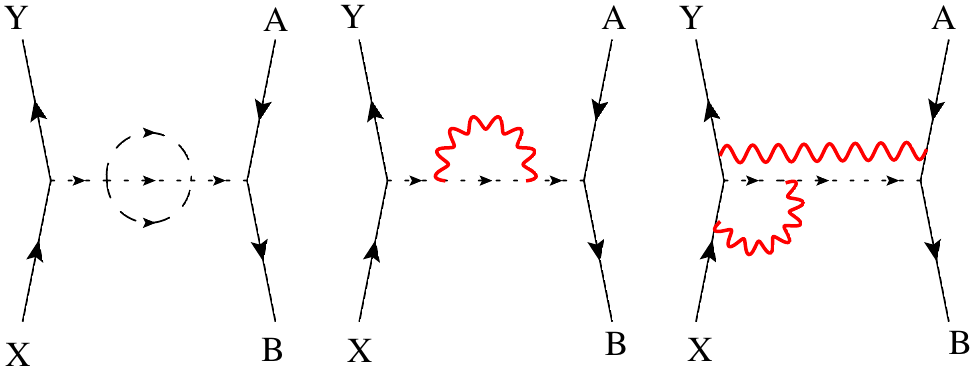}}}\,.
\end{equation}
As $\epsilon, \epsilon^\prime \to 0$, stripping off this product from the amplitude yields the correlator where $\langle \Phi^*(x) \Phi(y)\rangle$ is dressed with Wilson lines in the $p_X,p_Y,p_A,p_B$ directions:
\begin{equation}
M(p_X,p_Y,p_A,p_B) \to \epsilon \epsilon^\prime \int_{x,y} 
 e^{i(p_X {+} p_Y) x} e^{i(p_A {+} p_B)y} \langle \left(W_{p_X}^{q} \Phi W_{p_Y}^{*(q{+}1)}\right)(x) \left(W_{p_A}^{{-}Q} \Phi W_{p_B}^{*({-}Q{-}1)}\right)(y) \rangle
\end{equation}
The fact that inequivalent ``dressings" of the local operator with Wilson lines are possible simply reflects the many different ways we can couple $\Phi$ to external probes; since the probes themselves are charged and emit long-range gauge fields, the amplitudes (and hence the extracted correlator) does depend on the choices that are made.
Thus, while correlation functions for local charged operators don't exist, dressed version of these correlators exist, for both gauge theory and gravity, to all orders in $g$ and $\sqrt{G_N}$.

There is a deeper difficulty with gravity, which makes even these quasi-local ``Wilson-line dressed" correlators ambiguous at a non-perturbatively tiny level, of $O(exp(-M_{Pl}^2/s))$. As we saw in our example above, in order to be able to identify the piece of the amplitude for the heavy probes that is unambiguously associated with the coupling to the operator $\Phi$, it was important that the coupling to the probe broke some global symmetry of the problem. But we expect that gravity breaks all global symmetries, and in particular, we can't say that e.g. the $X Y^* A^* B$ amplitude is arbitrarily small; there is some (perhaps virtual black-hole mediated) rate for this process of $O(exp(-M_{Pl}^2/s))$ that pollutes any attempt to associate this amplitude with ``the" (Wilson-line dressed) correlator of interest, making it impossible to pick out a piece proportional to $\epsilon \epsilon^\prime$ as $\epsilon, \epsilon^\prime \to 0$. 

Summarizing more informally, in both gauge theories and gravity we don't have meaningful correlators of local charged operators, for the (relatively trivial) reason that we can't ignore the long-range gauge and gravitational fields. This can already be seen perturbatively in $g^2, G_N$, but to all order in these couplings, there are dressed versions of local operators that take care of the long-range fields at infinity, smoothly deforming the local correlators we have when $g^2,G_N=0$. But in gravity, due to exponentially small effects of $O(exp(-Area/G_N))$, associated with black-hole physics, even these dressed versions of local operators don't make precise sense. This is a concrete sense in which any notion of spacetime becomes ambiguous in quantum gravity, highlighting that e.g. the breakdown of locality in the context of the black-hole information paradox is an effect of $O(exp(-S_{BH}))$, and is otherwise invisible to every order in $G_N$.

\subsection{Weinberg-Witten}

The interpretation of correlators in terms of massive amplitudes allows us to re-interpret some familiar facts about massive amplitudes we have already encountered to other well-known facts about QFT's. Consider the Weinberg-Witten theorem~\cite{WW}, which in this way of thinking is essentially identical to Yang's theorem. Recall the discussion of consistent couplings of a massive spin $S$ particle to massless particles. Note that since conserved currents and stress tensors measure the charge and the momentum on single particle states respectively, we will be interested in the interaction of the massive state with two opposite helicity massless-particles $h_1=-h_2$.\footnote{Recall that all momenta are out going, so for $p_1$ and $p_2$ to represent the same particle, $h_1=-h_2$. } Our analysis showed that $S+h_2-h_1$ and $S+h_1-h_2$ must always be greater or equal to 0, this tells us that for $S=1$, $|h_1|=|h_2|\leq\frac{1}{2}$, i.e. massless particles with spin $>\frac{1}{2}$ cannot couple to a Lorentz covariant conserved current. Similarly for $S=2$, $|h_1|=|h_2|\leq1$, and massless particles with spin $>1$ cannot couple to a conserved stress-tensor. This is precisely the Weinberg-Witten theorem.

\subsection{Form Factors Example: Stress Tensor/Gluons}
From Weinberg-Witten theorem we know that the stress tensor can only couple to massless particles of spin $\leq 1$, thus we will consider form factors of a stress tensor and three gluons. Identifying the stress tensor as a massive spin-2 state, we will map this to a four-point amplitude involving one massive and three massless states:
\begin{equation}
\begin{gathered}
\begin{tikzpicture}[scale=0.7]
\draw[photon,thin] (0,0) -- (2,1) node[right] {$-$};
\draw[photon,thin] (0,0) -- (2,0) node[right] {$+$};
\draw[photon,thin] (0,0) -- (2,-1) node[right] {$+$};
\draw[fill=white] (0, 0) circle (0.3);
\path (0,0) node {$\times$};
\path (0,0) node[left=6pt] {$T$};
\end{tikzpicture}
\end{gathered}
\longrightarrow
\begin{gathered}
\begin{tikzpicture}[scale=.7]
\draw[scalar] (-1.73,-1)node[left] {$T_1$} -- (0,0);
\draw[photon,thin] (-1.73,1)node[left] {$2^-$} -- (0,0) ;
\draw[photon,thin] (1.73,-1)node[right] {$4^+$} -- (0,0) ;
\draw[photon,thin] (1.73,1)node[right] {$3^+$} -- (0,0) ;
\draw[fill=gray!20] (0, 0) circle (0.5);
\end{tikzpicture}
\end{gathered}
\end{equation}
Let us consider the $t$-channel massless residue. Since the gluon is ``charged" under the stress tensor, for the one massive two massless coupling, one should consider opposite helicity gluons. The $t$-channel residue can then be written as:
\eq
(\lambda_P)^4\frac{[p4]^2}{m^3}\frac{[3P]^3}{[P2][23]}=\frac{(\lambda_2)^4m[23]}{\langle43\rangle\langle24\rangle}\,,
\eqe
where again, the equality holds for $\langle23\rangle=0$. This leads us to the following simple expression for the form factor:
\eq
\langle \tilde{T}(1)|2^-3^+4^+\rangle=\frac{(\lambda_2)^4m}{\langle43\rangle\langle32\rangle\langle24\rangle}
\eqe
It is straight forward to check that the above result matches all three factorisation channels, as expected from its cyclic invariant form, up to the over all factor of $(\lambda_2)^4$ that takes care of the excess helicity weight and the stress tensor's SL(2,C) indices. We can straight forwardly extend to two stress tensors coupled to two gluons: 
\eq
\langle \tilde{T}(1)\tilde{T}(2)|3^-4^+\rangle=(\lambda_3)^4\left(\frac{([4|p_1)^2([3|p_2)^2}{t}+\frac{([3|p_1)^2([4|p_2)^2}{u}\right)\,.
\eqe

There is an elephant in the room that we have not yet addressed. So far we have been considering conserved operators as massive spinning states. But conserved operators are a tiny subset of an infinite number tensor operators, for which all must have well defined form factors (and in the next section momentum space correlation functions). Furthermore, we should be able to see there must be a kinematic distinction between conserved operators and non-conserved operators, such that higher-spin conserved currents for an interacting theory can be ruled out, \`a la Coleman–Mandula theorem~\cite{Coleman}.

As an exercise let's consider a theory with two scalars $(\phi, \bar{\phi})$  and the operators $\mathcal{O}_{1\mu}= \phi\overleftrightarrow{\partial}_{\mu} \bar{\phi}$ and $\mathcal{O}_{2\mu}= \phi\partial_\mu \bar{\phi}$. The first is a conserved current while the second is not. Let us now consider the three-point form factor for 
\eq
\langle \tilde{\mathcal{O}}_{1\alpha\dot{\alpha},\beta\dot{\beta}}|p_1p_2\rangle\sim (p_1-p_2)^{\alpha\dot{\alpha}},\quad \langle \tilde{\mathcal{O}}_{2\alpha\dot{\alpha},\beta\dot{\beta}}|p_1p_2\rangle\sim p_1^{\alpha\dot{\alpha}}\,.
\eqe
Converting the above result into pure undotted SL(2,C) indices by contracting with $(p_1+p_2)$ one finds:
\eqa
\langle \tilde{\mathcal{O}}_{1}|p_1p_2\rangle\sim[12]\lambda^{\{\alpha_1}_1\lambda_2^{\alpha_2\}},\quad\langle\tilde{\mathcal{O}}_{2}|p_1p_2\rangle\sim [12]\lambda_1^{\alpha_1}\lambda_2^{\alpha_2}=\frac{1}{2}[12]\left(\lambda_1^{\{\alpha_1}\lambda_2^{\alpha_2\}}+\langle12\rangle\varepsilon^{\alpha_2\alpha_1}\right)\,.
\eqae
Not surprisingly the form factor for $\mathcal{O}_{2}$ can be further decomposed into a combination of $S=2,1$ and $0$ states. Thus we see that a general operator simply corresponds to a linear combination of lower spin states. In position space this is a statement that a general current, for example, can  
\eq
\mathcal{O}^\mu=(\eta^{\mu\nu}-\frac{\partial^\mu\partial^\nu}{\Box})\mathcal{O}_\nu+\frac{\partial^\mu\partial^\nu}{\Box}\mathcal{O}_\nu\equiv\hat{\mathcal{O}}^\mu+\frac{\partial^\mu\partial^\nu}{\Box}\mathcal{O}_\nu\,
\eqe
where $\hat{\mathcal{O}}^\mu$ is the conserved piece. Note that while there is a conserved piece in a general operator, the projection introduces non-locality and are thus distinct from a genuine conserved operator. This non-locality is present in all the lower spin components in the projection.

Let us look at this distinction more closely in the context of general form factors. For an interacting theory, the form factor will in general have poles whose residue reveals the existence of a non-trivial S-matrix:
\eq
\includegraphics[scale=0.5]{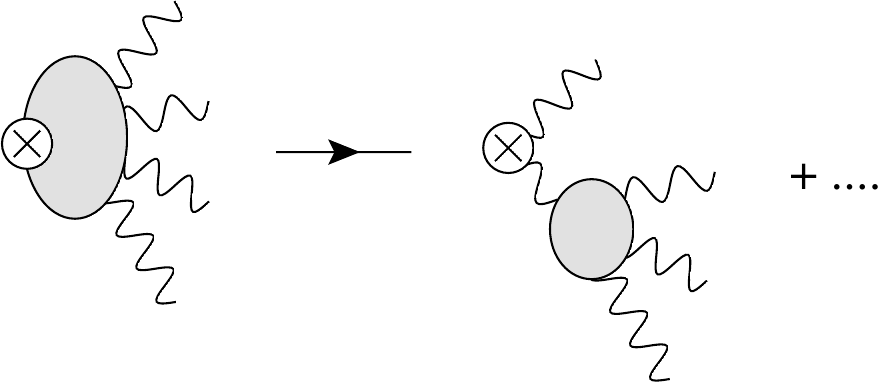}\,.
\eqe
Let us consider the particles to be massless, and take the momenta of the operator to be soft. Then just like the usual Weinberg's soft theorems for S-matrix, the form factor will be dominated by diagrams where one has the operator attached to the external leg
\eq
\sum_i\quad\vcenter{\hbox{\includegraphics[scale=0.5]{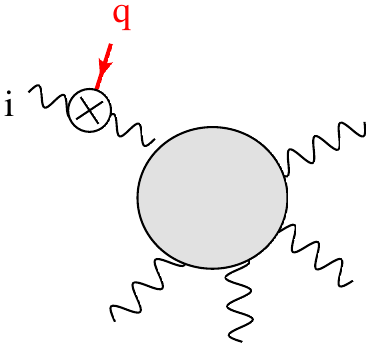}}}=\sum_i\frac{n(p_i,q)}{p_i\cdot q}M_n
\eqe
where $q$ is the soft momenta of the operator, $n(p_i,q)$ is the numerator function. If the operator is a tensor, then $n(p_i,q)$ should carry the corresponding Lorentz indices. Conserved tensor is reflected in that the form factor must vanish when contract with $q^\mu$. If we have a conserved current, then we can have $n(p_i,q)^\mu=e_i p^\mu_i$, where $e_i$ is the charge of each external state. The requirement of conservation then simply corresponds to the requirement of charge conservation. Similarly for conserved stress tensor we have  $n(p_i,q)^{\mu\nu}=\kappa p^\mu_ip^\nu_i$, and the conservation condition is simply stem from momentum conservation if the coupling $\kappa$ is universal. Note that for higher spins, $S>2$, there are no local solutions for $n(p_i,q)^{\mu_1\cdots \mu_S}$  such that the conserved quantity is respected. This is the Coleman-Mandula theorem! The assumptions that went into this argument is the existent of a non-trivial S-matrix, the analyticity of the form factor which can be interpreted as a massive S-matrix, and Lorentz invariance. The fact that the argument is closely related to Weinberg's soft theorems for gauge bosons is not a surprise in view of our usual intuition that if a conserved tensor exists in an interacting theory, then we can always weakly gauge it and have non-trivial S-matrix involving the gauge boson.

Note that while one can always project out a conserved piece for non-conserved tensors, the corresponding form factor will include non-local pieces. Indeed in this case we can have, for example, $n(p_i,q)^\mu=\frac{q^\mu \tilde{n}(p_i,q)}{q^2}=\frac{q^\mu \tilde{n}(p_i,q)}{m^2}$. This non-locality is again reflected in the singularity of the $m^2\rightarrow 0$ limit. This of course is an artifact of our projection, since there will be lower spin contributions coming along that will contain the same singularity and conspire to cancel, producing a smooth $m^2\rightarrow 0$ limit.

\subsection{Current and Stress-Tensor Correlators}
Let's consider the two and three-point correlation functions for stress-tensors in a conformal theory. In momentum space, the tree-level correlator are computed by gluing tree-level amplitudes with one massive leg and two massless legs. For conformal theories, the available tensor structures are constrained by conformal symmetry. In momentum space, this constraint is simply a reflection of the uniqueness of the three-point amplitude, which is fixed by the spin of the massive state and the helicities of the massless legs.  

For example the two point function receives contribution from:
\eqa
&&\langle T_{\alpha_1\alpha_2\alpha_3\alpha_4} T_{\beta_1\beta_2\beta_3\beta_4} \rangle =\quad \vcenter{\hbox{\includegraphics[scale=0.4]{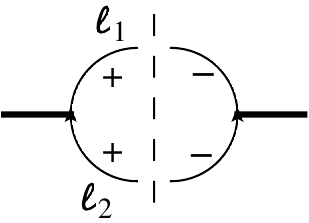}}} I_2\left[S^{\ell_1\ell_2}_{\alpha_1\alpha_2}S^{\ell_1\ell_2}_{\alpha_3\alpha_4}S^{\ell_1\ell_2}_{\beta_1\beta_2}S^{\ell_1\ell_2}_{\beta_3\beta_4}\right]\nonumber\\
&+&\vcenter{\hbox{\includegraphics[scale=0.4]{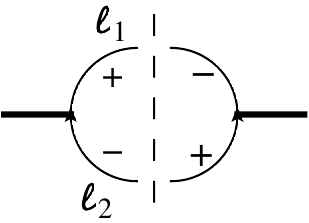}}} I_2\left[\prod_{i=1}^4 S^{\ell_1\ell_2}_{\beta_i\alpha_i}\right]+\vcenter{\hbox{\includegraphics[scale=0.4]{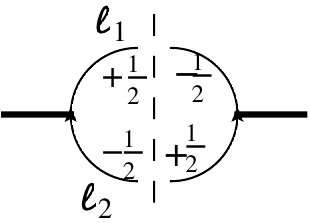}}} I_2\left[S^{\ell_1\ell_2}_{\alpha_1\alpha_2}S^{\ell_1\ell_2}_{\beta_1\beta_2}S^{\ell_1\ell_2}_{\alpha_3\beta_3}S^{\ell_1\ell_2}_{\alpha_4\beta_4}\right]
\eqae
where we've listed the contributions from different internal helicity configuration and $I_2[X]$ is defined as:
\eq
I_2[X]\equiv \int d^4\ell\frac{X}{\ell^2(\ell-k)^2}
\eqe
where $k$ is the momenta of the stress tensor. The operator $S^{\ell_1\ell_2}_{\alpha_1\alpha_2}$ is a shorthand notation for $\ell_{1\alpha_1\dot{\beta}}\ell_{2\alpha_2}\,^{\dot{\beta}}$. Note that it is understood that the expression must be symmetrized over $\{\alpha_i\}$ and $\{\beta_i\}$ separately, as well as over exchanging $\alpha_i\leftrightarrow \beta_i$, which takes into account the conjugate helicity configurations. For the scalar and and equal helicity fermion contributions, their tensor structure are identical to that of equal helicity gauge field.

For the three-point function one has:
\eqa
&&\langle T_{\alpha_1\alpha_2\alpha_3\alpha_4} T_{\beta_1\beta_2\beta_3\beta_4} T_{\gamma_1\gamma_2\gamma_3\gamma_4} \rangle =\vcenter{\hbox{\includegraphics[scale=0.4]{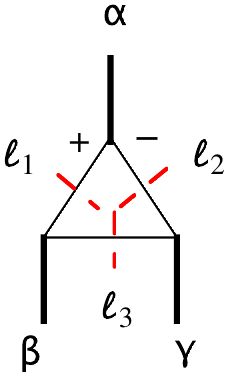}}}\left\{I_3\left[S^{\ell_1\ell_2}_{\beta_1\alpha_1}S^{\ell_1\ell_2}_{\beta_2\alpha_2}S^{\ell_1\ell_3}_{\beta_3\gamma_1}S^{\ell_1\ell_3}_{\beta_4\gamma_2}S^{\ell_3\ell_2}_{\gamma_3\alpha_3}S^{\ell_3\ell_2}_{\gamma_4\alpha_4}\right]\right.\nonumber\\
&+&\left.I_3\left[S^{\ell_1\ell_2}_{\beta_1\alpha_1}S^{\ell_1\ell_2}_{\beta_2\alpha_2}S^{\ell_2\ell_3}_{\alpha_3\beta_3}S^{\ell_2\ell_3}_{\alpha_4\beta_4}S^{\ell_2\ell_3}_{\gamma_1\gamma_2}S^{\ell_2\ell_3}_{\gamma_3\gamma_4}\right]\right\}+\vcenter{\hbox{\includegraphics[scale=0.4]{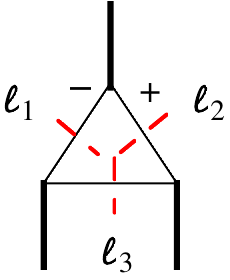}}}\left\{I_3\left[S^{\ell_1\ell_2}_{\alpha_1\gamma_1}S^{\ell_1\ell_2}_{\alpha_2\gamma_2}S^{\ell_1\ell_3}_{\alpha_3\beta_1}S^{\ell_1\ell_3}_{\alpha_4\beta_2}S^{\ell_3\ell_2}_{\beta_3\gamma_3}S^{\ell_3\ell_2}_{\beta_4\gamma_4}\right]\right.\nonumber\\
&+&\left.I_3\left[S^{\ell_1\ell_2}_{\alpha_1\gamma_1}S^{\ell_1\ell_2}_{\alpha_2\gamma_2}S^{\ell_1\ell_3}_{\alpha_3\gamma_3}S^{\ell_1\ell_3}_{\alpha_4\gamma_4}S^{\ell_1\ell_3}_{\beta_1\beta_2}S^{\ell_1\ell_3}_{\beta_3\beta_4}\right]\right\}+\vcenter{\hbox{\includegraphics[scale=0.4]{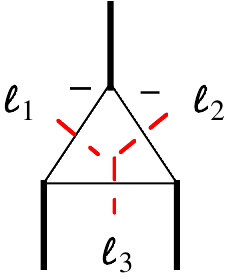}}}\left\{I_3\left[S^{\ell_1\ell_3}_{\alpha_1\beta_1}S^{\ell_1\ell_3}_{\alpha_2\beta_2}S^{\ell_2\ell_3}_{\alpha_3\beta_3}S^{\ell_2\ell_3}_{\alpha_4\beta_4}S^{\ell_2\ell_3}_{\gamma_1\gamma_2}S^{\ell_2\ell_3}_{\gamma_3\gamma_4}\right]\right.\nonumber\\
&+&\left.I_3\left[S^{\ell_1\ell_3}_{\beta_1\beta_2}S^{\ell_1\ell_3}_{\beta_3\beta_4}S^{\ell_1\ell_3}_{\alpha_1\gamma_1}S^{\ell_1\ell_3}_{\alpha_2\gamma_2}S^{\ell_2\ell_3}_{\alpha_3\gamma_3}S^{\ell_2\ell_3}_{\alpha_4\gamma_4}\right]\right\}+\vcenter{\hbox{\includegraphics[scale=0.4]{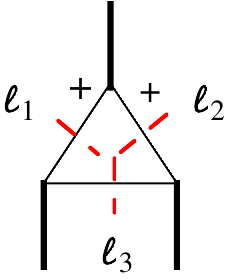}}}\left\{I_3\left[S^{\ell_1\ell_2}_{\alpha_1\alpha_2}S^{\ell_1\ell_2}_{\alpha_3\alpha_4}S^{\ell_1\ell_2}_{\beta_1\gamma_1}S^{\ell_1\ell_2}_{\beta_2\gamma_2}S^{\ell_2\ell_3}_{\gamma_3\beta_3}S^{\ell_2\ell_3}_{\gamma_4\beta_4}\right]\right.\nonumber\\
&+&\left.I_3\left[S^{\ell_1\ell_2}_{\alpha_1\alpha_2}S^{\ell_1\ell_2}_{\alpha_3\alpha_4}S^{\ell_1\ell_2}_{\beta_1\gamma_1}S^{\ell_1\ell_2}_{\beta_2\gamma_2}S^{\ell_1\ell_3}_{\beta_3\gamma_3}S^{\ell_1\ell_3}_{\beta_4\gamma_4}\right]\right\}+\vcenter{\hbox{\includegraphics[scale=0.4]{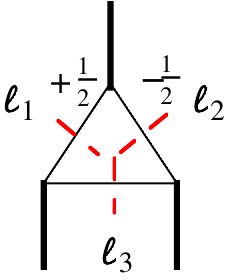}}}\left\{I_3\left[S^{\ell_1\ell_2}_{\alpha_1\alpha_2}S^{\ell_1\ell_2}_{\beta_3\alpha_3}S^{\ell_1\ell_3}_{\beta_1\beta_2}S^{\ell_2\ell_3}_{\gamma_1\gamma_2}S^{\ell_2\ell_3}_{\gamma_3\gamma_4}S^{\ell_2\ell_3}_{\alpha_4\beta_4}\right]\right.\nonumber\\
&+&\left.I_3\left[S^{\ell_1\ell_2}_{\alpha_1\alpha_2}S^{\ell_1\ell_2}_{\beta_3  \alpha_3}S^{\ell_1\ell_3}_{\beta_1\beta_2}S^{\ell_1\ell_3}_{\beta_4 \gamma_3 }S^{\ell_2\ell_3}_{\gamma_1\gamma_2}S^{\ell_2\ell_3}_{\alpha_4 \gamma_4}\right]\right\}+\vcenter{\hbox{\includegraphics[scale=0.4]{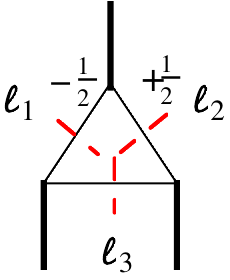}}}\left\{I_3\left[S^{\ell_1\ell_2}_{\alpha_1\alpha_2}S^{\ell_2\ell_1}_{\gamma_3\alpha_3}S^{\ell_2\ell_3}_{\gamma_1\gamma_2}S^{\ell_1\ell_3}_{\beta_1\beta_2}S^{\ell_1\ell_3}_{\beta_3\beta_4}S^{\ell_1\ell_3}_{\alpha_4\gamma_4}\right]\right.\nonumber\\
&+&\left.I_3\left[S^{\ell_1\ell_2}_{\alpha_1\alpha_2}S^{\ell_2\ell_1}_{\gamma_3  \alpha_3}S^{\ell_2\ell_3}_{\gamma_1\gamma_2}S^{\ell_2\ell_3}_{\gamma_4 \beta_3 }S^{\ell_1\ell_3}_{\beta_1\beta_2}S^{\ell_1\ell_3}_{\alpha_4 \beta_4}\right]\right\}
\eqae
and $I_3[X]$ is defined as:
\eq
I_3[X]\equiv \int d^4\ell\frac{X}{\ell_1^2(\ell_1-k_2)^2(\ell_1+k_1)^2}
\eqe
where $k_1,k_2$ are the momenta carried by the $\alpha_i$ and $\beta_i$ indexed stress-tensor respectively. Again symmetrisation interns of $\{\alpha_i\}$, $\{\beta_i\}$ and $\{\gamma_i\}$ are implied and the equal helicity fermion on the one of the vertices as well as internal scalars do not produce new tensor structures.

\section{Outlook}

Relativistic quantum mechanics governs the laws of nature at low
enough energies so that physics can be described in flat space, with a
finite number of interacting particles. ``Quantum field theory" is the
standard textbook approach to this physics, where, as useful
theoretical constructs, ``local quantum fields" are introduced, along with
the attendant baggage of field redefinition and gauge redundancies,
in order to allow a description of the physics in a way compatible
with relativistic locality and unitarity. But the on-shell approach to
scattering amplitudes suggests that this may not be the only way---that
we might instead be able to describe relativistic quantum mechanics
without local quantum fields, directly in terms of the physical
particles. \footnote{It is amusing that the on-shell program is often
contrasted with the standard approach using Feynman diagrams, since
Feynman's primary physical motivation for introducing his diagrams to
begin with was to get rid of quantum fields---and he was famously
disappointed to learn, via Dyson's proof, that his diagrams were so
closely related to field theory after all!}

In this paper we have taken the first steps to extending the ideas of
this on-shell approach to cover particles of all masses and spins in
four dimensions. The purely kinematical part of our discussion has
been fundamentally trivial---but trivializing the kinematics allows to
understand the structure of the physics as following seamlessly from
the foundational principles of Poincare Invariance, Locality and
Unitarity in a satisfying way.

We have seen many aspects of this understanding throughout this paper.
The structure of three particle amplitudes, for any mass and spin, is
fixed by Poincare invariance.  For massless particles, there is a
peculiarity for high enough spin---the three particle amplitudes are
superficially ``non-local" in the sense of having poles; while this doesn't
show up in $(3,1)$ signature Minkowski space where these amplitudes
vanish, it does mean that consistent factorization at four points is
non-trivial, and indeed, all but the usual massless theories we know
and love, of interacting spin $(0,1/2,1,3/2,2)$, are ruled out by
these considerations. We learn that we can only have a single massless
spin two particle, with universal couplings, that the massless spin
one particles must have the structure of Yang-Mills theories, and spin
3/2 requires supersymmetry. Furthermore the mere existence of a
consistent amplitude coupling to gravitons rules out all higher spin
massless particles. 

Similarly there is still a superficial ``non-locality"
associated with the coupling of a single massive particle to massless
particles with spin---the ``$x-$ factor"---which again makes
factorization non-trivial. Unlike
the case for massless particles, we {\it can} (non-trivially) find
consistently factorizing four-particle amplitudes for any choice of
three-particle couplings, (with the usual restrictions on consistent
couplings to massless spin one and spin 2 particles). But for massive particles of high
enough spin, these consistently factorizing amplitudes are badly
behaved at high energies---growing with powers of $(p_i \cdot p_j/m^2)$,
so that the massless limit cannot be taken smoothly. This tells us
that even massive particles of high enough spin cannot be separated
by a parametrically large gap from other particles---massive particles
with high spin cannot be ``elementary". Finally, three particle
amplitudes involving all massive particles are local, but naturally
have powers of $1/m$. Thus, theories of massive particles can only
smoothly interpolate to massless amplitudes at high energies for
special choices of spectra and couplings; conversely, starting from
massless helicity amplitudes at high energies, we can ``unify" subsets
of these amplitudes into massive ones in some cases. This can be done
for spin 1 and spin 3/2 particles, representing the on-shell avatars
of the Higgs and super-Higgs mechanism, but we can see that gravity
can't be ``Higgsed" in this way.

In the context of this summary it is perhaps also worth briefly
describing the on-shell understanding of the most famous  general
consequences of relativistic quantum mechanics: the existence of
antiparticles and the spin-statistics connection.

The existence of antiparticles is essentially hardwired into
the on-shell formalism, since by fiat we are considering analytic
functions of Lorentz-invariant kinematical variables, with consistent
factorization on all possible channels. To be a little more
explicit on these ancient points,  we can ask how causality is encoded
in the $S$-matrix in any theory, with or without Lorentz invariance.
At tree-level, causality tells us that the amplitude can only have
simple poles as a function of energy variables. If the particles have
a dispersion relation of the form $E = \omega(\vec{p})$, the poles can
be either be of the form $1/(E + \omega(\vec{p}))$, or also $1/(E -
\omega(\vec{p}))$ if the interaction Hamiltonian allows particle
production. But in a Lorentz invariant theory, neither $(E +
\omega(\vec{p}))$ nor $(E - \omega(\vec{p}))$ are individually invariant, so
Lorentz invariance and causality forces us to have poles of the form
$\frac{1}{(E^2 - \omega(\vec{p})^2)} = \frac{1}{p^2 - m^2}$. This is how we see
that causality demands this familiar pole structure at tree-level,
which as a byproduct also forces the existence of non-zero amplitudes
for the production of degenerate particles and antiparticles. 

The on-shell understanding of the connection between spin and
statistics is slightly more interesting, and makes use of the universality of 
coupling to gravity. Indeed we
saw vividly that the structure of the four-particle amplitude for
gravi-compton scattering off particles of general mass and spin is completely fixed, and in particular,
forces the correct spin-statistics connection. This deeply relies on
the non-triviality of how residues in different channels are
consistent with each other, forcing the ``$s$" and ``$u$"
channels---related by particle interchange---to have fixed relative
signs. It is not surprising that an on-shell understanding of a
classic fact related to locality and unitarity should be related to
coupling to gravity---after all it is precisely the ability to ``weakly
gauge" gravity that gives a physical probe (via the existence of an
energy momentum tensor) of the locality of quantum field theory. We also described how other famous general results in field theory, such as the Weinberg-Witten and Coleman-Mandula theorems, are interpreted in directly on-shell terms.

Moving beyond tree-scattering, we also took some first steps for computing amplitudes at one-loop,
where the on-shell picture is especially powerful, as seen in the
speed and transparency of the computation for electron $(g-2)$ and the
QCD beta function. While not discussed in this paper, chiral
anomalies, together with the possibility of cancelling them via the
Green-Schwartz mechanism,  also have a beautiful on-shell
understanding, arising from the necessity to interpret poles in
one-loop amplitudes fixed by generalized unitarity \cite{yutindave}.

But of course, much more importantly than providing a conceptually
transparent and technically straightforward understanding of standard
results, we hope that the formalism introduced in this paper removes
the trivial barriers to exploring the new frontier of massive
scattering amplitudes, which is filled with fascinating physical
questions. We close by listing just a small number of these.

We have focused
almost entirely on the computation of tree-level three- and
four-particle amplitudes, so one completely obvious question is the
extension of e.g. BCFW recursion relations to any number of external
particles, especially for Higgsed Yang-Mills theories. Of course for
massless particles the BCFW shift must be performed for massless
particles of appropriate helicity in order to ensure the absence of
poles at infinity, so the obvious challenge is that the massive
amplitudes unify both the ``good" and ``bad" helicity combinations
into a single object.

Another clear goal is the systematic computation of all the  massive
amplitudes in the Standard Model, starting at tree level but moving to
multi-loop level. It is worth mentioning at least one exciting
motivation for this undertaking. Future Higgs factories---like the CEPC
or TLEP---can also run on the $Z$-pole, producing between $10^9-10^11$ $Z$
particles. Making full use of this data will require a computation of
$Z$-couplings at three to four loop accuracy. And unlike QCD calculations of
backgrounds at the LHC, for which the perturbative computations must
ultimately be convolved with non-perturbative information such PDF's
and hadron fragmentation functions to connect with experiment, these
precision electroweak calculations are unaffected by hadronic
uncertainties at the needed level of precision, so any theoretical
predictions can be unambiguously connected to exquisitely precise
experimental measurements!

It is also clearly of interest to investigate massive amplitudes in
supersymmetric theories, this should of course be
especially interesting in the context of the ${\cal N}=4$ SYM on the
Coulomb branch. Now even our first look at the on-shell avatar of the
Higgs and Super-Higgs mechanisms, showed that the Higgsed amplitudes
are {\it more} unified than their massless counterparts. Thus we
should expect that all the natural objects encountered for massless
amplitudes---such as tree amplitudes, leading singularities and
on-shell diagrams, which are separated into different ``$k$"
sectors---are somehow unified into more interesting objects. Amongst
other things the extension of BCFW to the Higgsed theories might be
most natural in the massive ${\cal N}=4$ on-shell diagram formulation.
And of course it would be fascinating to see if the
Grassmannian/Amplituhedron structures underlying the theory and the
origin of the moduli space is somehow extended/deformed away from the
origin.

All of the physics we have discussed in this paper has revolved around the consistency of long-distance physics: the on-shell focus on factorization and cuts at tree and loop level is meant to ensure that infrared singularities needed by locality and unitarity are correctly accounted for, and this fixes the structure of the amplitudes. For theories with growing amplitudes in the ultraviolet, needing a UV completion, it is very natural to ask the same questions: can the physics of UV completion also be determined from the consistency conditions of locality and unitarity? If the UV completion has a weak coupling, the question becomes perfectly sharply posed, and in the context of unitarizing the Fermi interaction or $WW$ scattering, searching for a tree-level UV completion correctly led to the prediction of massive W particles and Higgses as the completion of the weak interactions. Turning to the even more famous problem of UV completion for gravity scattering amplitudes, we encounter a well-known novelty. As will be discussed at greater length in \cite{stringpaper}, any weakly coupled UV completion for gravity amplitudes, (or for that matter, also Yang-Mills or $\phi^3$ theory, any theory with non-trivial three-particle amplitudes),  must involve an infinite tower of particles with infinitely increasing spins, as of course familiar from string theory. It is a tantalizing prospect to try and ``derive string theory" in this way, as giving the only possible consistent tree scattering amplitudes for gravitons coupled to the infinite tower of massive higher spin particles necessary for UV completion.  But consideration of amplitudes involving massive higher spin particles is necessary for any possible uniqueness, since as shown in \cite{stringpaper}, deformations of the string scattering amplitudes with only gravitons as external particles, compatible with all the standard rules,  have been identified. This is not at all surprising. Since we know the presence of gravity makes massless higher spin particles impossible, the coexistence of gravity unified with an infinite tower of massive higher spin particles must involve the strongest consistency conditions imaginable. Again, the massive amplitude formalism we have discussed in this paper trivializes kinematical issues so that important physics points can be studied with an unobstructed view, and with this in hand we will return to string theory and the challenge of UV completion in \cite{stringpaper}.

\section*{Acknowledgements} We are grateful to Zvi Bern, Rutger Boels, Nathaniel Craig, Lance Dixon and David Kosower for interesting
discussions and detailed comments on the draft. We also thank  Ming-Zhi Chung, Neil Christensen and  Alexander Ochirov for comments for the revised version of the draft. We would especially like to thank the referee's tremendous effort in carefully working through the manuscript, giving us valuable feed backs. The work of NAH is supported by the DOE under grant  DOE DE-SC0009988. Y-t Huang is supported by MoST grant 106-2628-M-002 -012 -MY3. Y-t Huang would like to thank the Institute for Advanced Study for its hospitality during the completion of this work. T.-C. Huang is supported by the Orr Fellowship.

\appendix

\newpage
\section{Conventions}\label{Conventions} 
In this paper we follow the mostly minus convention $(+,-,-,-)$, and an on-shell momentum satisfies $p^2=m^2$. The SL(2,C) and SU$_L$(2) indices are raised and lowered as
\eq
\psi_\alpha=\psi^\beta \varepsilon_{\alpha\beta},\quad \psi^\alpha=\varepsilon^{\alpha\beta}\psi_\beta,\quad \varepsilon^{\alpha\beta}\varepsilon_{\beta\gamma}=\delta^\alpha_\gamma 
\eqe
where we use $\varepsilon_{\alpha\beta}=-\varepsilon^{\alpha\beta}=\left(\begin{array}{cc}0 & -1 \\1 & 0\end{array}\right)$. The spinor contraction can be converted to vector contraction following 
\eq
p_1^{\alpha\dot\alpha}p_{2\alpha\dot\alpha}=2p_1^\mu p_{2\mu}\,,
\eqe
and hence for massive momenta, $p^{\alpha\dot\alpha}p_{\alpha\dot\alpha}=2m^2$. The vector indices are converted to spinorial ones as:
\eq
\gamma^\mu=\left(\begin{array}{cc}0 & \sigma^\mu_{\alpha \dot{\beta}} \\ \bar{\sigma}^{\mu\dot{\alpha} \beta} & 0\end{array}\right)\rightarrow (\displaystyle{\not}p+m)=\left(\begin{array}{cc} m \delta^{\beta}_\alpha & p_{\alpha\dot{\beta}} \\ p^{\dot{\alpha}\beta} & m\delta^{\dot{\alpha}}_{\dot{\beta}}\end{array}\right)
\eqe

\section{SU(2) Irreps as Symmetric Tensors}\label{SU(2)App} 
In this appendix we review, mostly to set notation, the elementary treatment of representations of $SU(2)$ as symmetric tensors, and briefly discuss some of its standard applictions, such as a transparent determination of spherical harmonics. 
The standard treatment of representations of $SU(2)$ is the one encountered by most undergraduates in beginning quantum mechanics courses. Since we can mutually diagonalize $\vec{J}^2$ and $J_z$, eigenstates of these operators are labeled by
 $|s,j_{\hat{z}}\rangle$, where the $\hat{z}$ reminds us that we have chosen to diagonalize the operator $J_z$, and we have $\vec{J}^2 |s,j_{\hat{z}}\rangle = s(s+1) |s,j_{\hat{z}} \rangle, J_z |s,j_{\hat{z}} \rangle = m |s,j_{\hat{z}}\rangle$. The irrep is $(2s+1)$ dimensional with $j_{\hat{z}}$ taking all the values $-s \leq j_{\hat{z}} \leq +s$. The spin information in a general state $|\psi \rangle$ is then entirely contained in specifying $\langle s,j_{\hat{z}}|\psi \rangle$. 

But for our purposes it is more convenient to describe an irrep of $SU(2)$ as a completely symmetric $SU(2)$ tensor with $2j$ indices:
\begin{equation}
\psi_{i_1 \cdots i_{2 s}}
\end{equation}
where $i$ is the $SU(2)$ index. The inner product $\langle \chi | \psi \rangle$ between two states is given by 
\begin{equation}
\langle \chi | \psi \rangle = \varepsilon^{\i_1 j_1} \cdots \varepsilon^{i_{2 s} j_{2s}} (\chi_{\i_1 \cdots i_{2s}})^* \psi_{j_1 \cdots j_{2s}}
\end{equation}
Saying that $\psi$ is an $SU(2)$ tensor is just the statement that the rotation generators $\vec{J}$ act as  
\begin{equation}
(\vec{J} \psi)_{i_1 \cdots i_{2s}} = (\frac{1}{2} \vec{\sigma})_{i_1}^{j_1} \psi_{j_1 \cdots i_{2s}} + \cdots  + (\frac{1}{2} \vec{\sigma})_{i_{2s}}^{j_{2s}} \psi_{i_1 \cdots j_{2s}}
\end{equation}
Note that the dimensionality of ths space is precisely $2\times3\times \cdots \times (2j+1)/(1 \times 2 \times \cdots \times 2 j) = (2j+1)$ as desired. 
Using that $\vec{\sigma}_{i}^j \cdot \vec{\sigma}_{k}^{l} = 2 \delta^{j}_{k} \delta^{l}_{i} - \delta^j_i \delta^l_k$, we trivially see that $(\vec{J}^2 \psi)_{i_1 \cdots i_{2s}} = s (s+1) \psi_{i_1 \cdots i_{2s}}$. If we choose to diagonalize $\sigma_z$ with eigenstates $(\sigma_z)_i^j \zeta^{\hat{z}, \pm}_j = \pm \zeta^{\hat{z},\pm}_i$, then the spin $s$ tensor that is an eigenstate of $J_z$ with eigenvalue $j_{\hat{z}}$ is 
\begin{equation}
\psi^{s,j_{\hat{z}}} = (\zeta^{\hat{z},+})^{s + j_{\hat{z}}} (\zeta^{\hat{z},-})^{j - j_{\hat{z}}} 
\end{equation}
where here and in what follows, since the tensor indices on $\psi$ are always symmetrized there is no need to write them explicitly when no confusion can arise. We can also express the same fact in a different way, telling us how to extract $\langle s,j_{\hat{z}}|\psi \rangle$ from the tensor $\psi_{i_1, \cdots, i_{2s}}$: 
\begin{equation}
\zeta_i \equiv \alpha_+ \zeta^{\hat{z},+}_i + \alpha_- \zeta^{\hat{z},-}_i; \, \zeta^{i_1} \cdots \zeta^{i_{2s}} \psi_{i_1 \cdots i_{2s}} = \sum_{j_{\hat{z}}} \alpha_+^{s + j_{\hat{z}}} \alpha_-^{s - j_{\hat{z}}} \langle s,j_{\hat{z}}|\psi \rangle
\end{equation}

The tensor representation makes it trivial to give explicit  expressions for finite rotations, and  expand the eigenstate $\psi^{s,j_{\hat{n}}}$ for a general direction $\hat{n}$ pointing in the usual $(\theta, \phi)$ direction, as a linear combination of $\psi^{s,j_{\hat{z}}}$'s.  We only need to know the relation for spin $1/2$:
\begin{equation}
\left(\begin{array}{c} \zeta^{\hat{n},+} \\ \zeta^{\hat{n},-} \end{array} \right) = \left(\begin{array}{cc} c  &  - s^* \\ s & c  \end{array} \right) \left(\begin{array}{c} \zeta^{\hat{z},+} \\ \zeta^{\hat{z},-} \end{array} \right)\, \, {\rm where} \, c \equiv {\rm cos} \frac{\theta}{2}, \, s \equiv {\rm sin} \frac{\theta}{2} e^{i \phi}
\end{equation}
We can then look at 
\begin{eqnarray}
\psi^{s,j_{\hat{n}}} &=& \left( \zeta^{\hat{n},+}\right)^{s + j_{\hat{n}}} \left(\zeta^{\hat{n},-}\right)^{s - j_{\hat{n}}} \nonumber \\ &=&  \left(c \zeta^{\hat{z},+} - s \zeta^{\hat{z},-} \right)^{s + j_{\hat{n}}}
\left(s^* \zeta^{\hat{z},+} + c \zeta^{\hat{z},-}\right)^{s - j_{\hat{n}}} \nonumber \\ &=& \sum_{j_{\hat{z}}} R^s_{j_{\hat{n}},j_{\hat{z}}}(\theta, \phi) \psi^{s,j_{\hat{z}}}
\end{eqnarray}
with 
\begin{equation}
R^s_{j_{\hat{n}},j_{\hat{z}}}(\theta,\phi) = \sum_{m_{\pm},m_+ + m_- = s + j_{\hat{z}}}  \left(\begin{array}{c} s + j_{\hat{n}} \\ m_+ \end{array} \right) \left(\begin{array}{c} s - j_{\hat{n}} \\ m_-\end{array} \right)  (c)^{m_+} (-s)^{s + j_{\hat{n}} - m_+} (c)^{s - j_{\hat{n}} - m_-} (s^*)^{ m_-}
\end{equation}
The tensor formalism also makes it trivial to construct spherical harmonics, which naturally arise in building irreps of $SU(2)$ which are polynomials in a 3-vector $\vec{x}$. Of course we are used to converting $\vec{x}$ to $SU(2)$ indices by dotting with the $\sigma$ matrices, but this gives us an object $\vec{\sigma}_i^j \cdot \vec{x}$ with an upstairs and downstairs index, while for the purposes of building irreps we would like to work with symmetric tensors and all downstairs indices. So it is natural to look instead at $x_{i j} = \epsilon_{i k} x^{k}_j$; explicitly we have 
\begin{equation}
x_i^j = \left(\begin{array}{cc} z & x - i y \\ x + i y & -z \end{array} \right), \, x_{i j} = \left(\begin{array}{cc} - (x - i y) & z \\ z & (x + i y) \end{array} \right)
\end{equation}
We would like to make symmetric rank $2 s$ tensors from a product of $s$ $x_{ij}$'s. But we don't need to do the symmetrizations explicitly; again because of the symmetrization all the information is contained in 
\begin{equation}
\zeta^{i_1} \zeta_{j_1} \cdots \zeta^{i_s} \zeta_{j_s} x_{i_1 j_1} \cdots x_{i_s j_s} = (\zeta \zeta x)^s 
\end{equation}
Putting $\zeta_i = (\alpha_+,\alpha_-)$ and so $\zeta^i = (-\alpha_-, \alpha_+)$, expanding the above gives us the generating function for spherical harmonics.  Letting $\vec{x}$ be the unit vector with $(x+iy) = {\rm sin} (\theta) e^{i \phi}$ and $z = {\rm cos}(\theta)$, we have 
\begin{equation}
(\zeta \zeta x)^s = \left(\alpha_+^2 {\rm sin} (\theta) e^{i \phi}  - 2 \alpha_+ \alpha_- {\rm cos}(\theta)  - \alpha_-^2 {\rm sin}(\theta) e^{-i \phi}\right)^s  
\equiv \sum_{j_{\hat{z}}} \alpha_+^{s+j_{\hat{z}}} \alpha_-^{s - j_{\hat{z}}} Y_{s, j_{\hat{z}}}(\theta,\phi) 
\end{equation}
\section{Explicit Kinematics} 
For massless particles, we have 
\begin{equation}
\lambda_{\alpha} = \sqrt{2 E} \left(\begin{array}{c} c \\ s \end{array} \right), \tilde{\lambda}_{\dot{\alpha}} = \sqrt{2 E} \left(\begin{array}{c} c \\ s^* \end{array} \right)
\end{equation}
For massive particles, we can write 
\begin{equation}
\lambda_{\alpha}^{I} = \left(\begin{array}{cc} \sqrt{E+p} c & - \sqrt{E - p} s^* \\ \sqrt{E+p} s & \sqrt{E- p} c \end{array} \right), \tilde{\lambda}_{I \dot{\alpha}} =  \left(\begin{array}{cc} \sqrt{E+p} c & - \sqrt{E - p} s^* \\ \sqrt{E+p} s & \sqrt{E- p} c \end{array} \right)
\end{equation}
We can write this equivalently as 
\begin{eqnarray}
\lambda_{\alpha}^I &=& \sqrt{E+p} \zeta^+_\alpha(p) \zeta^{- I}(k) + \sqrt{E-p} \zeta_\alpha^{-}(p) \zeta^{+ I}(k) \nonumber \\ 
\tilde{\lambda}_{\dot{\alpha}}^I &=& \sqrt{E+p} \tilde{\zeta}^{-}_{\dot{\alpha}}(p) \zeta^{+I}(k) + \sqrt{E-p} \tilde{\zeta}^{+}_{\dot{\alpha}}(p) \zeta^{-I}(k)
\end{eqnarray}
where 
\begin{equation}
\zeta^{+}_{\alpha} = \left(\begin{array}{c} c \\ s \end{array} \right),  \tilde{\zeta}^-_{\dot{\alpha}} =  \left(\begin{array}{c} c \\ s^* \end{array} \right); \, \, \zeta^{-}_{\alpha} = \left(\begin{array}{c} -s^* \\ c \end{array} \right), 
\tilde{\zeta}^+_{\dot{\alpha}} =  \left(\begin{array}{c} -s \\ c \end{array} \right)
\end{equation}
We can read off the specific spin components as in the previous appendix, since by using the above expressions for $\lambda_{\alpha}^I, \tilde{\lambda}_{\dot{\alpha}}^I$ we can expand for any particle: 
\begin{equation}
M^{\{I_1 \cdots I_{2S}\}} =\sum_{j_z} \left( (\zeta^+)^{S + j_z} (\zeta^-)^{S - j_z}\right)^{\{I_1 \cdots I_{2S}\}} M(j_z)
\end{equation}

\section{Comparison with Feynman Diagrams for Compton Scattering}\label{FeynDia} 
Here we directly construct Compton scattering from Feynman rules, and converting into our notations. We begin with
\eqa
\vcenter{\hbox{\includegraphics[scale=0.5]{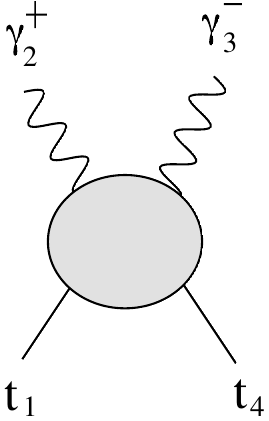}}}:\quad\epsilon_2^\mu\epsilon_3^\nu\bar{v}_1\left(\frac{\gamma^\nu(\displaystyle{\not}P_{12}+m)\gamma^\mu}{s-m^2}+\frac{\gamma^\mu(\displaystyle{\not}P_{13}+m)\gamma^\nu}{u-m^2}\right)u_4\,,
\eqae
where $P_{ij}=p_i+p_j$. Peeling off $u_4$ and $\bar{v}_1$, we obtain two $4\times 4$ numerator factor each given by:
\eq
n_s=\left(\begin{array}{cc} m \epsilon_{3\alpha\dot{\gamma}}\epsilon_{2}^{\dot{\gamma}\delta}& \quad\epsilon_{3\alpha\dot{\beta}}(P_{21})^{\dot{\beta}\gamma}\epsilon_{2\gamma\dot{\delta}} \\ \epsilon_{3}^{\dot{\alpha}\beta}(P_{12})_{ \beta\dot{\gamma}}\epsilon_2^{\dot{\gamma}\delta} &  m \epsilon_3^{\dot{\alpha}\gamma}\epsilon_{2\gamma\dot{\delta}}\end{array}\right),\quad n_u=\left(\begin{array}{cc} m \epsilon_{2\alpha\dot{\gamma}}\epsilon_{3}^{\dot{\gamma}\delta}& \quad\epsilon_{2\alpha\dot{\beta}}(P_{13})^{\dot{\beta}\gamma}\epsilon_{3\gamma\dot{\delta}} \\ \epsilon_{2}^{\dot{\alpha}\beta}(P_{13})_{\beta\dot{\gamma}}\epsilon_3^{\dot{\gamma}\delta} &  m \epsilon_2^{\dot{\alpha}\gamma}\epsilon_{3\gamma\dot{\delta}}\end{array}\right)\,.
\eqe
Substituting the explicit polarization vectors one finds:
\eq
n_s=\frac{\left(\begin{array}{cc} m \lambda_{3\alpha}[\tilde{q}2]q^\delta & \lambda_{4\alpha}[\tilde{q}|P_{12}|q\rangle\tilde{\lambda}_{2\dot{\delta}} \\ \tilde{q}^{\dot{\alpha}}\langle3|p_{1}|2]q^{\delta} & m\tilde{q}^{\dot{\alpha}}\langle1q\rangle\tilde{\lambda}_{4\dot{\delta}}\end{array}\right)}{\langle2q\rangle[3\tilde{q}]},\quad n_u=\frac{\left(\begin{array}{cc} m q_{\alpha}[2\tilde{q}]\lambda_3^\delta & q_{\alpha}[2|p_{1}|3\rangle\tilde{q}_{\dot{\delta}} \\ \tilde{\lambda}_2^{\dot{\alpha}}\langle q|P_{12}|\tilde{q}]\lambda_3^{\delta} & m\tilde{\lambda}_2^{\dot{\alpha}}\langle q3\rangle\tilde{q}_{\dot{\delta}}\end{array}\right)}{\langle2q\rangle[3\tilde{q}]}
\eqe
where $q,\tilde{q}$ are the reference spinors for the polarization vectors. The elements in the $4\times4$ matrix is in different SL(2,C) representations. We again judicially multiply factors of $p/m$ to convert it into our preferred basis, which has leg $1$ in the undotted basis, and leg $4$ in the dotted basis. That is, we mupltiply:
\eq
\left(\begin{array}{cc} \frac{p_4^{\dot{\alpha}\alpha}}{m} & \delta^{\dot{\alpha}}_{\dot{\beta}}\end{array}\right)\left(\begin{array}{cc}\mathcal{O}_\alpha\,^\delta & \mathcal{O}_{\alpha\dot{\delta}} \\ \mathcal{O}^{\dot{\beta}\delta} &  \mathcal{O}^{\dot{\beta}}\,_{\dot{\delta}}\end{array}\right)\left(\begin{array}{c}\delta^{\beta}_{\delta} \\ \frac{p_1^{\dot{\delta}\beta}}{m}\end{array}\right)
\eqe
where the $\mathcal{O}$s are stand ins for matrix elements of $n_s$, $n_u$. Summing up the terms and choosing $q=\lambda_3$ and $\tilde{q}=\tilde{\lambda}_2$, one finds:
\eq
\frac{\langle 3|p_1|2](\tilde{\lambda}_2^{\dot{\alpha}}\lambda_3^\beta-p_{4\alpha}^{\dot{\alpha}}\lambda_3^{\alpha} p_{1}^{\beta}\,_{\dot{\delta}}\tilde\lambda_{2}^{\dot\delta}/m^2)}{(u-m^2)(s-m^2)}\,.
\eqe
Contracting with external $\lambda,\tilde{\lambda}$s we recover eq.(\ref{GenSCom}).
\section{The High Energy Limit of Massive Three-Point Amplitude}
\label{HELimit}
Let us consider the HE limit of the three-point massive vector amplitude
\eq
\frac{gf^{abc}}{m_am_bm_c}\left[\langle\bold{1}\bold{2}\rangle [\bold{1}\bold{2}]\langle \bold{3}|p_1{-}p_2|\bold{3}]+\text{cyc.}\right]
\eqe

First consider the component amplitude ($1^-2^-3^+$). Its high enerly limit is given by: 
\eq\label{MHVHE}
\begin{gathered}
\begin{tikzpicture}[scale=0.5]
\draw[gluon] (-1, 1.73) node[left] {$2_b^-$}-- (0, 0);
\draw[gluon] (-1, -1.73) node[left]{$1_a^-$} -- (0, 0);
\draw[gluon] (0, 0) -- (2, 0)node[right] {$3_c^+$};
\end{tikzpicture}
\end{gathered}
\rightarrow \frac{gf^{abc}\left(\langle12\rangle [\tilde{\eta}_1\tilde{\eta}_{2}]\langle\eta_3|p_1{-}p_2|3]+\langle2\eta_3\rangle [\tilde\eta_23]\langle1|p_2{-}p_3|\tilde\eta_{1}]+\langle\eta_31\rangle [3\tilde\eta_{1}]\langle 2|p_3{-}p_1|\tilde\eta_{2}]\right)}{m_am_bm_c}
\eqe
Since in the high energy limit we will be interested in the MHV configuration, we have:
\begin{equation}
\tilde{\lambda}_1=\langle 23\rangle\tilde{\xi},\quad \tilde{\lambda}_2=\langle 31\rangle\tilde{\xi},\quad \tilde{\lambda}_3=\langle 12\rangle\tilde{\xi}\,,
\end{equation}
and eq.(\ref{MHVHE}) simplifies to:
\eqa
&&\frac{gf^{abc}\left(\langle2\eta_3\rangle [\tilde\eta_23]\langle1|p_2{-}p_3|\tilde\eta_{1}]+\langle\eta_31\rangle [3\tilde\eta_{1}]\langle 2|p_3{-}p_1|\tilde\eta_{2}]\right)}{m_am_bm_c}\nonumber\\
&=&2\frac{gf^{abc}}{m_c}\left(\frac{\langle\eta_32\rangle\langle12\rangle^2}{\langle23\rangle}+\frac{\langle\eta_31\rangle\langle12\rangle^2}{\langle31\rangle}\right)=2gf^{abc}\frac{\langle12\rangle^3}{\langle23\rangle\langle31\rangle}
\eqae
where we have repeatedly used identities such as $[\tilde\eta_13]=\frac{\langle12\rangle}{\langle23\rangle}[\tilde\eta_11]=m_a\frac{\langle12\rangle}{\langle23\rangle}$, which holds for MHV kinematics. 

A more interesting component would be ($1^02^-3^0$). Keeping in mind that extracting the longitudinal term correspond to choosing $\lambda^{\{I}\tilde\lambda^{J\}}\rightarrow \lambda\tilde\lambda{-}\eta\tilde\eta$, the relevant terms are:
\eqa\label{MHVHE2}
\begin{gathered}
\begin{tikzpicture}[scale=0.5]
\draw[gluon] (-1, 1.73) node[left] {$2_b^-$}-- (0, 0);
\draw[scalardashed] (-1, -1.73) node[left]{$1_a$} -- (0, 0);
\draw[scalardashed] (0, 0) -- (2, 0)node[right] {$3_c$};
\end{tikzpicture}
\end{gathered}
&\rightarrow& \frac{gf^{abc}}{m_am_bm_c}\bigg\{  \langle12\rangle[1\tilde\eta_2](\langle3\eta_1\rangle[\tilde\eta_13]-\langle3\eta_2\rangle[\tilde\eta_23]-\langle1\eta_3\rangle[\tilde\eta_31]+\langle2\eta_3\rangle[\tilde\eta_32])\nonumber\\
 &+&\langle23\rangle[\tilde\eta_23](\langle1\eta_2\rangle[\tilde\eta_21]-\langle1\eta_3\rangle[\tilde\eta_31]-\langle\eta_12\rangle[2\tilde\eta_1]+\langle\eta_13\rangle[3\tilde\eta_1])\nonumber\\
 &-&\left.\left(\frac{1}{2}\langle23\rangle[3\tilde\eta_2](\langle\eta_31\rangle[\tilde\eta_31]+\langle\eta_13\rangle[\tilde\eta_13])-\frac{1}{2}\langle21\rangle[1\tilde\eta_2](\langle\eta_31\rangle[\tilde\eta_31]+\langle\eta_13\rangle[\tilde\eta_13])\right)\right\}\,.\nonumber\\
\eqae
Substituting explicit representation for $[\tilde\eta_ij]$ for MHV kinematics, one finds:
\eq
\frac{gf^{abc}}{m_am_c}\frac{\langle12\rangle\langle23\rangle}{\langle31\rangle}\left(m_b^2-m^2_c-m^2_a\right)\,.
\eqe

\section{Examples for 1 Massive 3 Massless Amplitudes}
\label{FurtherExp}

For three-point amplitudes, since the all massless and one massive two massless amplitudes are unique, this tells us that the massless residue for the 1 massive 3 massless amplitude is unique. If the residue is non-local, then consistent factorization in the other channel may forces the theory to have a particular one massless two massive interaction. Here we present some examples.

We consider the four-point amplitude of arbitrary higher spin-$S$, two massless scalars and a graviton:
\eq\label{GeneralS}
M( \mathbf{1}^{S}2^03^{+2}4^0)\,.
\eqe  
We can now look at the massless residue for $s$-channel,
\eq
\vcenter{\hbox{\includegraphics[scale=0.5]{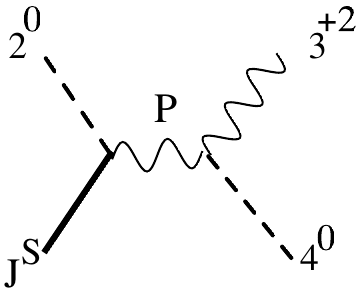}}}\quad \frac{(\lambda_2)^{S}(\lambda_P)^{S}[2P]^S}{m^{2S-1}}\times \frac{[3P]^2[34]^2}{[4P]^2M_{pl}}=\frac{(\lambda_2)^{S}([2|p_1)^{S-2}}{m^{2S-5}} \frac{[34]^2(\lambda_4)^2}{\langle23\rangle^2 M_{pl}}\,,
\eqe
where $M_{pl}$ is the Plank mass. Note that we have double poles $1/\langle23\rangle^2$, which is a general feature for couplings involving gravitons. The presence of double poles indicate that we have access to information in other channel. Let's start with $S=2$, dressing the residue with $1/s$ propagator, we find:
\eq\label{Spin2}
\frac{m}{M_{pl}}\frac{1}{s}(\lambda_2)^{2}(\lambda_4)^2 \frac{[34]^2}{\langle23\rangle^2}=\frac{m}{M_{pl}}\frac{(\lambda_2)^2(\lambda_4)^2[34][23]}{\langle32\rangle \langle43\rangle t}\quad\rightarrow\quad M( \mathbf{1}^{2}2^03^{+2}4^0)=\frac{m}{M_{pl}}\frac{(\lambda_2)^2(\lambda_2)^2[34][23]}{\langle32\rangle\langle34\rangle (u-m^2)}\,.
\eqe 
Note that the double pole has been converted into a $t$-channel massless and an $u$-channel massive pole $u-m^2$. The residue of the massive channel can be identified with $M_3(\mathbf{1}^{S=2}3^{+2}\mathbf{P}^{S=2})\times M_3(\mathbf{P}^{S=2}2^04^0)$, where $M_3(\mathbf{1}^{S=2}3^{+2}\mathbf{P}^{S=2})$ is the minimally coupling between a graviton and massive spin-2 states. Indeed using minimal coupling in the $u$-channel, we find the following residue:
\eq
\vcenter{\hbox{\includegraphics[scale=0.4]{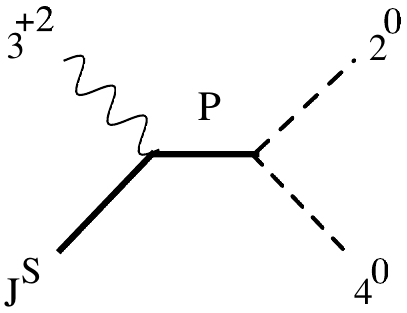}}}\quad x^2_{13}\times [24]^2(\lambda_2)^{2}(\lambda_4)^2\sim(\lambda_2)^2(\lambda_4)^2\frac{[34][23]}{\langle43\rangle\langle23\rangle}\,,
\eqe
which indeed matches that of eq.(\ref{Spin2}). This is a general feature for amplitudes of eq.(\ref{GeneralS}), consistent factorization will require the presence of a three point minimal coupling for graviton to two massive states. Consider $S=3$, the $s$-channel residue can be represented in a way that it can readily be completed:
\eq
(\lambda_2)^{S}([2|p_1)^{S-2} \frac{[34]^2(\lambda_4)^2}{\langle23\rangle^2}\bigg|_{\langle34\rangle=0}=(\lambda_2)^{3}(\lambda_4)^3\left(\frac{[34]^2[32]}{\langle23\rangle\langle24\rangle}-\frac{[42][34]^2[23]}{\langle23\rangle t}\right)\,,
\eqe  
Indeed putting back the $s$-channel propagator and writing $-t\rightarrow(u-m^2)$, we find the form factor given as:
\eq
M( \mathbf{1}^{3}2^03^{+2}4^0)=(\lambda_2)^{3}(\lambda_4)^3\left(\frac{[34][32]}{\langle23\rangle\langle24\rangle\langle43\rangle}+\frac{[42][34][23]}{\langle23\rangle \langle43\rangle(u-m^2)}\right)\,.
\eqe
It is not difficult to see that the massive residue of this amplitude contains the minimum coupling for the spin-3 states:
\eq
x^2_{13}(\lambda_2)^{3}(\lambda_4)^3[24]^3\sim(\lambda_2)^{3}(\lambda_4)^3\frac{[23][34][24]}{\langle23\rangle\langle43\rangle}\,.
\eqe

\end{document}